\documentclass[11pt]{article}
\usepackage[margin=1in]{geometry}
\usepackage{url, dsfont, dsfont}
\usepackage[round]{natbib}
\usepackage[colorlinks, bookmarksopen, bookmarksnumbered, citecolor=red, urlcolor=red]{hyperref}
\usepackage{verbatim}
\usepackage{xr}
\usepackage{xr-hyper}
\usepackage{xcolor}
\hypersetup{colorlinks=true, linkcolor={red}, citecolor={blue}, urlcolor={blue}}
\usepackage{soul}
\AtBeginDocument{\hypersetup{pdfborder={0 0 0}}}
\usepackage{tikz}
\usetikzlibrary{arrows.meta, positioning}

\usepackage{moreverb, subcaption, threeparttable, multicol, multirow, lscape, bm, soul, amsthm, amsfonts, amsmath, amssymb}
\usepackage[title]{appendix}
\usepackage{mathtools, color, xcolor}
\usepackage{tabularx, ltablex, floatrow, inputenc}
\usepackage[]{graphicx}
\usepackage{color, threeparttable, multirow, multicol, setspace}
\usepackage{bbm, caption, tikz, colortbl, dcolumn, float, enumitem, booktabs, makecell}

\providecommand{\bd}[1]{\boldsymbol{#1}}
\newcommand{\Ex}{\mathbb{E}}
\newcommand{\Vx}{\mathbb{V}}

\newcommand{\mc}{\mathcal}
\newcommand{\mb}{\mathbf}
\newcommand{\independent}{\perp\!\!\!\perp}

\newtheorem{theorem}{Theorem}
\newtheorem{proposition}[theorem]{Proposition}%
\newtheorem{assumption}{Assumption}
\newtheorem{remark}{Remark}

\providecommand{\keywords}[1]{\small\textbf{{Keywords}} #1}

\title{\bf Estimation and Inference for Win Measures with Multiple Ordinal Endpoints Subject to Missingness}

\author{Yi Liu$^{*1}$, Huiman Barnhart$^{2,3}$, Sean O'Brien$^{2,3}$, \\ Yuliya Lokhnygina$^{2,3}$, and Roland A. Matsouaka$^{2,3}$ 
\bigskip
\\
\normalsize 
$^{1}$Department of Statistics, North Carolina State University, Raleigh, NC, USA \\
\normalsize 
$^{2}$Duke Clinical Research Institute, Durham, NC, USA \\
\normalsize 
$^{3}$Department of Biostatistics and Bioinformatics, Duke University, Durham, NC, USA 
\bigskip
\\
\normalsize 
$^*$Corresponding author: Yi Liu, \texttt{yliu297@ncsu.edu}
}
 
\date{}

\begin{document}

\maketitle

\begin{abstract}

Win measures, including the win ratio, win odds, net benefit, and desirability of outcome ranking, are increasingly used in randomized clinical trials with multiple hierarchical ordinal endpoints. In practice, however, one or more  component endpoints may have missing data. The standard pairwise-comparison approach, which treats pairs with missing outcomes as ties, can  produce biased estimates, even if the data are missing completely at random. Although inverse probability of censoring weighting methods have been developed for censored survival endpoints, corresponding methods for addressing missing hierarchical ordinal endpoints are not yet available. To address this gap, we develop inverse probability weighted (IPW) and augmented IPW (AIPW) estimators for win measures with hierarchical ordinal endpoints subject to missing data, allowing missingness to depend on treatment assignment and baseline covariates. The IPW estimator corrects bias by reweighting complete observed outcomes using joint non-missingness probabilities involved in estimating the joint cell probabilities that define the win measures. The AIPW estimator additionally incorporates outcome modeling, potentially improving estimation precision and achieving double robustness. For inference, we derive closed-form variance estimators for both methods based on influence functions. Simulation studies show that the standard approach can be substantially biased, whereas the proposed IPW and AIPW estimators exhibit small bias and near-nominal coverage under their respective identifying and model conditions. Applications to the SCOUT-CAP and ACTT-1 trials illustrate the practical utility of the proposed methods. An R package \texttt{WinMO} is provided for implementation. 

\end{abstract} \hspace{10pt}


\newpage
\doublespacing 

\section{Introduction}\label{sec:intro}
Composite endpoints are widely used in randomized clinical trials to assess treatment effect while capturing multiple clinically relevant endpoints within a single analysis. Combining different endpoints into a single composite endpoint using, for instance, the time-to-first-event approaches can increase event rates, reduce the required sample size, and bypass the need for multiple comparison adjustments \citep{nascimento2026making}. However, this approach can be limiting. Analysis of time-to-first-event endpoints only considers time-to-event endpoints; it weights them all equally (or assigns equal priority) and focuses on the first occurring event while ignoring any subsequent events \citep{sankoh2014use, siquier2025use}. In many clinical trials, endpoints are inherently multivariate and may follow a natural priority structure, based on clinical importance or severity \citep{overbey2025navigating}. For example, a study may consider that mortality is more clinically important than hospitalization (or severe adverse) events which, in return, is considered more important than symptom improvement. Hence, the  win measures (or win statistics) provide an appealing and natural tool to evaluate treatment effects according to a pre-specified hierarchy of endpoints, congruent with clinical priorities and can accommodate all types of endpoints \citep{pocock2024win, barnhart2025trial}. 

Commonly used win measures include the win ratio (WR) \citep{pocock2012win} and win odds (WO) \citep{song2023win, dong2023win, dong2026win}, which quantify the treatment effect on the relative scale, as well as the net benefit (NB), which summarizes the treatment effect on the additive scale. Another related metric is the desirability of outcome ranking (DOOR), originally proposed by \cite{evans2016using} for a single ordinal endpoint and later extended by \cite{barnhart2025sample} to settings with multiple endpoints.  

Despite the increased adoption of win measures in clinical studies \citep{hamasaki2024design, zheng2024use, shu2026desirability}, valid and efficient estimation of win measures often faces the recurring challenge of coarse data \citep{heitjan1993ignorability}, especially missing data. In practice, one or more endpoint components are often missing because of intermittent assessments, loss to follow-up, protocol deviations, or noncompliance. Missing data may occur at different levels of the  hierarchy of endpoints, complicating the pairwise comparison procedure that underlies standard win measure analyses \citep{buyse2010generalized}. The standard approach, based on $U$-statistic theory for inference, estimates win, loss, and tie probabilities through pairwise comparisons, where each pair comprises a treated and control participants. In the standard approach, if there are missing data on a particular endpoint when comparing a pair of subjects, the comparison is often treated as a tie \citep{Buyse2025GPC}, even though its true status may be different, had the missing data been observed. As a result, the pairwise-comparison estimator of win measures can be biased even if data are missing completely at random (MCAR), the most strict missingness mechanism, because the bias is induced by systematically classifying comparisons with missing data as ties. As illustrated by an example in \cite{li2024elusiveness}, this standard approach can substantially bias estimation of the WR under MCAR and, more fundamentally, bias estimation of the win, loss, and tie probabilities. Consequently, other win measures that are functions of these probabilities are also expected to be biased. 

Another strategy sometimes used in practice is complete-case analysis \citep{cui2025wins}. With complete-case analysis, any participant with missingness in any one component of the hierarchical endpoints are removed completely from the analyses, regardless of whether there are complete data in other component endpoints. Then pairwise comparisons are conducted only among the remaining participants with complete data in all endpoints. The complete-case analysis is valid only under the MCAR assumption \citep{rubin1976inference, wang2023missing}. This requirement is often implausible in clinical studies, where missingness may depend on treatment assignment and baseline covariates. Moreover, by discarding partially observed participants, complete-case analysis reduces the effective sample size substantially and can compromise estimation efficiency.

Motivated by the issue of missing data and the limitations of current standard approach, we study a setting that, to our knowledge, has not been considered in prior work: estimation and inference for a broad class of win measures, including WR, WO, NB, and DOOR, with multiple hierarchical ordinal (or binary) endpoints subject to missingness where the missing data mechanism may depend on treatment assignment or baseline covariates. We propose an inverse probability weighted (IPW) estimator that draws on ideas from causal inference and missing data analysis for a single outcome \citep{rubin1974estimating}. {Building upon the IPW approach, we also propose a doubly robust augmented IPW (AIPW) estimator, which incorporates additional outcome models based on treatment and covariates to improve robustness and make more efficient use of available covariate information.} The proposed framework naturally extends to any measure that can be expressed as a smooth function of the win, loss, and tie probabilities. 



Several methods have been recently developed to address the issue of missing data in win measures. \cite{dong2020inverse} proposed inverse probability censoring weighting (IPCW) for WR estimation under non-informative censoring. \cite{dong2021adjusting} extended the method to covariate-adjusted IPCW (CovIPCW) to account for covariate-dependent censoring. However, both IPCW and CovIPCW are applicable only to time-to-event endpoints and cannot be used for other types of endpoints. \cite{wang2025adjusted} proposed using IPW-adjusted win measures, but focused on covariate imbalance rather than missing data.  \cite{wang2025restricted} developed a parametric model-based multiple imputation approach, but considered settings with right-censored endpoint(s) evaluated up to a prespecified and finite time horizon. \cite{liu2026estimation} proposed a nonparametric maximum likelihood estimator (called the ``$\mc S$-score method'') for two hierarchical endpoints under censoring and missingness, but this framework is restricted to settings with a survival endpoint in the first hierarchy and a non-survival endpoint in the second. \cite{shu2026doubly} proposed a doubly robust AIPW estimator for DOOR under a single ordinal score, but did not consider other win measures, multiple ordinal endpoints or the presence of missing data. \cite{cao2025covariate} developed IPW, overlap-weighted, and augmented estimators for win measures, but focused only on a single ordinal endpoint and did not address the impact of or issues related to missing data.

Different from these approaches, our key contributions are outlined as follows. First, our IPW estimator reweights observed joint endpoints by the propensity of non-missingness. We achieve this by decomposing the pairwise win, loss, and tie probabilities into products of joint cell probabilities across endpoint components. The decomposition reduces the original pairwise comparison problem into estimation of a collection of cell probabilities at each layer of the endpoint hierarchy. We then plug-in the IPW-estimated cell probabilities to estimate the target win measures, since they are defined as smooth functions of cell probabilities. The IPW estimator can be interpreted as a a weighted observed-case estimator at each hierarchy level, in which individualized weights  are assigned to observed outcomes so that the resulting reweighted comparisons remove the bias in the unweighted estimator. 

To further improve efficiency and robustness, our second contribution is the doubly robust AIPW estimator based on outcome-model augmentation and semiparametric efficiency theory \citep{robins1994estimation, tsiatis2007semiparametric, kennedy2016semiparametric}. This estimator augments the IPW estimator with a treatment-specific outcome regression model based on baseline covariates and applies the resulting predictions to all participants. Because baseline covariates and treatment assignment are assumed to be fully observed, the approach enables prediction of hierarchical outcomes for all participants. Although the AIPW estimator requires specification of an additional outcome model, we show that it remains consistent if either the missingness model or the outcome model is correctly specified, thereby enjoying the double robustness property. Moreover, when both nuisance models are correctly specified, the AIPW estimator attains the semiparametric efficiency bound for each joint cell probability under the corresponding reduced observed-data model described in Appendix \ref{subapp:theory-aipw}.

For statistical inference, we develop influence-function-based \citep{tsiatis2007semiparametric} asymptotic variance estimators for both the IPW and AIPW methods. These variance estimators account for the three sources of uncertainty that are related to the  estimation of the win measures: the postulated nuisance-function used to determine the parameters of the propensity score and outcome regression models as well as Taylor's expansion with respect to cell probabilities to estimate the win, loss, and tie probabilities that define the win measures. The resulting inference procedures are valid, computationally stable, and avoid intensive resampling. A user-friendly R package \texttt{WinMO} is available (\url{https://github.com/yiliu1998/WinMO}) to implement our proposed methods for up to three hierarchical ordinal endpoints.

The remainder of the paper is organized as follows. Section~\ref{sec:method} introduces preliminaries of win measures and the proposed IPW and AIPW estimators with their inference procedures. Section~\ref{sec:simu} reports simulation studies assessing our proposed methods versus standard approach where comparisons with missing components are treated as ties. Section~\ref{sec:data} presents real data applications from two real clinical trials: SCOUT-CAP and ACTT-1. Finally, Section \ref{sec:concludes} concludes the paper with comments and remarks.  

\section{Methodology}\label{sec:method}

\subsection{Problem set-up and assumptions}\label{subsec:setup}

We consider a randomized clinical trial with a binary treatment $A \in \{0,1\}$, where $A=1$ denotes the active treatment and $A=0$ the control, standard care, or placebo treatment. Data are collected on participants' baseline covariates $X_1,\dots,X_p$ and a pre-specified hierarchy of $K$ post-randomization ordinal outcomes $\mb Y =(Y_1,\dots,Y_K)$, ordered from highest to lowest clinical priority, where $p\ge 0$ and $K\geq 1$ are positive integers.

For each $k=1,\dots,K$ and $a=0,1$, we assume the support of $Y_k$ is $\{1,\dots,\ell_k\}$ for some integer $\ell_k\geq 1$, and that larger values of $Y_k$ correspond to better outcomes. 
Each outcome $Y_k$ is subject to missingness, i.e., we only observe the coarsened information $(R_k, R_kY_k)$, where $R_{k} = I(Y_{k}\text{ is observed})$ to indicate the outcome $Y_k$ for $k=1,\dots,K$ is observed for a participant, where $I(\cdot)$ is the indicator function. We also denote $\mb X = (1,X_1,\dots,X_p)^\top$ a vector that comprises of a constant term 1 and  $p$ baseline observed covariates ($p>0$). Together, we have observations from an independently and identically distributed (i.i.d.) population, denoted by $\mc O = \{(\mb X_i, A_i, R_{1i}, R_{1i}Y_{1i},\dots, R_{Ki}, R_{Ki}Y_{Ki})\}_{i=1}^n$.


Let $\widetilde R_{1:k} = I(R_1=\dots=R_k=1)$ denote joint non-missingness of the first $k$ endpoints, for any $k=1,\dots,K$. {This cumulative indicator is defined by the hierarchical structure of the estimand and does not require the individual missingness indicators $(R_1,\dots,R_K)$ to follow a monotone missingness pattern. Thus, nonmonotone patterns such as $R_1=1$, $R_2=0$, and $R_3=1$ are allowed.} We make the following identification assumption.

\begin{assumption}[Missingness mechanism and positivity]\label{asp:miss-pos}
    For each treatment group $a=0,1$ and all $k=1,\dots,K$, the following conditions hold:
    \begin{enumerate}[label=(\roman*)]
        \item Conditional on treatment and covariates, any first $k$ joint non-missingness indicators are independent of the corresponding outcomes:
        $\widetilde R_{1:k} \independent (Y_1,\dots,Y_k)\mid(A=a,\mb X)$.
        \item There exists a positive constant $\eta < \infty$ such that
        $\pi_k(a,\mb X) = \Pr(\widetilde R_{1:k}=1 \mid A=a,\mb X) > 1/\eta,$
        for almost all $\mb X$.
    \end{enumerate}
\end{assumption}

In other words, Assumption~\ref{asp:miss-pos}(i) states that, for each $k$, joint non-missingness of the first $k$ endpoints is conditionally independent of these endpoints given treatment and covariates, while allowing the missingness mechanism to vary across $(A,\mb X)$. Although $(R_1,\dots,R_K)\in\{0,1\}^K$ gives rise to $2^K$ possible missingness patterns, it is sufficient to assume conditional independence only for the indicators of complete observations up to the $k$th endpoint, $\widetilde R_{1:k}$, for each $k=1,\dots,K$. This formulation is relatively parsimonious while still ensuring identification and validity of the proposed estimators in Sections~\ref{subsec:ipw} and~\ref{subsec:outaug}. The Assumption~\ref{asp:miss-pos}(i) is a special type of missing-at-random (MAR) assumption. Under the usual MAR framework, the missingness pattern for an outcome may depend on all observed information, including other outcomes on the hierarchy, if observed. In contrast, our assumption restricts the missingness mechanism to depend only on treatment and baseline covariates, and not on partially observed outcome information. It is less restrictive than MCAR but more restrictive than the general MAR assumption. 

Assumption~\ref{asp:miss-pos}(ii) imposes a mild positivity condition requiring the probability of jointly observing the first $k$ endpoints to be bounded away from zero. This prevents the IPW weights from becoming arbitrarily large at any hierarchical level. In finite samples, positivity violations may lead to unstable IPW estimators when some endpoint components are rarely observed. In practice, this assumption is reasonable as long as the trial does not have extremely low observation probabilities for any endpoint categories.

\subsection{Target win measures for hierarchical ordinal endpoints}\label{subsec:measures}

For each treatment group $a\in\{0,1\}$, define the joint cell probabilities
$$
p_a(i_1,\dots,i_K)=\Pr(Y_1=i_1,\dots,Y_K=i_K\mid A=a),\qquad i_k\in\{1,\dots,\ell_k\},
$$
with total mass $\sum_{i_1=1}^{\ell_1}\cdots\sum_{i_K=1}^{\ell_K}p_a(i_1,\dots,i_K)=1$.  
To simplify the presentation, define the treatment-specific marginal joint probabilities for the first $k$ endpoints: 
$$
P_a^{1:k}(i_1,\dots,i_k)=\sum_{i_{k+1}=1}^{\ell_{k+1}}\cdots\sum_{i_K=1}^{\ell_K} p_a(i_1,\dots,i_K) = \Pr(Y_1=i_1,\ldots,Y_k=i_k\mid A=a),\qquad k=1,\dots,K-1. 
$$
In other words, $P_a^{1:k}(i_1,\dots,i_k)$ fixes the first $k$ coordinates $(i_1,\dots,i_k)$ and sums over the supports of the remaining endpoints $(Y_{k+1},\dots,Y_K)$, so that it represents the joint probability of observing $(Y_1=i_1,\dots,Y_k=i_k)$ in treatment group $a$. 
Note that this notation implies two special cases: $P_a^1(i_1)=\sum_{i_2,\dots,i_K}p_a(i_1,\dots,i_K)$ and $P_a^{1:K}(i_1,\dots,i_K)=p_a(i_1,\dots,i_K)$.

The probability that a randomly chosen treated subject \textit{wins} against a randomly chosen control subject is obtained sequentially across the pre-specified hierarchy of endpoints. A comparison is first attempted at $Y_1$; if it is tied, the comparison moves to $Y_2$, and so on until $Y_K$. We denote the probability of win, loss, and tie by $p_W$, $p_L$ and $p_T$, respectively. 

Then, the probabilities of win, loss and tie can be expressed as follows,
\begin{align}\label{eq:win-pr}
p_W & = \underbrace{\sum_{i_1>i_1'} P_1^{1}(i_1)P_0^{1}(i_1')}_{\text{decided at }Y_1}
+ \underbrace{\sum_{i_1}\sum_{i_2>i_2'} P_1^{1:2}(i_1,i_2) P_0^{1:2}(i_1,i_2')}_{\text{decided at }Y_2}
+ \nonumber \\
& \quad \cdots
+ \underbrace{\sum_{i_1=\dots=i_{K-1}}\sum_{i_K>i_K'} P_1^{1:K}(i_1,\dots,i_K) P_0^{1:K}(i_1,\dots,i_K')}_{\text{decided at }Y_K}.
\end{align}
{In each term beyond the first endpoint, the comparison proceeds to endpoint $Y_k$ only for pairs that are tied on all preceding endpoints $Y_1,\ldots,Y_{k-1}$. Accordingly, the indices $i_1,\ldots,i_{k-1}$ represent the common endpoint values shared by the treated and control subjects on these preceding endpoints, rather than arbitrary combinations of values across the two subjects.}

The probability of a loss is obtained by reversing all inequalities in the above expression, i.e., 
\begin{align}\label{eq:loss-pr}
p_L & = {\sum_{i_1<i_1'} P_1^{1}(i_1)P_0^{1}(i_1')}
+ {\sum_{i_1}\sum_{i_2<i_2'} P_1^{1:2}(i_1,i_2)P_0^{1:2}(i_1,i_2')}
+ \nonumber\\
& \quad \cdots
+ {\sum_{i_1=\cdots=i_{K-1}}\sum_{i_K<i_K'} P_1^{1:K}(i_1,\dots,i_K) P_0^{1:K}(i_1,\dots,i_K')}.
\end{align}
The tie probability is then
\begin{align}\label{eq:tie-pr}
p_T = 1-p_W-p_L. 
\end{align}
Following \cite{barnhart2025sample}, the four different types of win measures are defined as
\begin{align}\label{eq:winstats}
   \text{Win ratio (WR):} \qquad &\theta_{\text{WR}} = \frac{p_W}{p_L};\nonumber\\
   \text{Net benefit (NB):} \qquad &\theta_{\text{NB}} = p_W - p_L;\nonumber\\
   \text{Win odds (WO):} \qquad &\theta_{\text{WO}} = \frac{p_W+0.5p_T}{p_L + 0.5p_T};\quad \text{and}\nonumber\\
   \text{Desirability of outcome ranking (DOOR):}\qquad & \theta_{\text{DOOR}} = p_W+0.5p_T. 
\end{align}

\subsection{Existing standard estimation approach}\label{subsec:counting}

When all endpoints are fully observed, the probabilities of win, loss, and tie can be estimated without bias by comparing outcomes between all possible pairs of participants in the treatment and control groups. Specifically, suppose there are $n_1$ participants on treatment ($A=1$) and $n_0$ participants on control ($A=0$). For each pair $(i,j)$, where $i$ indexes a treated patient and $j$ indexes a control patient, their outcome vectors are compared hierarchically. If the treated patient is better on the first endpoint, the pair is counted as a win; if the control patient is better, it is counted as a loss; if they are equal, the comparison moves to the next endpoint. If all endpoints are equal, the pair is counted as a tie.

Define the pairwise score
$$
C_{ij} =
\begin{cases}
1, & \text{if the treated patient wins}, \\
-1, & \text{if the control patient wins}, \\
0, & \text{if the outcomes are tied}.
\end{cases}
$$

The total number of wins and losses across all pairs are then
$$
W = \sum_{i=1}^{n_1}\sum_{j=1}^{n_0} I(C_{ij}=1), \quad
L = \sum_{i=1}^{n_1}\sum_{j=1}^{n_0} I(C_{ij}=-1),
$$
and the number of ties is $T = n_1 n_0 - W - L$.

Dividing by the total number of pairs gives the estimated probabilities of win, loss, and tie:
$$
\widehat{p}_W = \frac{W}{n_1 n_0}, \quad
\widehat{p}_L = \frac{L}{n_1 n_0}, \quad
\widehat{p}_T = \frac{T}{n_1 n_0}.
$$

Based on these estimated probabilities, the corresponding estimators for the four win-based measures are given by
\begin{align*}
\widehat{\theta}_{\text{WR}} &= \frac{\widehat{p}_W}{\widehat{p}_L}, \\
\widehat{\theta}_{\text{NB}} &= \widehat{p}_W - \widehat{p}_L, \\
\widehat{\theta}_{\text{WO}} &= \frac{\widehat{p}_W + 0.5\widehat{p}_T}{\widehat{p}_L + 0.5\widehat{p}_T}, \\
\widehat{\theta}_{\text{DOOR}} &= \widehat{p}_W + 0.5\widehat{p}_T.
\end{align*}

This pairwise comparison method is simple to implement and uses the entire dataset by forming all possible treated-control comparisons, thereby providing consistent estimates of win measures when data are complete. 

\subsection{Proposed inverse probability weighted estimator}\label{subsec:ipw}

In the presence of missing outcome data, pairwise comparisons can be biased if comparisons involving missing endpoint information are treated as ties. To address bias arising from missingness, we propose an IPW estimator based on the joint cell probabilities introduced in Section~\ref{subsec:measures}. 

We first model and estimate the propensity scores $\pi_k(a,\mb X)$, $k=1,\dots,K$, for the non-missingness indicators $\widetilde R_{1:k} = I(R_1=\dots=R_k=1)$ of the first $k$ endpoints, given $A=a$ and $\mb X$,  with $\widetilde R_{1:k}$ as response. The subjects are classified only according to whether the first $k$ endpoints are all observed. Consequently, subjects with missingness patterns such as $Y_1$ missing, $Y_2$ observed, and $Y_3$ missing, and those with $Y_1$ missing, $Y_2$ missing, and $Y_3$ observed, are treated identically since both yield $\widetilde R_{1:k}=0$ for $k\geq 3$.

A practical choice for the missingness propensity score model is the logistic regression model,
\[
\pi_k(a,\mb X; \bd\beta_k^{(a)})
=
\{1+\exp(-\mb X^\top\bd\beta_k^{(a)})\}^{-1},
\]
where $\bd\beta_k^{(a)}$ denotes the regression coefficients for treatment group $A=a$, estimated by maximum likelihood. Based on the fitted propensity scores, the Horvitz--Thompson (HT)-type IPW weight \citep{horvitz1952generalization} is defined by
\begin{align*}
    \omega_k(a,\mb X,\widetilde R_{1:k})
    =
    \frac{I(A=a)}{\Pr(A=a)}
    \cdot
    \frac{\widetilde R_{1:k}}{\pi_k(a,\mb X)},
    \qquad
    a=0,1,\quad k=1,\dots,K.
\end{align*}
In practice, this weight is estimated by plugging in the fitted model parameters:
\[
\widehat\omega_k(a,\mb X,\widetilde R_{1:k})
=
\frac{I(A=a)}{\widehat{\Pr}(A=a)}
\cdot
\frac{\widetilde R_{1:k}}
{\pi_k(a,\mb X;\widehat{\bd\beta}_k^{(a)})},
\qquad
\widehat{\Pr}(A=a)=n_a/n,
\]
where $n_a=\sum_{j=1}^n I(A_j=a)$. The corresponding HT-IPW estimator of a joint cell probability is
\[
\widehat P_a^{1:k,\mathrm{HT}}(i_1,\dots,i_k)
=
\frac{1}{n}
\sum_{j=1}^n
\widehat\omega_k
(a,\mb X_j,\widetilde R_{j,1:k})
I(Y_{1j}=i_1,\dots,Y_{kj}=i_k).
\]
{Although the HT-IPW estimator is consistent under correct specification of the non-missingness model, its estimated cell probabilities need not sum exactly to one within a treatment group and hierarchy level in finite samples. We therefore pursue a H\'ajek-type normalization for the proposed IPW estimator \citep{hajek1971comment}. Define}
\[
\widehat\omega_{k,\text{H\'ajek}}(a,\mb X_i,\widetilde R_{i,1:k})
=
\frac{
\widehat\omega_k(a,\mb X_i,\widetilde R_{i,1:k})
}{
n^{-1}\sum_{j=1}^n
\widehat\omega_k(a,\mb X_j,\widetilde R_{j,1:k})
}.
\]
{Using these normalized weights, our proposed IPW estimator of
$P_a^{1:k}(i_1,\dots,i_k)$ is given by
\begin{align}
\widehat P_a^{1:k,\mathrm{ipw}}(i_1,\dots,i_k)
&=\frac{1}{n}\sum_{j=1}^n\widehat\omega_{k,\text{H\'ajek}}
(a,\mb X_j,\widetilde R_{j,1:k})
I(Y_{1j}=i_1,\dots,Y_{kj}=i_k).
\label{eq:M1k-ipw-est}
\end{align}

Under correct specification of the non-missingness model and the regularity conditions given in Appendix~\ref{subapp:regularity}, our proposed
$\widehat P_a^{1:k,\mathrm{ipw}}(i_1,\dots,i_k)$
is consistent for
$P_a^{1:k}(i_1,\dots,i_k)$.
The formal consistency and influence-function results for the HT and H\'ajek IPW estimators are provided in Appendices \ref{subapp:theory-ipw-HT} and \ref{subapp:theory-ipw-Hajek}, respectively. 

The H\'ajek normalization improves finite-sample coherence within each treatment group and hierarchy level because the estimated cell probabilities are nonnegative and sum to one. However, we clarify that since the non-missingness models
$\pi_1(a,\mb X),\dots,\pi_K(a,\mb X)$
are fitted separately across hierarchy levels, the resulting estimates need not satisfy exact marginalization identities across levels in finite samples. Likewise, separate model fitting does not necessarily enforce the monotonicity $\widehat\pi_1(a,\mb X) \geq \cdots \geq \widehat\pi_K(a,\mb X)$. These coherence and monotonicity properties hold for the corresponding population quantities and are recovered asymptotically under correct model specification and the stated regularity conditions, although finite-sample departures may occur. We assess these properties through diagnostic analyses empirically in our simulations and data analyses. 

\begin{remark}
    A possible way to enforce these properties in finite samples is to use sequential factorizations of the non-missingness and outcome distributions. In particular, the cumulative non-missingness probability $\pi_k(a,\mb X)$ can be expressed as a product of conditional observation probabilities across hierarchy levels, which guarantees $\pi_1(a,\mb X)\geq\cdots\geq\pi_K(a,\mb X)$ by construction. Similarly, the joint outcome distribution can be factorized as $P(Y_1)\prod_{j=2}^K P(Y_j\mid Y_1,\ldots,Y_{j-1})$, thereby enforcing exact marginal coherence across levels and potentially improving efficiency in some settings by retaining all participants informative for earlier endpoints. Such sequential constructions, however, require additional conditional models and may be less stable when some outcome histories are sparse. In contrast, the proposed estimators remain consistent under correct specification of the working models, and the corresponding coherence and monotonicity properties are recovered asymptotically. We therefore view sequential factorization as a useful alternative for enforcing finite-sample constraints rather than a requirement for validity of the proposed approach, and leave a systematic comparison of these constructions to future work. 
\end{remark} 
}

The proposed estimators of the win and loss probabilities are obtained by substituting the {H\'ajek IPW cell-probability estimators} into \eqref{eq:win-pr} and \eqref{eq:loss-pr}, respectively. Specifically,
\begin{align}
\widehat p_W^{\mathrm{ipw}}
&=
\sum_{i_1>i_1'}
\widehat P_1^{1,\mathrm{ipw}}(i_1)
\widehat P_0^{1,\mathrm{ipw}}(i_1')
\nonumber\\
&\quad+
\sum_{i_1}\sum_{i_2>i_2'}
\widehat P_1^{1:2,\mathrm{ipw}}(i_1,i_2)
\widehat P_0^{1:2,\mathrm{ipw}}(i_1,i_2')
+\cdots
\nonumber\\
&\quad+
\sum_{i_1=\cdots=i_{K-1}}
\sum_{i_K>i_K'}
\widehat P_1^{1:K,\mathrm{ipw}}(i_1,\dots,i_K)
\widehat P_0^{1:K,\mathrm{ipw}}(i_1,\dots,i_K'),
\label{eq:win-pr-ipw}
\end{align}
and
\begin{align}
\widehat p_L^{\mathrm{ipw}}
&=
\sum_{i_1<i_1'}
\widehat P_1^{1,\mathrm{ipw}}(i_1)
\widehat P_0^{1,\mathrm{ipw}}(i_1')
\nonumber\\
&\quad+
\sum_{i_1}\sum_{i_2<i_2'}
\widehat P_1^{1:2,\mathrm{ipw}}(i_1,i_2)
\widehat P_0^{1:2,\mathrm{ipw}}(i_1,i_2')
+\cdots
\nonumber\\
&\quad+
\sum_{i_1=\cdots=i_{K-1}}
\sum_{i_K<i_K'}
\widehat P_1^{1:K,\mathrm{ipw}}(i_1,\dots,i_K)
\widehat P_0^{1:K,\mathrm{ipw}}(i_1,\dots,i_K').
\label{eq:loss-pr-ipw}
\end{align}
The tie probability is estimated by
\begin{align}
\widehat p_T^{\mathrm{ipw}}
=
1-\widehat p_W^{\mathrm{ipw}}
-\widehat p_L^{\mathrm{ipw}}.
\label{eq:tie-pr-ipw}
\end{align}
The four win measures are then estimated by substituting
$(\widehat p_W^{\mathrm{ipw}},
\widehat p_L^{\mathrm{ipw}},
\widehat p_T^{\mathrm{ipw}})$
into \eqref{eq:winstats}.

{The proposed IPW estimator does not require a monotone missingness pattern in the usual longitudinal sense. At hierarchy level $k$, a participant contributes whenever the first $k$ endpoints are jointly observed, as indicated by $\widetilde R_{1:k}=I(R_1=\cdots=R_k=1)$. For example, if $Y_1$ and $Y_3$ are observed but $Y_2$ is missing, the participant contributes to estimation of $P_a^1(i_1)$ but not to joint cell probabilities involving $(Y_1,Y_2)$ or $(Y_1,Y_2,Y_3)$. Thus, nonmonotone missingness patterns are allowed, although an observed lower-priority endpoint is not used at a hierarchy level when any preceding endpoint required for that level is missing.}

Compared with a conventional complete-case analysis, which discards any participant with missingness in any component endpoint, the proposed IPW estimator retains partially observed participants whenever they remain informative for a given hierarchy level. For instance, a participant with $Y_1$ observed but $Y_2$ missing can still contribute to estimation of $P_a^1(i_1)$, even though that participant does not contribute to joint cell probabilities involving both $Y_1$ and $Y_2$. In this sense, the proposed layer-wise construction makes more efficient use of partially observed hierarchical outcome information that would otherwise be discarded under a conventional complete-case analysis.


\subsection{Improving robustness and efficiency by outcome augmentation}\label{subsec:outaug}

While the IPW estimator uses treatment and covariate information to correct bias arising from missingness under Assumption \ref{asp:miss-pos}(i), it does not fully utilize all available outcome information, since it essentially operates as a weighted observed-case estimator at each hierarchy level. In addition, the IPW estimator can be biased if the postulated missingness models to estimate  $\pi_k(a,\mb X)$, $k=1,\dots,K$, are misspecified.

To improve efficiency and increase robustness to model misspecification, we propose the following AIPW estimator for $P_a^{1:k}(i_1,\dots,i_k)$, which further incorporates outcome regression models based on covariates. 

Consider the following outcome regression model for the conditional cell probability: 
\begin{align}\label{eq:OR}
    \mu_k(a,\mb X;i_1,\dots,i_k) = \Ex\{I(Y_1=i_1,\dots,Y_k=i_k)\mid A=a, \mb X\} = \Pr(Y_1=i_1,\dots,Y_k=i_k\mid A=a, \mb X). 
\end{align}
A practical choice for the outcome model is a \textit{baseline-category multinomial logistic model} (equivalently, a log-linear model for multinomial cell probabilities with a reference cell, e.g., the clinically best or worst outcome). Specifically, for each treatment $A=a$,
\begin{align}\label{eq:loglin}
\log \Pr(Y_1=i_1,\dots,Y_k=i_k\mid A=a,\mb X)
=\mb X^\top\bd\gamma_{i_1\dots i_k}^{(a)}-C^{(a)}(\mb X),
\end{align}
where $\bd\gamma_{i_1\dots i_k}^{(a)}$ is the cell-specific coefficient vector (with intercept absorbed in $\mb X$), and
$$
C^{(a)}(\mb X)
=
\log\sum_{i_1,\dots,i_k}
\exp\{\mb X^\top\bd\gamma_{i_1,\dots,i_k}^{(a)}\}.
$$
is the normalizing term ensuring probabilities sum to one. We usually take the clinically best or worst outcome cell as the reference cell. For example, for $k=2$ with $Y_1\in\{1,\dots,\ell_1\}$ and $Y_2\in\{1,\dots,\ell_2\}$, we may use $(\ell_1,\ell_2)$ as the reference cell. Then for any $(i_1,i_2)\neq(\ell_1,\ell_2)$,
$$
\log\frac{\Pr(Y_1=i_1, Y_2=i_2\mid\mb X, A=a)}{\Pr(Y_1=\ell_1, Y_2=\ell_2\mid\mb X, A=a)} = \mb X^\top\bd\gamma^{(a)}_{i_1,i_2}, 
$$
where the reference cell coefficient $\bd\gamma^{(a)}_{\ell_1,\ell_2}$ is fixed at zero.

The above outcome model provides a convenient working model for the conditional joint cell probabilities and can be fitted using standard multinomial maximum likelihood. The model need not be saturated when the number of endpoints or categories is large relative to the sample size; instead one can use more parsimonious log-linear specifications, such as models including only main effects and lower-order interactions among endpoint components. This allows a trade-off between flexibility and stability in estimating the fitted probabilities $\widehat\mu_k(a,\mb X;i_1,\dots,i_k)$ used in the AIPW construction.


With the outcome modeling, we have the following AIPW estimator for cell probabilities:
\begin{align}\label{eq:M1k-aipw-est}
   &  \widehat P_a^{1:k,\text{aipw}}(i_1,\dots,i_k)\nonumber \\ & = \frac{1}{n}\sum_{j=1}^n\left[\widehat\omega_k(a,\mb X_j,\widetilde R_{1:k,j})\{I(Y_{1j}=i_1,\dots,Y_{kj}=i_k) - \widehat\mu_k(a,\mb X_j;i_1,\dots,i_k)\} + \widehat\mu_k(a,\mb X_j;i_1,\dots,i_k)\right]. 
\end{align}
It can be shown that the AIPW estimator has the following attractive double robustness property: for each $k\in\{1,\dots,K\}$ and $a\in\{0,1\}$, when either the model of the non-missingness propensity score $\pi_k$ or the conditional outcome model $\mu_k$ is correctly specified, the estimator is consistent to $P_a^{1:k}(i_1,\dots,i_k)$. We prove this property in Appendix \ref{subapp:theory-aipw}. 

Appendix \ref{subapp:theory-aipw} also establishes that, when both nuisance models are correctly specified, the AIPW estimator attains the semiparametric efficiency bound for each joint cell probability $P_a^{1:k}(i_1,\dots,i_k)$ under the corresponding reduced observed-data model. This model retains baseline covariates, treatment assignment, the cumulative observation indicator, and the cell indicator when all first $k$ endpoints are observed.

\begin{remark}
Substituting the AIPW cell-probability estimates into \eqref{eq:win-pr}--\eqref{eq:tie-pr} yields estimators of the win, loss, and tie probabilities, and hence of the four win measures in
\eqref{eq:winstats}. Their large-sample properties follow from the joint asymptotic behavior of the cell-probability estimators and the delta method under the stated regularity conditions. We do not claim semiparametric efficiency for the resulting win-measure estimators under the full observed-data model. Rather, the efficiency result for the individual cell probabilities under the reduced observed-data model provides a theoretical rationale for outcome augmentation. Consistent with this result, the simulation studies in Section~\ref{sec:simu} show that AIPW typically has smaller RMSE than IPW when both nuisance models are correctly specified.
\end{remark}

\subsection{Variance estimation}\label{subsec:varest}

To quantify the estimation uncertainty of the proposed IPW and AIPW estimators, we derive influence-function-based variance estimators for the win measures in \eqref{eq:winstats}. Let $\theta$ denote a statistical parameter, and let $\psi(\mc O;\theta)$ be a mean-zero function with finite second moment. We say that $\psi(\mc O;\theta)$ is an influence function for $\widehat\theta$ if
\[
\sqrt{n}(\widehat\theta-\theta)
=
\frac{1}{\sqrt n}\sum_{i=1}^n
\psi(\mc O_i;\theta)
+
o_p(1).
\]
Thus, the asymptotic variance of $\widehat\theta$ is
\[
\frac{1}{n}\Ex\{\psi(\mc O;\theta)^2\},
\]
with empirical estimator
\[
\widehat{\Vx}(\widehat\theta)
=
\frac{1}{n^2}\sum_{i=1}^n
\widehat\psi(\mc O_i;\theta)^2.
\]

We first derive the influence functions of the proposed IPW and AIPW estimators for the joint cell probabilities
$P_a^{1:k}(i_1,\ldots,i_k)$.
We then propagate these influence functions to $(p_W,p_L,p_T)$ using first-order Taylor expansions of the products defining the win and loss probabilities. Finally, we apply the delta method to the mappings defining WR, NB, WO, and DOOR.

For each $k=1,\ldots,K$, $a\in\{0,1\}$, and cell
$(i_1,\ldots,i_k)$, the cell-probability estimators can be represented through estimating equations involving estimated nuisance parameters. Their influence functions have the generic form
\[
\psi\{\mc O;P_a^{1:k}(i_1,\ldots,i_k)\}
=
\phi\{\mc O;P_a^{1:k}(i_1,\ldots,i_k)\}
-
\mb B_{a,k}^{\top}
\psi(\mc O;\bd\eta_{a,k}),
\]
where $\phi$ denotes the leading estimating-function contribution,
$\bd\eta_{a,k}$ collects the relevant nuisance parameters, and
$\mb B_{a,k}$ denotes the corresponding sensitivity derivative.

For the proposed IPW estimator using H\'ajek weights, the nuisance contribution arises from estimation of the non-missingness-model parameter
$\bd\beta_k^{(a)}$. For the AIPW estimator, the nuisance parameters additionally include those of the conditional outcome model, denoted collectively by $\bd\gamma^{(a)}${, as well as $e_a=\Pr(A=a)$ when it is estimated by $\widehat e_a=n_a/n$. The first-order contribution from estimating $e_a$ vanishes when the outcome regression model is correctly specified.} More details, including the explicit influence-function expressions under logistic non-missingness models and multinomial log-linear outcome models, are provided in Web Appendices~A.4 and A.5. 

\begin{remark}
Although the treatment allocation probability is known by design in a randomized trial, one may alternatively estimate a working treatment propensity score using baseline covariates for additional covariate adjustment. Appendix~\ref{subapp:treatment-ps} studies this optional extension and shows that, under correct specification of the treatment propensity model, estimating the treatment propensity score can potentially reduce the first-order asymptotic variance of the H\'ajek IPW estimator by accounting for chance covariate imbalance. When the AIPW outcome regression model is correctly specified, no additional first-order efficiency gain is obtained from estimating the treatment propensity score.
\end{remark}

In practice, we obtain the empirical influence function
$\widehat\psi\{\mc O_i;P_a^{1:k}(i_1,\ldots,i_k)\}$
by replacing the population quantities with their sample analogs and evaluating the fitted nuisance models at the observed data. The win and loss probabilities are finite sums of products of joint cell probabilities, as shown in \eqref{eq:win-pr}--\eqref{eq:loss-pr}. Their influence functions are therefore obtained by the product rule. Let
$\widehat\psi(\mc O_i;p_W)$ and
$\widehat\psi(\mc O_i;p_L)$
denote the resulting empirical influence functions. Since
\[
p_T=1-p_W-p_L,
\]
the empirical influence function for the tie probability is
\[
\widehat\psi(\mc O_i;p_T)
=
-\widehat\psi(\mc O_i;p_W)
-\widehat\psi(\mc O_i;p_L).
\]

Define
\[
\widehat{\boldsymbol p}
=
(\widehat p_W,\widehat p_L,\widehat p_T)^\top,
\qquad
\widehat\psi(\mc O_i;\boldsymbol p)
=
\begin{pmatrix}
\widehat\psi(\mc O_i;p_W)\\
\widehat\psi(\mc O_i;p_L)\\
\widehat\psi(\mc O_i;p_T)
\end{pmatrix}.
\]
The joint covariance estimator is
\[
\widehat{\Vx}(\widehat{\boldsymbol p})
=
\frac{1}{n^2}\sum_{i=1}^n
\widehat\psi(\mc O_i;\boldsymbol p)
\widehat\psi(\mc O_i;\boldsymbol p)^\top.
\]
The uncentered empirical second-moment form is used because the population influence functions have mean zero; centering the empirical influence functions is asymptotically equivalent.

Each win measure can be written as
\[
\theta=g(p_W,p_L,p_T).
\]
For
$\widehat\theta=g(\widehat p_W,\widehat p_L,\widehat p_T)$,
the delta-method variance estimator is
\[
\widehat{\Vx}(\widehat\theta)
=
\nabla g(\widehat{\boldsymbol p})^\top
\widehat{\Vx}(\widehat{\boldsymbol p})
\nabla g(\widehat{\boldsymbol p}),
\]
where the gradient is evaluated at $\widehat{\boldsymbol p}$.

For NB and DOOR, inference is conducted on the original scale, with the corresponding $95\%$ confidence interval (CI)
\[
\widehat\theta
\pm
1.96\sqrt{\widehat{\Vx}(\widehat\theta)}.
\]

For WR and WO, inference is conducted on the log scale because these are positive ratio-type measures. For
$\theta=g(p_W,p_L,p_T)>0$,
\[
\widehat{\Vx}\{\log(\widehat\theta)\}
=
\nabla\log g(\widehat{\boldsymbol p})^\top
\widehat{\Vx}(\widehat{\boldsymbol p})
\nabla\log g(\widehat{\boldsymbol p}),
\]
where
\[
\nabla\log g(\boldsymbol p)
=
\frac{1}{g(\boldsymbol p)}
\nabla g(\boldsymbol p).
\]
The corresponding $95\%$ CI is
\[
\exp\left[
\log(\widehat\theta)
\pm
1.96\sqrt{\widehat{\Vx}\{\log(\widehat\theta)\}}
\right].
\]

For WR, the log-scale gradient is
\[
\nabla\log g_{\mathrm{WR}}(\boldsymbol p)
=
\left(
\frac{1}{p_W},
-\frac{1}{p_L},
0
\right)^\top.
\]
For WO, the log-scale gradient is
\[
\nabla\log g_{\mathrm{WO}}(\boldsymbol p)
=
\left(
\frac{1}{p_W+0.5p_T},
-\frac{1}{p_L+0.5p_T},
\frac{0.5}{p_W+0.5p_T}
-
\frac{0.5}{p_L+0.5p_T}
\right)^\top.
\]
For NB and DOOR, the original-scale gradients are
\[
\nabla g_{\mathrm{NB}}(\boldsymbol p)
=
(1,-1,0)^\top,
\qquad
\nabla g_{\mathrm{DOOR}}(\boldsymbol p)
=
(1,0,0.5)^\top.
\]

\section{Simulation Studies}\label{sec:simu}

In this section, we evaluate the proposed IPW and AIPW methods through Monte Carlo simulations. Although our methods are formulated in general settings with baseline covariates and treatment entering the missingness models and, for AIPW, the outcome models, we first consider a setting without covariate data. This simplification allows for a fair comparison with the standard pairwise comparison approach (which does not accommodate covariate adjustment) and conveniently illustrates its limitations. 

In the absence of covariates, our method reduces to adjusting for missingness separately within each treatment group using the overall non-missingness probability as the propensity score. Equivalently, within each treatment group, all participants are assumed to share the same probability of being observed at a given endpoint hierarchy. Moreover, without covariates, the IPW and AIPW estimators coincide because the outcome regression models reduce to constants.

We first examine the performance of the IPW estimator in a setting where endpoints are generated from treatment-specific multinomial outcome distributions, in Section \ref{subsec:data-nocovar}. We then extend the simulations to covariate-adjusted settings in Section \ref{subsec:data-covar}, where both the missingness and outcome-generating mechanisms depend on baseline covariates. 

{Throughout, we compare the proposed IPW and AIPW estimators with three alternatives: (i) the standard pairwise comparison approach, denoted by ``Standard,'' which treats any pairwise comparison involving a missing endpoint as a tie \citep{Buyse2025GPC}; the Standard approach is implemented using the \texttt{BuyseTest} R package; (ii) a complete-case analysis (CC), which excludes participants with any missing endpoint; and (iii) a multiple imputation analysis (MI). For MI, we use a generic MAR-based procedure. Binary endpoints are imputed using logistic regression, whereas ordinal endpoints with more than two categories are imputed using multinomial logistic regression. When applicable, treatment assignment, baseline covariates, and other observed endpoints are included as predictors. We generate 25 imputed data sets, compute the win measures separately within each, and combine the resulting estimates and variance estimates using Rubin’s rules \citep{rubin1978multiple}.}

We evaluate each competing method using four performance metrics defined below. Let $\theta$ denote the true win measure and $\widehat\theta_i$ its estimate from the $i$th Monte Carlo data replicate, $i=1,\dots,M$.
\begin{itemize}
    \item \textbf{Bias}: the mean estimation bias over the $M$ replicates, defined as $M^{-1}\sum_{i=1}^M (\widehat\theta_i - \theta)$. A Bias value closer to 0 indicates higher consistency;
    \item \textbf{RMSE}: the root mean squared error, defined as $\sqrt{M^{-1}\sum_{i=1}^M (\widehat\theta_i - \theta)^2}$. Smaller RMSE values indicate lower bias and greater efficiency;
    \item \textbf{CP}: the coverage probability, defined as the proportion of the $M$ replicates where the true $\theta$ falls within the 95\% CI constructed from the estimated variance and point estimate (for WR and WO, log transformation is used). A CP closer to 0.95 reflects accurate variance estimation and reliable inference. To account for the Monte-Carlo errors across  $M=2,000$ replications, we considered a CP outside of the interval $0.95\pm 1.96\sqrt{0.95\times 0.05/2,000}=[0.94, 0.96]$ as significantly deviating from the nominal 0.95 level;
    \item \textbf{CIW}: the average width of the 95\% CIs across the $M=2,000$ replicates. A narrower CIW indicates smaller variance estimation. 
\end{itemize} 

\subsection{Setting I: two binary endpoints without covariate}\label{subsec:data-nocovar}

The first setting uses the parameters in the example by \cite{li2024elusiveness}. For each scenario described below, we generate $M=2,000$ independent datasets, each with a total sample size of $n = 500$. 

We first generate the binary treatment $A \sim \text{Bernoulli}(0.5)$ to mimic 1:1 randomization. The two endpoints $(Y_1, Y_2)$ take values in $\{(1,1), (1,0), (0,1), (0,0)\}$, representing outcome values from best to worst. We consider two scenarios reflecting different treatment effects. For each treatment arm, $(Y_1, Y_2)$ follows a multinomial distribution as in \cite{li2024elusiveness}: 
\begin{itemize}
    \item Control ($A=0$): $(0.313, 0.268, 0.048, 0.373)$; and
    \item Treatment ($A=1$): $(0.26, 0.12, 0.47, 0.15)$ yielding $\theta_{\text{WR}}=1$ (no treatment effect), or $(0.5, 0.1, 0.2, 0.2)$ yielding $\theta_{\text{WR}}=1.69$ (a notable treatment effect).
\end{itemize}
Furthermore, for $j=1,2$, we impose missingness on $Y_j$ with probability $p_m^{(j0)}$ for controls and $p_m^{(j1)}$ for treated participants. We consider the following seven missingness scenarios as indicated in Table \ref{tab:sim-missingness}:
\begin{table}[H]
\centering
\begin{tabular}{llcccc}
\toprule
& Missing data scenario & $p_m^{(10)}$ & $p_m^{(11)}$ & $p_m^{(20)}$ & $p_m^{(21)}$ \\
\midrule
I & No missing data 
& 0 & 0 & 0 & 0 \\

II$^*$ & Homo,  $Y_1$ only (20\%, both groups) 
& 0.2 & 0.2 & 0 & 0 \\

III$^*$& Homo,  $Y_2$ only (20\%, both groups) 
& 0 & 0 & 0.2 & 0.2 \\

IV & Homo,  $Y_1$ and $Y_2$ (20\%, both groups) 
& 0.2 & 0.2 & 0.2 & 0.2 \\

V & Hete,   $Y_1$ only (30\% treated, 10\% control) 
& 0.3 & 0.1 & 0 & 0 \\

VI & Hete,   $Y_2$ only (30\% treated, 10\% control) 
& 0 & 0 & 0.3 & 0.1 \\

VII & Hete,   $Y_1$ and $Y_2$ (30\% treated, 10\% control) 
& 0.3 & 0.1 & 0.3 & 0.1 \\
\bottomrule
\end{tabular}
\caption{Missingness configurations used in simulations.}\label{tab:sim-missingness}
\begin{tablenotes}\scriptsize
\item Homogeneous: missingness rate by treatment group is identical (MCAR under this setting); Heterogeneous: missingness rates differ by treatment group. Scenarios $^*$II and III are those considered by \cite{li2024elusiveness}. 
\end{tablenotes}
\end{table}
Scenarios II and III were considered in \cite{li2024elusiveness}. In scenarios II--IV, the missing data are MCAR, as both treatment groups have the same missingness rate that is independent of $A$. In contrast, scenarios V--VII follow a mechanism where the missingness probabilities differ between treatment groups. 

Table~\ref{tab:sim-main-I} reports the results for WR under no treatment effect. When no outcomes are missing, all four methods yield identical point estimates and inference, as expected. Under missingness, the Standard method is systematically biased because comparisons involving missing endpoints are classified as ties. For example, in scenarios II and III, its estimated WRs are 1.466 and 0.826, respectively, compared with the true value of 1. These results are close to the corresponding values of 1.454 and 0.815 reported by \cite{li2024elusiveness}. {In contrast, the CC and IPW estimators exhibit small bias and near-nominal coverage across the missingness scenarios. MI also substantially reduces the bias and often yields smaller RMSE and CIW, although some coverage departures occur under the more challenging heterogeneous missingness scenarios.}

Overall, our IPW estimator performs well across all missingness scenarios, with small bias and CPs close to the nominal level. Results for the notable treatment-effect scenario and for WO, NB, and DOOR are provided in Appendix \ref{subapp:I-results} and show similar patterns.

\begin{table}[H]

\centering
\small
\caption{Simulation results for WR under Setting I and no treatment effect (true WR = 1).}
\label{tab:sim-main-I}
\begin{tabular}{lrrcccc}
\toprule
\textbf{Missing data scenario} & \textbf{Method} & \textbf{Est. WR} &
\textbf{Bias} & \textbf{RMSE} & \textbf{CP} & \textbf{CIW} \\
\midrule
No missing data & Standard & 1.012 & 0.012 & 0.132 & 0.956 & 0.519 \\
 & CC & 1.012 & 0.012 & 0.132 & 0.956 & 0.519 \\
 & MI & 1.012 & 0.012 & 0.132 & 0.956 & 0.519 \\
 & IPW & 1.012 & 0.012 & 0.132 & 0.956 & 0.519 \\
\addlinespace
Homo,  $Y_1$ only (20\%, both groups) & Standard & 1.466 & 0.466 & 0.508 & \textbf{0.203} & 0.778 \\
 & CC & 1.009 & 0.009 & 0.150 & 0.947 & 0.580 \\
 & MI & 0.999 & -0.001 & 0.147 & 0.945 & 0.562 \\
 & IPW & 1.009 & 0.009 & 0.150 & 0.947 & 0.580 \\
\addlinespace
Homo,  $Y_2$ only (20\%, both groups) & Standard & 0.826 & -0.174 & 0.211 & \textbf{0.700} & 0.460 \\
 & CC & 1.014 & 0.014 & 0.149 & 0.954 & 0.584 \\
 & MI & 1.001 & 0.001 & 0.136 & 0.947 & 0.525 \\
 & IPW & 1.013 & 0.013 & 0.138 & 0.950 & 0.531 \\
\addlinespace
Homo,  $Y_1$ and $Y_2$ (20\%, both groups) & Standard & 1.157 & 0.157 & 0.229 & \textbf{0.844} & 0.661 \\
 & CC & 1.011 & 0.011 & 0.166 & 0.943 & 0.653 \\
 & MI & 0.990 & -0.010 & 0.147 & 0.940 & 0.569 \\
 & IPW & 1.010 & 0.010 & 0.151 & 0.943 & 0.595 \\
\addlinespace
Hete,  $Y_1$ only (10\% control, 30\% treated) & Standard & 1.480 & 0.480 & 0.519 & \textbf{0.174} & 0.791 \\
 & CC & 1.007 & 0.007 & 0.149 & 0.950 & 0.600 \\
 & MI & 0.991 & -0.009 & 0.144 & 0.944 & 0.571 \\
 & IPW & 1.007 & 0.007 & 0.149 & 0.950 & 0.600 \\
\addlinespace
Hete,  $Y_2$ only (10\% control, 30\% treated) & Standard & 0.814 & -0.186 & 0.217 & \textbf{0.687} & 0.456 \\
 & CC & 1.008 & 0.008 & 0.148 & 0.952 & 0.599 \\
 & MI & 0.987 & -0.013 & 0.130 & 0.957 & 0.519 \\
 & IPW & 1.006 & 0.006 & 0.130 & 0.960 & 0.529 \\
\addlinespace
Hete,  $Y_1$ and $Y_2$ (10\% control, 30\% treated) & Standard & 1.161 & 0.161 & 0.233 & \textbf{0.845} & 0.673 \\
 & CC & 1.015 & 0.015 & 0.180 & 0.945 & 0.707 \\
 & MI & 0.981 & -0.019 & 0.152 & \textbf{0.933} & 0.578 \\
 & IPW & 1.011 & 0.011 & 0.157 & 0.951 & 0.618 \\
\bottomrule
\end{tabular}
\begin{tablenotes}\scriptsize
\item RMSE: root mean square error; CP: coverage probability, with values outside the range [0.94, 0.96] shown in bold; CIW: confidence interval width; Homo: homogeneous missing-data rates across treatment groups; Hete: heterogeneous missing-data rates across treatment groups; CC: complete-case analysis; MI: multiple imputation.
\end{tablenotes}
\end{table}

\subsection{Setting II: two ordinal endpoints with covariates}\label{subsec:data-covar}

To illustrate our methods’ ability to incorporate baseline covariates in both the outcome models and the missingness mechanisms, we consider a simulation setting with treatment- and covariate-dependent missingness. In this setting, the IPW and AIPW estimators are no longer equivalent, and, when the outcome models are correctly specified, the AIPW estimator can be more efficient in theory. We use $M=2,000$ simulation replicates with sample size $n=1,000$ in each run.

We first generate two baseline covariates,
$X_1 \sim \mathcal{N}(0,1), X_2 \sim \text{Bernoulli}(0.5)$
and a binary treatment indicator $A \sim \text{Bernoulli}(0.5)$, independent of $(X_1, X_2)$. Two latent continuous outcomes are generated according to the linear models
$
Y_1^* = 1.5 + X_1 + 2X_2 + 0.6A + 0.2(A X_1) + 0.25(A X_2) + \varepsilon_1,
$
and
$
Y_2^* = 1.1 + 1.4X_1 + X_2 + 0.4A + 0.5(A X_1) + 0.75(A X_2) + \varepsilon_2,
$
where $\varepsilon_1$ and $\varepsilon_2$ are independent logistic random variables with mean zero and unit scale.

The observed ordinal outcomes are defined as
$Y_1 = \mathbb{I}(Y_1^* > 0),$ and
$$
Y_2 =
\begin{cases}
0, & Y_2^* \le -1, \\
1, & -1 < Y_2^* \le 1, \\
2, & Y_2^* > 1.
\end{cases}
$$
Missingness is introduced through outcome-specific response indicators that depend on baseline covariates. Specifically, for $k \in \{1,2\}$ and treatment group $a \in \{0,1\}$, we define
$$
\Pr(R_{ka} = 1 \mid X_1, X_2)
= \text{expit}\!\left(\alpha_{ka}
+ \gamma_{ka1} X_1 + \gamma_{ka2} X_2\right),
\qquad \text{ where expit}(x)=\{1+\exp(-x)\}^{-1}.
$$
The slope parameters are set to
$$
(\gamma_{101},\gamma_{102})=(0.5,1),\quad
(\gamma_{111},\gamma_{112})=(1,1),\quad
(\gamma_{201},\gamma_{202})=(1,0.5),\quad
(\gamma_{211},\gamma_{212})=(1,1).
$$
The intercepts $\alpha_{ka}$ are calibrated to achieve prespecified marginal missingness rates within each treatment group, i.e., $\Pr(R_{ka}=1)=1-p_m^{(ka)}$.
Finally, the observed outcomes satisfy
$$
Y_k \text{ is observed } \iff R_{ka} = 1 \text{ for } A = a,\quad k\in\{1,2\}.
$$
We consider multiple missing data configurations; the setting of $(p_m^{(10)},p_m^{(11)},p_m^{(20)},p_m^{(21)})$ is the same as Table \ref{tab:sim-missingness}. All missing data mechanisms depend on $(X_1, X_2)$ but not on the outcomes conditional on covariates, corresponding to Assumption \ref{asp:miss-pos}(i). 

For brevity, we only present results for the WR in this section; the results for the other win measures are deferred to Appendix \ref{subapp:II-results}. Table \ref{tab:sim-II-res-WR} compares the performance of the standard, {CC, MI,} IPW, and AIPW estimators across different scenarios of missing data. To further examine the robustness of the proposed estimators to model misspecifications, we also report the results under different specifications of the missingness or outcome models. For the IPW estimator, case A corresponds to a correctly specified propensity score model and case B to a misspecified propensity score model. For the AIPW estimator, case A corresponds to both models being correctly specified, case B to a misspecified missingness model, and case C to a misspecified outcome model. The correctly specified missingness and outcome models include both covariates $X_1$ and $X_2$ in the logistic and baseline-category logit models, respectively, whereas the misspecified model includes $X_2$ only.

The simulation results shown in Table \ref{tab:sim-II-res-WR} indicate that the Standard method is unbiased with valid CP only when there are no missing data. In this complete-data setting, the Standard method is equivalent to the IPW estimator, as expected. {When missing data are present, both the Standard and CC analyses exhibit substantial bias and poor CP in several scenarios. MI substantially reduces this bias and generally achieves near-nominal CP. However, MI can be less efficient than AIPW: its RMSEs and CIWs are generally larger than those of AIPW-A and AIPW-B.} Under correctly specified nuisance functions, both IPW-A and AIPW-A exhibit small bias and near-nominal CP across the different missing-data scenarios. Moreover, compared with IPW-A, AIPW-A generally yields smaller RMSEs and narrower CIs, confirming the efficiency gain from incorporating a correctly specified outcome regression model.

The double robustness of the proposed AIPW estimator is further supported by the model-misspecification results. In particular, comparing case B between IPW and AIPW is informative because the missingness model is misspecified in both cases. Under IPW-B, the estimator exhibits substantial bias in several scenarios, whereas AIPW-B continues to show small bias and CP close to 0.95. Comparing AIPW-B and AIPW-C further clarifies the roles of the two nuisance models. When only the missingness model is misspecified, AIPW-B performs similarly to AIPW-A. When only the outcome model is misspecified, AIPW-C remains approximately unbiased, consistent with double robustness, but generally loses the efficiency advantage of AIPW-A, as reflected by larger RMSEs and wider CIs.

As an additional finite-sample diagnostic, we found no propensity score monotonicity violations or negative raw tie-probability estimates across the seven scenarios, the five IPW/AIPW specifications, and all Monte Carlo replicates. The smallest raw tie-probability estimate was 0.2623.

Overall, these findings indicate that AIPW remains consistent when at least one nuisance model is correctly specified. They also show that correct specification of the outcome model is important for achieving efficiency gains over IPW, whereas MI provides good bias control but is generally less efficient than AIPW.

\begin{table}[H]

\centering
\footnotesize
\caption{Simulation results for WR under Setting II: bias and RMSE (true WR = 1.29).}
\label{tab:sim-II-res-WR}
\setlength{\tabcolsep}{2.8pt}
\renewcommand{\arraystretch}{1.02}
\begin{tabular}{p{6.5cm} cccccccc}
\toprule
 & \multicolumn{8}{c}{\textbf{Bias}} \\
\cmidrule(lr){2-9}
\textbf{Missing data scenario} & \textbf{Standard} & \textbf{CC} & \textbf{MI} & \multicolumn{2}{c}{\textbf{IPW}} & \multicolumn{3}{c}{\textbf{AIPW}} \\
\cmidrule(lr){5-6}\cmidrule(lr){7-9}
 & & & & \textbf{A} & \textbf{B} & \textbf{A} & \textbf{B} & \textbf{C} \\
\midrule
No missing data & 0.004 & 0.004 & 0.004 & 0.004 & 0.004 & 0.001 & 0.001 & 0.004 \\
Homo,  $Y_1$ only (20\%, both groups) & 0.157 & 0.157 & 0.004 & 0.002 & 0.148 & 0.002 & 0.002 & 0.004 \\
Homo,  $Y_2$ only (20\%, both groups) & 0.100 & 0.161 & 0.006 & 0.002 & 0.072 & 0.005 & 0.005 & 0.006 \\
Homo,  $Y_1$ and $Y_2$ (20\%, both groups) & 0.154 & 0.154 & 0.002 & 0.003 & 0.147 & 0.002 & 0.002 & 0.003 \\
Hete,  $Y_1$ only (10\% control, 30\% treated) & -0.075 & -0.075 & 0.006 & 0.004 & -0.024 & 0.003 & 0.003 & 0.003 \\
Hete,  $Y_2$ only (10\% control, 30\% treated) & 0.013 & -0.076 & 0.001 & 0.003 & -0.011 & -0.001 & -0.001 & -0.000 \\
Hete,  $Y_1$ and $Y_2$ (10\% control, 30\% treated) & -0.068 & -0.068 & 0.012 & 0.004 & -0.026 & 0.006 & 0.006 & 0.009 \\
\midrule
 & \multicolumn{8}{c}{\textbf{RMSE}} \\
\cmidrule(lr){2-9}
\textbf{Missing data scenario} & \textbf{Standard} & \textbf{CC} & \textbf{MI} & \multicolumn{2}{c}{\textbf{IPW}} & \multicolumn{3}{c}{\textbf{AIPW}} \\
\cmidrule(lr){5-6}\cmidrule(lr){7-9}
 & & & & \textbf{A} & \textbf{B} & \textbf{A} & \textbf{B} & \textbf{C} \\
\midrule
No missing data & 0.088 & 0.088 & 0.088 & 0.089 & 0.089 & 0.066 & 0.066 & 0.088 \\
Homo,  $Y_1$ only (20\%, both groups) & 0.195 & 0.195 & 0.096 & 0.098 & 0.188 & 0.074 & 0.074 & 0.098 \\
Homo,  $Y_2$ only (20\%, both groups) & 0.147 & 0.201 & 0.094 & 0.093 & 0.123 & 0.073 & 0.072 & 0.096 \\
Homo,  $Y_1$ and $Y_2$ (20\%, both groups) & 0.193 & 0.193 & 0.095 & 0.095 & 0.185 & 0.075 & 0.074 & 0.098 \\
Hete,  $Y_1$ only (10\% control, 30\% treated) & 0.124 & 0.124 & 0.098 & 0.100 & 0.106 & 0.076 & 0.076 & 0.098 \\
Hete,  $Y_2$ only (10\% control, 30\% treated) & 0.099 & 0.124 & 0.091 & 0.094 & 0.095 & 0.070 & 0.069 & 0.092 \\
Hete,  $Y_1$ and $Y_2$ (10\% control, 30\% treated) & 0.121 & 0.121 & 0.099 & 0.097 & 0.104 & 0.078 & 0.078 & 0.101 \\
\midrule
& \multicolumn{8}{c}{\textbf{CP}} \\
\cmidrule(lr){2-9}
\textbf{Missing data scenario} & \textbf{Standard} & \textbf{CC} & \textbf{MI} & \multicolumn{2}{c}{\textbf{IPW}} & \multicolumn{3}{c}{\textbf{AIPW}} \\
\cmidrule(lr){5-6}\cmidrule(lr){7-9}
 & & & & \textbf{A} & \textbf{B} & \textbf{A} & \textbf{B} & \textbf{C} \\
\midrule
No missing data & 0.950 & 0.950 & 0.950 & 0.950 & 0.950 & 0.951 & 0.951 & 0.950 \\
Homo,  $Y_1$ only (20\%, both groups) & \textbf{0.719} & \textbf{0.719} & 0.950 & 0.951 & \textbf{0.730} & 0.955 & 0.956 & \textbf{0.973} \\
Homo,  $Y_2$ only (20\%, both groups) & \textbf{0.845} & \textbf{0.698} & 0.946 & 0.949 & \textbf{0.896} & 0.945 & 0.947 & 0.949 \\
Homo,  $Y_1$ and $Y_2$ (20\%, both groups) & \textbf{0.712} & \textbf{0.712} & 0.957 & \textbf{0.962} & \textbf{0.749} & 0.953 & 0.952 & \textbf{0.972} \\
Hete,  $Y_1$ only (10\% control, 30\% treated) & \textbf{0.871} & \textbf{0.871} & 0.946 & \textbf{0.938} & \textbf{0.929} & 0.948 & 0.948 & \textbf{0.964} \\
Hete,  $Y_2$ only (10\% control, 30\% treated) & 0.957 & \textbf{0.874} & 0.952 & 0.953 & 0.945 & 0.952 & 0.958 & 0.959 \\
Hete,  $Y_1$ and $Y_2$ (10\% control, 30\% treated) & \textbf{0.887} & \textbf{0.887} & \textbf{0.939} & 0.954 & 0.940 & 0.943 & 0.945 & \textbf{0.961} \\
\midrule
 & \multicolumn{8}{c}{\textbf{CIW}} \\
\cmidrule(lr){2-9}
\textbf{Missing data scenario} & \textbf{Standard} & \textbf{CC} & \textbf{MI} & \multicolumn{2}{c}{\textbf{IPW}} & \multicolumn{3}{c}{\textbf{AIPW}} \\
\cmidrule(lr){5-6}\cmidrule(lr){7-9}
 & & & & \textbf{A} & \textbf{B} & \textbf{A} & \textbf{B} & \textbf{C} \\
\midrule
No missing data & 0.349 & 0.349 & 0.349 & 0.349 & 0.349 & 0.257 & 0.257 & 0.349 \\
Homo,  $Y_1$ only (20\%, both groups) & 0.455 & 0.455 & 0.374 & 0.380 & 0.448 & 0.292 & 0.292 & 0.424 \\
Homo,  $Y_2$ only (20\%, both groups) & 0.415 & 0.457 & 0.358 & 0.364 & 0.393 & 0.279 & 0.281 & 0.373 \\
Homo,  $Y_1$ and $Y_2$ (20\%, both groups) & 0.453 & 0.453 & 0.374 & 0.381 & 0.449 & 0.293 & 0.292 & 0.423 \\
Hete,  $Y_1$ only (10\% control, 30\% treated) & 0.379 & 0.379 & 0.375 & 0.379 & 0.392 & 0.295 & 0.296 & 0.412 \\
Hete,  $Y_2$ only (10\% control, 30\% treated) & 0.388 & 0.379 & 0.358 & 0.365 & 0.367 & 0.278 & 0.280 & 0.369 \\
Hete,  $Y_1$ and $Y_2$ (10\% control, 30\% treated) & 0.381 & 0.381 & 0.376 & 0.379 & 0.392 & 0.295 & 0.297 & 0.415 \\
\bottomrule
\end{tabular}
\begin{tablenotes}\scriptsize
\item Homo: the two treatment groups have the same (homogeneous) marginal missing-data rate; Hete: the two treatment groups have different (heterogeneous) marginal missing-data rates.
\item Standard: missing comparisons are treated as ties; CC: complete-case analysis; MI: multiple imputation using 25 imputed data sets; IPW: H\'ajek-normalized inverse probability weighting; IPW-A: correctly specified missingness model; IPW-B: misspecified missingness model; AIPW-A: both missingness and outcome models correctly specified; AIPW-B: misspecified missingness model; AIPW-C: misspecified outcome model.
\end{tablenotes}
\end{table}

\subsection{Additional experiments on Horvitz--Thompson weighting and sensitivity to the missingness assumption}

To further investigate the practical performance and limitations of the proposed methods, we conducted two additional simulation experiments. The first compares the proposed H\'ajek IPW estimator with its Horvitz--Thompson version, and the second examines settings in which the missingness mechanism further depends on the outcome and therefore violates Assumption~\ref{asp:miss-pos}. Details of both experiments are provided in Web Appendices~B.3 and B.4.

In the first experiment (Appendix~\ref{subapp:hajek}), we compared H\'ajek and Horvitz--Thompson weighting under selected scenarios from Setting~II. The two approaches produced similar bias and RMSE in the settings considered. However, the H\'ajek estimator ensured that the estimated cell probabilities summed to one within each treatment group and hierarchy level up to numerical precision and generally showed improved CI coverage, with comparable CIs. These findings support our use of H\'ajek weighting as the primary IPW formulation. As expected, neither weighting scheme is robust to misspecification of the non-missingness model, so the doubly robust AIPW estimator remains important when protection against nuisance-model misspecification is desired.

In the second experiment (Appendix~\ref{subapp:miss-vio}), we examined sensitivity to an outcome-dependent missingness mechanism in which observation of $Y_2$ depended on the observed value of $Y_1$, thereby violating Assumption~\ref{asp:miss-pos}. The Standard and CC methods exhibited substantial bias and poor coverage, whereas MI generally yielded small bias and near-nominal coverage. Although IPW and AIPW are not guaranteed to be consistent under this mechanism, the covariate-adjusted specifications showed relatively small bias in the scenarios considered, with AIPW generally providing smaller RMSEs and narrower CIs than MI. These results underscore the importance of the missingness assumption and the value of MI as a complementary sensitivity analysis when missingness may depend on previously observed outcomes. 

\section{Data Applications}\label{sec:data}

To illustrate our proposed methods, we consider two clinical trial data with ordinal endpoint(s) subject to missingness, SCOUT-CAP and ACTT-1. 

\subsection{Study I: SCOUT-CAP trial}\label{subsec:scout}

The SCOUT-CAP trial is a multi-center randomized double-blind placebo-controlled clinical trial \citep{williams2022short}. It compares a short-course (5-day) versus standard-course (10-day) oral $\beta$-lactam strategy for outpatient treatment of non-severe community-acquired pneumonia in young children who demonstrated early clinical improvement during the first several days of therapy. A total of 380 participants were enrolled from outpatient clinic, urgent care, or emergency settings across multiple U.S. sites and randomized in a 1:1 ratio to discontinue antibiotics after 5 days (and switch to placebo on days 6--10) or to continue the same antibiotic through day 10. Outcomes were measured at the first outcome assessment visit (OAV1; study days 6--10) and a later visit (OAV2; study days 19--25). Similar to \citep{williams2022short}, we analyze the outcomes at OAV1. 

To illustrate our methods, we consider two analyses with two different sets of endpoints. The first set has only one ordinal outcome that is the primary outcome with 8 categories, constructed by collapsing and combining the categories of three ordinal outcomes into smaller number of categories. To get a sense on what would have happened if the these ordinal outcomes were not combined, we also performed the analysis on these three original outcomes, based on the following hierarchy: (i) adequate clinical response (ACR) (Yes/No), (ii) resolution of pneumonia symptoms (RPS) (Persistent/Resolved), and (iii) maximal antibiotic-associated adverse effects (MAAE) (None, Mild, Moderate or Severe). In both analyses, we included age (an indicator of $\geq 2$ years), sex and race in the analysis as baseline covariates. 

\begin{table}[H]
\centering
\footnotesize
\begin{tabular}{lccccccc}
  \toprule
Variable (with $n(\%)$) 
& \makecell{Overall\\$(n=380)$} 
& \makecell{Standard Course \\ (Controls, $n=191)$ } 
& \makecell{Short Course \\ (Treated, $n=189)$ } \\ 
  \midrule
\multicolumn{4}{c}{\textbf{Baseline covariates}} \\ 
Age $\geq 2$ years & 269 (70.8\%) & 135 (70.7\%) & 134 (70.9\%) \\
Sex = female & 186 (48.9\%) & 91 (47.6\%) & 95 (50.3\%) \\
Race \\
~~~Black & 99 (26.1\%) & 51 (26.7\%) & 48 (25.4\%)  \\
~~~White & 234 (61.6\%) & 116 (60.7\%) & 118 (62.4\%)   \\
~~~Others & 47 (12.4\%) & 24 (12.6\%) & 23 (12.2\%)   \\
\midrule
\multicolumn{4}{c}{\textbf{Endpoints}} \\
\textbf{Single ordinal endpoint}  & & & \\
~~~ACR with antibiotic-associated adverse
effects \\
~~~~~~None & 204 (53.7\%) & 107 (56.0\%) & 97 (51.3\%) \\ 
~~~~~~Mild & 89 (23.4\%) & 42 (22.0\%) & 47 (24.9\%) \\ 
~~~~~~Moderate & 24 (6.3\%) & 10 (5.2\%) & 14 (7.4\%) \\ 
~~~~~~Severe & 2 (0.5\%) & 2 (1.1\%) & 0 (0\%) \\ 
~~~~~~Persistent symptoms  & 24 (6.3\%) & 13 (6.8\%) & 11 (5.8\%) \\ 
~~~No ACR \\
~~~~~~With ED or clinic visit  & 3 (0.8\%) & 1 (0.5\%) & 2 (1.1\%) \\ 
~~~~~~With hospitalization  & 0 (0\%) & 0 (0\%) & 0 (0\%) \\ 
~~~~~~Death & 0 (0\%) & 0 (0\%) & 0 (0\%) \\ 
Missing data count & 34 (8.9\%) & 16 (8.4\%) & 18 (9.5\%) \\ 
\addlinespace
\textbf{Three hierarchical endpoints}  & & & \\
~~~\textbf{ACR} & \\
~~~~~~No & 3 (0.8\%) & 1 (0.5\%) & 2 (1.1\%) \\ 
~~~~~~Yes & 358 (94.2\%) & 180 (94.2\%) & 178 (94.2\%) \\ 
~~~Missing data count for ACR & 19 (5.0\%) & 10 (5.2\%) & 9 (4.8\%) \\ 
\addlinespace
~~~\textbf{RPS} & \\
~~~~~~Persistent symptoms & 28 (7.4\%) & 15 (7.9\%) & 13 (6.9\%) \\ 
~~~~~~Resolved & 331 (87.1\%) & 168 (88.0\%) & 163 (86.2\%) \\ 
~~~Missing data count for RPS & 21 (5.5\%) & 8 (4.2\%) & 13 (6.9\%) \\ 
\addlinespace
~~~\textbf{MAAE}  \\
~~~~~~None & 219 (57.6\%) & 113 (59.2\%) & 106 (56.1\%) \\ 
~~~~~~Mild & 99 (26.1\%) & 47 (24.6\%) & 52 (27.5\%) \\ 
~~~~~~Moderate & 28 (7.4\%) & 13 (6.9\%) & 15 (7.9\%) \\
~~~~~~Severe & 3 (0.8\%) & 2 (1.1\%) & 1 (0.5\%) \\
~~~Missing data count for MAAE & 31 (8.2\%) & 16 (8.4\%) & 15 (7.9\%) \\ 
\addlinespace
\textbf{Missingness for ACR, RPS and MAAE hierarchy}\\
~~~Missing data count for ACR & 19 (5.0\%) & 10 (5.2\%) & 9 (4.8\%) \\ 
~~~Any missing data count for ACR or RPS & 24 (6.3\%) & 10 (5.2\%) & 14 (7.4\%) \\ 
~~~Any missing data count for ACR, RPS, or MAAE & 36 (9.5\%) & 17 (8.9\%) & 19 (10.1\%) \\
\bottomrule
\end{tabular}
\caption{Summary statistics for the SCOUT-CAP trial data.} 
\label{tab:sumstat-stu1}
\begin{tablenotes}
\scriptsize
\item Numbers are presented as $n$ (\%). Percentages are calculated using non-missing observations for each variable and may not sum to column totals due to missing data.
\item All endpoints are ordinal with numerical scores, ordered from best to worst (top to bottom).
\item Abbreviations: adequate clinical response (ACR); resolution of pneumonia symptoms (RPS); maximal antibiotic-associated adverse effects (MAAE); emergency department (ED).
\end{tablenotes}
\end{table}
The summary statistics of the SCOUT-CAP trial data are shown in Table~\ref{tab:sumstat-stu1}. The baseline covariates are well balanced between the standard-course and short-course groups. In the pooled sample, we observe modest but non-negligible missing data across the three component endpoints (ACR, RPS, and MAAE), where about 10\% of participants have at least one missing data. The rate of missing data is about 9\% for the single ordinal score and ranges from approximately 5\% to 10\% across the three component endpoints. 

We fitted logistic regression models for non-missingness propensity scores, where we included the baseline covariates, treatment, and covariate-by-treatment interaction terms. Including covariate-by-treatment interactions is equivalent to fitting separate regression models within each treatment group. For the single ordinal score, we modeled missingness using logistic regression. For the three hierarchical endpoints, following our theory and the hierarchical structure, we separately modeled the non-missingness of ACR, the joint non-missingness of ACR and RPS, and the joint non-missingness of ACR, RPS, and MAAE (3 models in total). In this example, none of the regression terms was statistically significant in any of these models (all p-values $\geq 0.1$), suggesting that missingness was not strongly associated with the measured covariates or treatment. This finding is consistent with an approximately MCAR mechanism, at least with respect to the observed variables. 

Figure \ref{fig:ps-scoutcap} in Appendix \ref{subapp:data-ps} presents histograms of the non-missingness propensity score  distributions by treatment group under the four logistic regression models described above. For all models used to construct the IPW weights, the minimum estimated propensity score is 0.72, and most scores are concentrated around 0.9, suggesting no apparent violation of the positivity assumption (Assumption \ref{asp:miss-pos}(ii)) in this example. Additional finite-sample diagnostics showed that all estimated win, loss, and raw tie probabilities from the IPW analyses were within $[0,1]$. For the single ordinal score, the H\'ajek-normalized IPW cell probabilities summed to one within each treatment group up to numerical precision, with maximum absolute deviations of $2.22\times10^{-16}$ both without and with covariate adjustment. The corresponding raw tie-probability estimates were 0.4223 and 0.4209, respectively. For the three hierarchical endpoints, the H\'ajek-normalized IPW cell probabilities again summed to one within each treatment group and hierarchy level up to numerical precision, with maximum absolute deviations of $2.22\times10^{-16}$ without covariate adjustment and $3.33\times10^{-16}$ with covariate adjustment. No propensity-score monotonicity violation was observed for the no-covariate IPW analysis. For the covariate-adjusted IPW analysis, the separately fitted propensity models exhibited some monotonicity violations in this finite sample, with a maximum ordering excess of 0.01433. Nevertheless, all estimated win, loss, and raw tie probabilities remained within $[0,1]$, with raw tie-probability estimates of 0.4187 and 0.4187 for the IPW analyses without and with covariate adjustment, respectively. 

\begin{table}
\centering
\scriptsize
\caption{Data analysis results for SCOUT-CAP trial data.}
\label{tab:stu1-main}
\begin{tabular}{llcccc}
\toprule
\multicolumn{6}{c}{\textbf{Single ordinal score}} \\
\addlinespace
\textbf{Measure} & \textbf{Method} & \textbf{Estimate} & \textbf{SE} & \textbf{95\% CI} & \textbf{p-value} \\
\midrule
\multirow{6}{*}{WR} 
   & Standard & 1.147 & 0.218 & (0.790, 1.664) & 0.471 \\ 
   & {CC} & {1.147} & {0.218} & {(0.790, 1.664)} & {0.471} \\
   & {MI} & {1.173} & {0.224} & {(0.807, 1.704)} & {0.404} \\
   & IPW (no covariate) & 1.147 & 0.218 & (0.790, 1.664) & 0.471 \\ 
   & IPW (with covariates) & 1.164 & 0.221 & (0.802, 1.688) & 0.425 \\ 
   & AIPW (with covariates) & 1.167 & 0.217 & (0.811, 1.680) & 0.407 \\
\addlinespace

\multirow{6}{*}{WO} 
   & Standard & 1.082 & 0.119 & (0.873, 1.342) & 0.471 \\ 
   & {CC} & {1.082} & {0.119} & {(0.873, 1.342)} & {0.471} \\
   & {MI} & {1.097} & {0.121} & {(0.883, 1.362)} & {0.404} \\
   & IPW (no covariate) & 1.082 & 0.119 & (0.873, 1.342) & 0.471 \\ 
   & IPW (with covariates) & 1.092 & 0.120 & (0.880, 1.354) & 0.425 \\  
   & AIPW (with covariates) & 1.093 & 0.118 & (0.885, 1.350) & 0.408 \\
\addlinespace

\multirow{6}{*}{NB} 
   & Standard & 0.033 & 0.045 & (-0.056, 0.122) & 0.471 \\ 
   & {CC} & {0.039} & {0.055} & {(-0.068, 0.147)} & {0.471} \\
   & {MI} & {0.046} & {0.055} & {(-0.062, 0.154)} & {0.403} \\
   & IPW (no covariate) & 0.039 & 0.055 & (-0.068, 0.147) & 0.471 \\ 
   & IPW (with covariates) & 0.044 & 0.055 & (-0.064, 0.151) & 0.424 \\ 
   & AIPW (with covariates) & 0.045 & 0.054 & (-0.061, 0.150) & 0.407 \\
\addlinespace

\multirow{6}{*}{DOOR} 
   & Standard & 0.516 & 0.023 & (0.472, 0.561) & 0.471 \\ 
   & {CC} & {0.520} & {0.027} & {(0.466, 0.573)} & {0.471} \\
   & {MI} & {0.523} & {0.028} & {(0.469, 0.577)} & {0.403} \\
   & IPW (no covariate) & 0.520 & 0.027 & (0.466, 0.573) & 0.471 \\ 
   & IPW (with covariates) & 0.522 & 0.027 & (0.468, 0.576) & 0.424 \\  
   & AIPW (with covariates) & 0.522 & 0.027 & (0.470, 0.575) & 0.407 \\
\midrule

\multicolumn{6}{c}{\textbf{Three hierarchical endpoints}$^*$} \\
\addlinespace
\textbf{Measure} & \textbf{Method} & \textbf{Estimate} & \textbf{SE} & \textbf{95\% CI} & \textbf{p-value} \\
\midrule
\multirow{6}{*}{WR} 
   & Standard & 1.163 & 0.219 & (0.804, 1.681) & 0.423 \\ 
   & {CC} & {1.177} & {0.225} & {(0.810, 1.711)} & {0.393} \\
   & {MI} & {1.150} & {0.214} & {(0.799, 1.657)} & {0.452} \\
   & IPW (no covariate) & 1.172 & 0.222 & (0.809, 1.699) & 0.400 \\ 
   & IPW (with covariates) & 1.182 & 0.224 & (0.816, 1.713) & 0.377 \\
   & AIPW (with covariates) & 1.182 & 0.224 & (0.815, 1.714) & 0.378 \\ 
\addlinespace

\multirow{6}{*}{WO} 
   & Standard & 1.091 & 0.119 & (0.882, 1.350) & 0.423 \\ 
   & {CC} & {1.098} & {0.121} & {(0.886, 1.362)} & {0.392} \\
   & {MI} & {1.084} & {0.117} & {(0.878, 1.339)} & {0.452} \\
   & IPW (no covariate) & 1.097 & 0.121 & (0.884, 1.361) & 0.402 \\ 
   & IPW (with covariates) & 1.102 & 0.121 & (0.889, 1.366) & 0.377 \\  
   & AIPW (with covariates) & 1.102 & 0.121 & (0.888, 1.367) & 0.378 \\ 
\addlinespace

\multirow{6}{*}{NB} 
   & Standard & 0.037 & 0.047 & (-0.054, 0.129) & 0.422 \\ 
   & {CC} & {0.047} & {0.055} & {(-0.060, 0.154)} & {0.392} \\
   & {MI} & {0.040} & {0.054} & {(-0.065, 0.146)} & {0.452} \\
   & IPW (no covariate) & 0.046 & 0.055 & (-0.062, 0.154) & 0.401 \\ 
   & IPW (with covariates) & 0.048 & 0.055 & (-0.059, 0.156) & 0.376 \\ 
   & AIPW (with covariates) & 0.048 & 0.055 & (-0.059, 0.156) & 0.377 \\
\addlinespace

\multirow{6}{*}{DOOR} 
   & Standard & 0.519 & 0.023 & (0.473, 0.565) & 0.422 \\ 
   & {CC} & {0.523} & {0.027} & {(0.470, 0.577)} & {0.392} \\
   & {MI} & {0.520} & {0.027} & {(0.468, 0.573)} & {0.452} \\
   & IPW (no covariate) & 0.523 & 0.027 & (0.469, 0.577) & 0.401 \\ 
   & IPW (with covariates) & 0.524 & 0.027 & (0.471, 0.578) & 0.376 \\ 
   & AIPW (with covariates) & 0.524 & 0.027 & (0.470, 0.578) & 0.377 \\ 
\bottomrule
\end{tabular}

\begin{tablenotes}\scriptsize
\item SE: standard error; CI: confidence interval; CC: complete-case analysis; MI: multiple imputation.
\item $^*$The three hierarchical endpoints are adequate clinical response (ACR), resolution of pneumonia symptoms (RPS), and maximal antibiotic-associated adverse effects (MAAE).
\end{tablenotes}
\end{table}

In Table~\ref{tab:stu1-main}, we report estimates of four win measures (WR, WO, NB, and DOOR) under the standard analysis, which treats any comparison involving missing data as a tie; {CC; MI; the IPW estimator without covariates; and the covariate-adjusted IPW and AIPW estimators. We include CC and MI as commonly used approaches for handling missing data, providing practical benchmarks for comparison with the proposed methods.} We include the IPW estimator without covariates to illustrate that the proposed weighting approach can also be applied when users choose not to incorporate baseline covariates, while providing a direct comparison with the standard analysis independent of covariate adjustment.

Although all methods lead to the same conclusion regarding statistical significance at the 0.05 level, our proposed estimators generally yield smaller p-values than the standard analysis for all measures across both analyses (single and three endpoints), with overall larger p-value reductions from the two covariate-adjusted approaches (IPW and AIPW). Moreover, the AIPW estimator generally produces SEs and CI widths that are comparable to or slightly smaller than those from IPW. In light of our simulation results, this also suggests that the specified outcome models are effective in reducing outcome variation and improving estimation efficiency. 


We note that without covariate adjustment, the IPW estimators for WR and WO based on the single ordinal score yield exactly the same results as the standard estimator. This equivalence is expected: when covariates are omitted, the IPW weights are constant within each treatment group, and for ratio measures (WR and WO), which are on the multiplicative scale, these constants cancel between their numerator and denominator. Consequently, the point estimates coincide with those from the standard analysis. However, differences can arise when the weights vary across individuals through covariate-dependent missingness modeling. In contrast, because NB and DOOR are on the additive scale, the constant propensity scores do not cancel, and we thus observe differences between their unadjusted IPW and standard estimates. Therefore, WR and WO are just special cases, and the underlying estimated win, loss, and tie probabilities by standard and IPW (no covariate) estimators are still different.


\subsection{Study II: ACTT-1 trial}\label{subsec:covid}

The ACTT-1 trial is a double-blinded, placebo-controlled trial for evaluating the effect of remdesivir for the treatment of coronavirus disease 2019 (COVID-19) \citep{remdesivir2020nejm}. A total of 1,062 patients were randomized at baseline (with 541 assigned to the treated group [remdesivir] and 521 to the placebo). 

To illustrate our proposed methods, we consider an ordinal severity endpoint of 8-levels (1 through 8 with higher values indicating more severe condition) defined in \citep{remdesivir2020nejm}  on day 27. We use baseline age, sex, race and the baseline ordinal score as covariates. The baseline ordinal score is the same ordinal endpoint taken at baseline, but only takes values in the worst four categories $\{4,5,6,7\}$ due to inclusion criteria. 
There are 11 missing values in the baseline severity score. Because handling missing covariates is outside the scope of this paper, we exclude observations with missing severity scores, yielding a total sample size of $n=1,051$. 

Table \ref{tab:sumstat-stu2} presents summary statistics for the baseline covariates and the endpoint, defined as the ordinal score at day 27. In this example, the overall proportion of missing endpoint data is small (approximately 9\%) and is similar between the two treatment groups.

\begin{table}[ht]
\centering
\small
\begin{tabular}{lcccccc}
  \toprule
Variable (with mean (SD) or $n(\%)$) & \makecell{Overall\\$(n=1,051)$} & \makecell{Placebo \\ $(n=518)$ } & \makecell{Treated \\ $(n=533)$ } \\ 
  \midrule
\multicolumn{4}{c}{\textbf{Baseline covariates}} \\
Age & 58.9 (15.0) & 59.1 (15.4) & 58.7 (14.6) \\
Sex = female  & 373 (64.5\%) & 188 (63.7\%) & 185 (65.3\%) \\
Race \\
~~~~White
& 562 (53.5\%) & 287 (55.4\%) & 275 (51.6\%) \\
~~~~Black or African American
& 219 (20.8\%) & 114 (22.0\%) & 105 (19.7\%) \\
~~~~Asian
& 135 (12.8\%) & 56 (10.8\%) & 79 (14.8\%) \\
~~~~Others
& 135 (12.8\%) & 61 (11.8\%) & 74 (13.9\%) \\
{Baseline ordinal score (value)} \\
~~~~Hospitalized, no supplemental oxygen, needs care (4)
& 138 (13.1\%) & 63 (12.2\%) & 75 (14.1\%) \\
~~~~Hospitalized, requiring supplemental oxygen (5)
& 435 (41.4\%) & 203 (39.2\%) & 232 (43.5\%) \\
~~~~Hospitalized, noninvasive ventilation or high-flow oxygen (6)
& 193 (18.4\%) & 98 (18.9\%) & 95 (17.8\%) \\
~~~~Hospitalized, invasive ventilation or ECMO (7)
& 285 (27.1\%) & 154 (29.7\%) & 131 (24.6\%) \\
\midrule
\multicolumn{4}{c}{\textbf{Endpoint}} \\
\textbf{Ordinal score at day 27 (value)} \\
~~~~Not hospitalized, no activity limitation (1/2)
& 647 (67.5\%) & 305 (63.7\%) & 342 (71.2\%)  \\
~~~~Hospitalized, no oxygen, no ongoing care (3)
& 7 (0.7\%) & 3 (0.6\%) & 4 (0.8\%)  \\
~~~~Hospitalized, no oxygen, ongoing care (4)
& 38 (4.0\%) & 19 (4.0\%) & 19 (4.0\%) \\
~~~~Hospitalized, requiring supplemental oxygen (5)
& 42 (4.4\%) & 22 (4.6\%) & 20 (4.2\%) \\
~~~~Hospitalized, high-flow oxygen or noninvasive ventilation (6)
& 18 (1.9\%) & 13 (2.7\%) & 5 (1.0\%) \\
~~~~Hospitalized, invasive ventilation or ECMO (7)
& 73 (7.6\%) & 41 (8.6\%) & 32 (6.7\%) \\
~~~~Death (8)
& 134 (14.0\%) & 76 (15.9\%) & 58 (12.1\%) \\
Missing data count 
& 92 (8.8\%) & 39 (7.5\%) & 53 (9.9\%) \\
   \bottomrule
\end{tabular}
\begin{tablenotes}
\scriptsize
\item SD: standard deviation. 
\item Numbers are presented as $n$ (\%). Percentages are calculated using non-missing observations and may not sum to column totals due to missing data.
\item The baseline severity score and the endpoint (ordinal score at day 27) are ordinal with prespecified numerical scores; outcome categories are ordered from best to worst (top to bottom).
\item ECMO: extra corporeal membrane oxygenation. 
\end{tablenotes}
\caption{Summary statistics for the ACTT-1 trial data.} 
\label{tab:sumstat-stu2}
\end{table}

We fitted a logistic regression model for the non-missingness indicator including all baseline covariates, treatment, and their interactions. The coefficient for the ``severity score = 6'' category was statistically significant (p-value $= 0.02$), whereas all other coefficients were not statistically significant (p-value $> 0.05$). 

Figure \ref{fig:ps-actt1} in Appendix \ref{subapp:data-ps} presents the histogram of the estimated non-missingness propensity scores by treatment groups. The distributions show good overlap between treatment groups. The minimum estimated propensity score across all participants is 0.74, suggesting no apparent violation of the positivity assumption. {Additional diagnostics showed that all estimated cell probabilities and raw win, loss, and tie probabilities were within $[0,1]$ across the three IPW/AIPW specifications. For the H\'ajek-normalized IPW estimator, the cell probabilities summed to one within each treatment group up to numerical precision, with maximum absolute deviations of $6.66\times10^{-16}$ without covariate adjustment and $2.22\times10^{-16}$ with covariate adjustment. Across all three specifications, the raw tie-probability estimates ranged from 0.4801 to 0.4837. A propensity-score monotonicity check was not applicable because this application involved a single endpoint.}

Table~\ref{tab:stu2-main} reports the win measure analyses for the day-27 ordinal score in the ACTT-1 trial. Across all four win measures (WR, WO, NB, and DOOR), the Standard, {CC, MI,} IPW (no covariates), and IPW (with covariates) methods provide statistically significant evidence of a treatment benefit at the 0.05 level, whereas the AIPW estimator yields p-values greater than 0.05. As expected, and consistent with the results for the single ordinal score in the SCOUT-CAP trial, the standard analysis and the IPW analysis without covariates coincide exactly for the ratio-type measures WR and WO. 

Incorporating baseline covariates into the IPW and AIPW estimators generally produces larger p-values for all win measures. Although the AIPW estimator yields smaller SEs, its point estimates are more attenuated than those from the IPW estimators; for WR, the estimates are 1.365, 1.343, and 1.217 for IPW without covariates, covariate-adjusted IPW, and covariate-adjusted AIPW, respectively. We therefore conducted additional diagnostics in Appendix \ref{subapp:actt1-aipw}. In Appendix \ref{subsubapp:actt1-aipw}, collapsing sparse outcome categories 3--4 or 3--7 produced nearly unchanged AIPW WR estimates of 1.215 and 1.218, respectively, suggesting that sparse categories were not driving the shift. When only baseline severity was retained in the outcome model, the AIPW estimate remained close to the primary result at 1.231, whereas excluding baseline severity and retaining age, sex, and race increased the estimate to 1.318. Baseline severity also showed clear associations with the observed day-27 ordinal outcome in both treatment groups, supporting its important prognostic role. In Appendix \ref{subsubapp:actt1-simulation}, we conducted an ACTT-1-motivated simulation to assess whether a discrepancy of the observed magnitude could arise in finite samples. The IPW and AIPW estimators remained approximately unbiased, achieved near-nominal coverage, and had uncertainty comparable to that observed in ACTT-1. The replicate-specific absolute difference, $\left|\widehat{\theta}_{\mathrm{WR}}^{\mathrm{ipw}}-\widehat{\theta}_{\mathrm{WR}}^{\mathrm{aipw}}\right|$, was at least 0.10 in 28.5\% of simulation replicates and at least 0.15 in 10.6\% of replicates. Taken together, these results suggest that the observed IPW--AIPW difference in ACTT-1 is not attributable to sparse outcome categories or obvious model instability, but is consistent with the prognostic information incorporated through the AIPW outcome model together with plausible finite-sample variability and the observed difference is plausible in a single sample. Because the true treatment effect is unknown in the observed data, we cannot determine which point estimate is closer to the underlying truth. However, the simulation study indicates that the AIPW estimator can be a more efficient estimator, as its outcome regression component accounts for the strong association between a predictive baseline covariate and the outcome.

\begin{table}[H]
\centering
\caption{Win analysis results for the day-27 ordinal score in the ACTT-1 trial.}\label{tab:stu2-main}
\begin{tabular}{lrcccc}
\toprule
\textbf{Measure} & \textbf{Method} & \textbf{Estimate} & \textbf{SE} & \textbf{95\% CI} & \textbf{p-value} \\
\midrule
\multirow{6}{*}{WR} 
   & Standard & 1.365 & 0.165 & (1.077, 1.730) & 0.010 \\ 
   & {CC} & {1.365} & {0.165} & {(1.077, 1.730)} & {0.010} \\
   & {MI} & {1.338} & {0.161} & {(1.057, 1.694)} & {0.016} \\
   & IPW (no covariate) & 1.365 & 0.165 & (1.077, 1.730) & 0.010 \\ 
   & IPW (with covariates) & 1.343 & 0.163 & (1.060, 1.704) & 0.015 \\ 
   & AIPW (with covariates) & 1.217 & 0.138 & (0.974, 1.520) & 0.084 \\ 
\midrule
\multirow{6}{*}{WO} 
   & Standard & 1.174 & 0.073 & (1.039, 1.326) & 0.010 \\ 
   & {CC} & {1.174} & {0.073} & {(1.039, 1.326)} & {0.010} \\
   & {MI} & {1.162} & {0.072} & {(1.029, 1.312)} & {0.016} \\
   & IPW (no covariate) & 1.174 & 0.073 & (1.039, 1.326) & 0.010 \\ 
   & IPW (with covariates) & 1.164 & 0.072 & (1.030, 1.314) & 0.015 \\ 
   & AIPW (with covariates) & 1.107 & 0.065 & (0.987, 1.242) & 0.083 \\ 
\midrule
\multirow{6}{*}{NB} 
   & Standard & 0.067 & 0.026 & (0.016, 0.117) & 0.010 \\ 
   & {CC} & {0.080} & {0.031} & {(0.019, 0.140)} & {0.010} \\
   & {MI} & {0.075} & {0.031} & {(0.014, 0.135)} & {0.015} \\
   & IPW (no covariate) & 0.080 & 0.031 & (0.019, 0.140) & 0.010 \\ 
   & IPW (with covariates) & 0.076 & 0.031 & (0.015, 0.136) & 0.014 \\ 
   & AIPW (with covariates) & 0.051 & 0.029 & (-0.007, 0.108) & 0.083 \\ 
\midrule
\multirow{6}{*}{DOOR} 
   & Standard & 0.533 & 0.013 & (0.508, 0.558) & 0.010 \\ 
   & {CC} & {0.540} & {0.015} & {(0.510, 0.570)} & {0.010} \\
   & {MI} & {0.537} & {0.015} & {(0.507, 0.568)} & {0.015} \\
   & IPW (no covariate) & 0.540 & 0.015 & (0.510, 0.570) & 0.010 \\ 
   & IPW (with covariates) & 0.538 & 0.015 & (0.508, 0.568) & 0.014 \\ 
   & AIPW (with covariates) & 0.525 & 0.015 & (0.497, 0.554) & 0.083 \\ 
\bottomrule
\end{tabular}
\begin{tablenotes}\footnotesize
    \item SE: standard error estimate; CI: confidence interval; CC: complete-case analysis; MI: multiple imputation.
\end{tablenotes}
\end{table}

\section{Concluding Remarks}\label{sec:concludes}

In this paper, we propose two estimators for win measures in randomized clinical trials involving multiple hierarchical ordinal (including binary) endpoints subject to missingness. Under a conditional independence assumption allowing missingness to depend on treatment and baseline covariates, we propose an IPW estimator that reweights the contributions of pairwise comparisons and corrects the bias of the Standard approach.

Building on the IPW estimator, we develop an AIPW estimator that incorporates outcome regression and enjoys double robustness. When both nuisance models are correctly specified, it attains the semiparametric efficiency bound for each joint cell probability under the corresponding reduced observed-data model. {We do not claim semiparametric efficiency for the final win measures under the full observed-data model. Nevertheless, our simulations demonstrate the potential of outcome augmentation to improve estimation precision. Deriving estimators that fully exploit partially observed endpoint information under the full observed-data model remains an important direction for future research.
}

{A limitation of the IPW construction is that estimation at hierarchy level $k$ uses only participants for whom the first $k$ endpoints are jointly observed. Consequently, an observed lower-priority endpoint does not contribute to estimation at that level if any preceding endpoint is missing. This restriction arises from the hierarchical cell-probability construction rather than from a monotone-missingness assumption. The outcome-regression component of the AIPW estimator partially mitigates this limitation by using baseline covariates to predict the relevant cell probabilities for all participants.} Under the stated regularity conditions, the influence-function-based asymptotic variance estimator in Section~\ref{subsec:varest} provides valid uncertainty quantification for both the IPW and AIPW estimators.

{
Through extensive simulations, we show that, under the considered missingness mechanisms, our IPW and AIPW methods generally yield more reliable win-measure estimates than the Standard approach, which treats comparisons involving missing values as ties, and complete-case analysis. MI provides a complementary approach for handling missing hierarchical endpoints, particularly under an MAR assumption. MI and IPW rely on their respective imputation and missingness models, whereas AIPW remains valid when either the missingness model or the outcome model is correctly specified. In our simulations, MI generally yielded small bias and near-nominal coverage, making it a useful practical benchmark. When both AIPW nuisance models were correctly specified, AIPW generally achieved smaller RMSEs and narrower CIs than IPW and MI. Finally, applications to the SCOUT-CAP and ACTT-1 clinical trials illustrate the practical utility of our methods in settings with non-negligible missing outcome data.
}

We acknowledge several limitations that motivate future research. First, our current proposals focus on ordinal (or binary) endpoints and do not directly accommodate continuous endpoints, which are also commonly observed in practice. Extending the framework to hierarchies containing continuous components would be meaningful but non-trivial because it would require modeling joint distributions involving continuous or mixed discrete and continuous variables. Density-estimation methods, such as kernel density estimation \citep{silverman2018density}, may provide a useful starting point.

Second, the proposed estimators may be affected by data sparsity when ordinal endpoints have many categories or when the number of endpoints is large. Finite-sample performance may deteriorate when some joint categories have very small or zero cell probabilities. Our current R package supports up to three endpoints. A practical remedy may be to merge sparse categories before applying the proposed methods; however, principled and possibly data-adaptive procedures for handling sparsity remain an important direction for future work.

{Third, Assumption~\ref{asp:miss-pos} excludes missingness mechanisms in which the observation of a later endpoint depends on previously observed outcomes. If dropout depends on such outcomes, the IPW and AIPW estimators are not generally guaranteed to be consistent. Our additional sensitivity simulations in Appendix~\ref{subapp:miss-vio} illustrate this limitation and show that MI can provide a useful complementary analysis in such settings. Therefore, when outcome-dependent dropout is clinically plausible, the results should be interpreted together with appropriate sensitivity analyses. If endpoints are collected in a prespecified temporal order, a separate monotone-missingness model may be considered, but this is a distinct assumption and is not required for the cumulative indicators used in our framework.}

Finally, several broader extensions are worth pursuing. The general strategy may be adapted to observational studies with treatment propensity scores \citep{shu2026desirability, cao2025covariate}, settings in which treatment allocation probabilities are estimated even under randomization \citep{liu2025coadvise, gao2024does}, overlap weighting \citep{li2018balancing, matsouaka2025overlap, liu2024average, li2025variance}, stratified win measures \citep{dong2018stratified}, multi-center clinical trials \citep{zhuang2025assessment}, clustered randomized trials \citep{fang2025sample}, and targeted subgroup analyses \citep{liu2025targeted, liu2024multi, liu2026privacy}. To further reduce reliance on parametric nuisance models, future work may investigate sample splitting and cross-fitting with flexible machine-learning methods \citep{van2007super, chernozhukov2018double, wang2025rate, westling2024inference}.

\subsection*{Acknowledgments}

Yi Liu was supported by the National Heart, Lung, and Blood Institute (NHLBI) of the National Institutes of Health (NIH) under Award Number T32HL079896. The content of this paper is solely the responsibility of the authors and does not necessarily represent the official views of the NIH. The Duke Clinical Research Institute Biostatistics \& Data Science Research Fund also supported this work.

\subsection*{Data Availability Statement}

The real data analyzed in Section~\ref{sec:data} are not publicly available but may be requested from the corresponding trial investigators.

\subsection*{Disclosure Statement}

The authors declare no conflicts of interest. We acknowledge the use of OpenAI's ChatGPT generative AI model solely for language polishing, consistency checking, grammatical editing, and summarizing simulation results generated in R. No other large language models were used in the preparation of this manuscript. The authors take full responsibility for the final content of the manuscript. 

\bibliography{refs}

@inproceedings{liu2026privacy,
  title={Privacy-Protected Causal Survival Analysis Under Distribution Shift},
  author={Liu, Yi and Levis, Alexander and Zhu, Ke and Yang, Shu and Gilbert, Peter and Han, Larry},
  booktitle={International Conference on Learning Representations},
  volume={2026},
  pages={94685--94722},
  year={2026}
}

@article{heitjan1993ignorability,
  title={Ignorability and coarse data: Some biomedical examples},
  author={Heitjan, Daniel F},
  journal={Biometrics},
  volume={49},
  number={4},
  pages={1099--1109},
  year={1993},
  publisher={JSTOR}
}

@article{shu2026doubly,
  title={Doubly robust estimation of desirability of outcome ranking {(DOOR)} probability with application to {MDRO} studies},
  author={Shu, Shiyu and Hamasaki, Toshimitsu and Evans, Scott and Komarow, Lauren and van Duin, David and Diao, Guoqing},
  journal={Statistics in Biopharmaceutical Research},
  number={just-accepted},
  pages={1--10},
  year={2026},
  url={https://doi.org/10.1080/19466315.2026.2697866},
  publisher={Taylor \& Francis}
}

@article{shu2026desirability,
  title={Desirability of outcome ranking {(DOOR)} analysis for multivariate survival outcomes with application to {ACTT-1} trial},
  author={Shu, Shiyu and Diao, Guoqing and Hamasaki, Toshimitsu and Evans, Scott},
  journal={Clinical Trials},
  volume={23},
  number={1},
  pages={23--32},
  year={2026},
  publisher={SAGE Publications Sage UK: London, England}
}

@article{evans2016using,
  title={Using outcomes to analyze patients rather than patients to analyze outcomes: a step toward pragmatism in benefit: risk evaluation},
  author={Evans, Scott R and Follmann, Dean},
  journal={Statistics in Biopharmaceutical Research},
  volume={8},
  number={4},
  pages={386--393},
  year={2016},
  publisher={Taylor \& Francis}
}

@article{siquier2025use,
  title={Use and misuse of composite endpoints in randomised clinical trials},
  author={Siquier-Padilla, Joan and Gonzalez-Manzanares, Rafael and Rossello, Xavier},
  journal={Heart},
  volume={112},
  number={18},
  pages={1004--1013},
  year={2026},
  publisher={BMJ Publishing Group Ltd and British Cardiovascular Society}
}

@article{sankoh2014use,
  title={Use of composite endpoints in clinical trials},
  author={Sankoh, Abdul J and Li, Haihong and D'Agostino Sr, Ralph B},
  journal={Statistics in Medicine},
  volume={33},
  number={27},
  pages={4709--4714},
  year={2014},
  publisher={Wiley Online Library}
}

@article{overbey2025navigating,
  title={Navigating composite endpoints: Methods and recommendations for trial design and interpretation},
  author={Overbey, Jessica R and Mentz, Robert J and Allen-Savietta, Cora},
  journal={Journal of Cardiac Failure},
  year={2025},
  volume={32},
  number={8},
  pages={1428--1427},
  publisher={Elsevier}
}

@article{nascimento2026making,
  title={Making sense of composite endpoints: efficiency, meaning and clinical relevance in modern cardiovascular trials},
  author={Nascimento, Bruno R and Marino, B{\'a}rbara CA and Marino, Marcos Antonio},
  journal={Heart},
  year={2026},
  volume={112},
  pages={1004--1013},
  publisher={BMJ Publishing Group Ltd and British Cardiovascular Society}
}

@incollection{hamasaki2024design,
  title={Design of clinical trials with the desirability of outcome ranking methodology},
  author={Hamasaki, Toshimitsu and He, Yijie and Wu, Qihang and Evans, Scott R},
  booktitle={Biostatistics in Biopharmaceutical Research and Development: Clinical Trial Design, Volume 1},
  pages={137--159},
  year={2024},
  publisher={Springer}
}

@article{dong2018stratified,
  title={The stratified win ratio},
  author={Dong, Gaohong and Qiu, Junshan and Wang, Duolao and Vandemeulebroecke, Marc},
  journal={Journal of Biopharmaceutical Statistics},
  volume={28},
  number={4},
  pages={778--796},
  year={2018},
  publisher={Taylor \& Francis}
}

@article{fang2025sample,
  title={Sample size determination for win statistics in cluster-randomized trials},
  author={Fang, Xi and Cao, Zhiqiang and Li, Fan},
  journal={arXiv preprint arXiv:2510.22709},
  year={2025}
}

@article{liu2026estimation,
  title={Estimation and inference of the win ratio for two hierarchical endpoints subject to censoring and missing data},
  author={Liu, Yi and Barnhart, Huiman and O’Brien, Sean and Lokhnygina, Yuliya and Matsouaka, Roland A},
  journal={Journal of Biopharmaceutical Statistics},
  pages={1--28},
  year={2026},
  publisher={Taylor \& Francis}
}

@article{li2025variance,
  title={Variance estimation for weighted average treatment effects},
  author={Li, Huiyue and Liu, Yi and Zhou, Yunji and Liu, Jiajun and Fu, Dezhao and Matsouaka, Roland A},
  journal={Statistics in Biosciences},
  pages={1--73},
  year={2025},
  publisher={Springer}
}

@article{dong2026win,
  title={Win statistics (win ratio, win odds, and net benefit): Noncollapsibility and standardization for randomized clinical trials},
  author={Dong, Gaohong and Gamalo-Siebers, Margaret and Cui, Ying and Huang, Bo and Luo, Xiaolong and Tian, Lu},
  journal={Journal of Biopharmaceutical Statistics},
  number={just-accepted},
  pages={1--17},
  year={2026},
  url={https://doi.org/10.1080/10543406.2026.2626058},
  publisher={Taylor \& Francis}
}

@article{wang2023missing,
  title={Missing data imputation for a multivariate outcome of mixed variable types},
  author={Wang, Tuo and Zilinskas, Rachel and Li, Ying and Qu, Yongming},
  journal={Statistics in Biopharmaceutical Research},
  volume={15},
  number={4},
  pages={826--837},
  year={2023},
  publisher={Taylor \& Francis}
}

@article{cao2025covariate,
  title={Covariate-adjusted win statistics in randomized clinical trials with ordinal outcomes},
  author={Cao, Zhiqiang and Zuo, Scott and Baumann, Mary Ryan and Plourde, Kendra and Heagerty, Patrick and Tong, Guangyu and Li, Fan},
  journal={arXiv preprint arXiv:2508.20349},
  year={2025}
}

@article{matsouaka2025overlap,
  title={Overlap, matching, or entropy weights: what are we weighting for?},
  author={Matsouaka, Roland A and Liu, Yi and Zhou, Yunji},
  journal={Communications in Statistics--Simulation and Computation},
  volume={54},
  number={7},
  pages={2672--2691},
  year={2025},
  publisher={Taylor \& Francis}
}

@book{silverman2018density,
  title={Density estimation for statistics and data analysis},
  author={Silverman, Bernard W},
  year={2018},
  publisher={Routledge}
}

@article{song2023win,
  title={The win odds: statistical inference and regression},
  author={Song, James and Verbeeck, Johan and Huang, Bo and Hoaglin, David C and Gamalo-Siebers, Margaret and Seifu, Yodit and Wang, Duolao and Cooner, Freda and Dong, Gaohong},
  journal={Journal of Biopharmaceutical Statistics},
  volume={33},
  number={2},
  pages={140--150},
  year={2023},
  publisher={Taylor \& Francis}
}

@article{buyse2010generalized,
  title={Generalized pairwise comparisons of prioritized outcomes in the two-sample problem},
  author={Buyse, Marc},
  journal={Statistics in Medicine},
  volume={29},
  number={30},
  pages={3245--3257},
  year={2010},
  publisher={Wiley Online Library}
}

@book{Buyse2025GPC,
  title     = {Handbook of Generalized Pairwise Comparisons: Methods for Patient-Centric Analysis},
  editor    = {Marc Buyse and Johan Verbeeck and Everardo D. Saad and Mickaël De Backer and Vaiva Deltuvaite-Thomas and Geert Molenberghs},
  year      = {2025},
  publisher = {Chapman and Hall/CRC},
  address   = {New York},
  isbn      = {9781003390855},
  doi       = {10.1201/9781003390855}
}

@article{williams2022short,
  title={Short-vs standard-course outpatient antibiotic therapy for community-acquired pneumonia in children: the {SCOUT-CAP} randomized clinical trial},
  author={Williams, Derek J and Creech, C Buddy and Walter, Emmanuel B and Martin, Judith M and Gerber, Jeffrey S and Newland, Jason G and Howard, Lee and Hofto, Meghan E and Staat, Mary A and Oler, Randolph E and others},
  journal={JAMA pediatrics},
  volume={176},
  number={3},
  pages={253--261},
  year={2022}
}

@article{remdesivir2020nejm,
author = {John H. Beigel  and Kay M. Tomashek  and Lori E. Dodd  and Aneesh K. Mehta  and Barry S. Zingman  and Andre C. Kalil  and Elizabeth Hohmann  and Helen Y. Chu  and Annie Luetkemeyer  and Susan Kline  and Diego Lopez de Castilla  and Robert W. Finberg  and Kerry Dierberg  and Victor Tapson  and Lanny Hsieh  and Thomas F. Patterson  and Roger Paredes  and Daniel A. Sweeney  and William R. Short  and Giota Touloumi  and David Chien Lye  and Norio Ohmagari  and Myoung-don Oh  and Guillermo M. Ruiz-Palacios  and Thomas Benfield  and Gerd Fätkenheuer  and Mark G. Kortepeter  and Robert L. Atmar  and C. Buddy Creech  and Jens Lundgren  and Abdel G. Babiker  and Sarah Pett  and James D. Neaton  and Timothy H. Burgess  and Tyler Bonnett  and Michelle Green  and Mat Makowski  and Anu Osinusi  and Seema Nayak  and H. Clifford Lane },
title = {Remdesivir for the Treatment of {COVID}-19 — Final Report},
journal = {New England Journal of Medicine},
volume = {383},
number = {19},
pages = {1813-1826},
year = {2020},
doi = {10.1056/NEJMoa2007764},

URL = {https://www.nejm.org/doi/full/10.1056/NEJMoa2007764},
eprint = {https://www.nejm.org/doi/pdf/10.1056/NEJMoa2007764}
}

@incollection{kennedy2016semiparametric,
  title={Semiparametric theory and empirical processes in causal inference},
  author={Kennedy, Edward H},
  booktitle={Statistical causal inferences and their applications in public health research},
  pages={141--167},
  year={2016},
  publisher={Springer}
}

@phdthesis{zheng2024use,
  title={The Use of the Win Ratio Method in Clinical Trials},
  author={Zheng, Sirui},
  year={2024},
  school={Liverpool School of Tropical Medicine}
}

@article{pocock2024win,
  title={The win ratio in cardiology trials: lessons learnt, new developments, and wise future use},
  author={Pocock, Stuart J and Gregson, John and Collier, Timothy J and Ferreira, Joao Pedro and Stone, Gregg W},
  journal={European Heart Journal},
  volume={45},
  number={44},
  pages={4684--4699},
  year={2024},
  publisher={Oxford University Press UK}
}

@article{wang2025adjusted,
  title={Adjusted win ratio using the inverse probability of treatment weighting},
  author={Wang, Duolao and Zheng, Sirui and Cui, Ying and He, Nengjie and Chen, Tao and Huang, Bo},
  journal={Journal of Biopharmaceutical Statistics},
  volume={35},
  number={1},
  pages={21--36},
  year={2025},
  publisher={Taylor \& Francis}
}

@article{liu2024average,
  title={Average treatment effect on the treated, under lack of positivity},
  author={Liu, Yi and Li, Huiyue and Zhou, Yunji and Matsouaka, Roland A},
  journal={Statistical Methods in Medical Research},
  volume={33},
  number={10},
  pages={1689--1717},
  year={2024},
  publisher={SAGE Publications Sage UK: London, England}
}

@article{liu2025targeted,
  title={Targeted Data Fusion for Region-Specific Survival Effects in the AMP HIV Prevention Trials},
  author={Liu, Yi and Levis, Alexander W and Zhu, Ke and Yang, Shu and Gilbert, Peter B and Han, Larry},
  journal={arXiv preprint arXiv:2501.18798},
  year={2025}
}

@article{dong2020inverse,
  title={The inverse-probability-of-censoring weighting (IPCW) adjusted win ratio statistic: an unbiased estimator in the presence of independent censoring},
  author={Dong, Gaohong and Mao, Lu and Huang, Bo and Gamalo-Siebers, Margaret and Wang, Jiuzhou and Yu, GuangLei and Hoaglin, David C},
  journal={Journal of Biopharmaceutical Statistics},
  volume={30},
  number={5},
  pages={882--899},
  year={2020},
  publisher={Taylor \& Francis}
}

@article{dong2021adjusting,
  title={Adjusting win statistics for dependent censoring},
  author={Dong, Gaohong and Huang, Bo and Wang, Duolao and Verbeeck, Johan and Wang, Jiuzhou and Hoaglin, David C},
  journal={Pharmaceutical Statistics},
  volume={20},
  number={3},
  pages={440--450},
  year={2021},
  publisher={Wiley Online Library}
}

@article{gao2024does,
  title={When does adjusting covariate under randomization help? A comparative study on current practices},
  author={Gao, Ying and Liu, Yi and Matsouaka, Roland},
  journal={BMC Medical Research Methodology},
  volume={24},
  number={1},
  pages={250},
  year={2024},
  publisher={Springer}
}

@article{wang2025rate,
  title={Rate doubly robust estimation for weighted average treatment effects},
  author={Wang, Yiming and Liu, Yi and Yang, Shu},
  journal={Journal of Causal Inference},
  volume={13},
  number={1},
  pages={20240073},
  year={2025},
  publisher={De Gruyter}
}

@article{robins1995analysis,
  title={Analysis of semiparametric regression models for repeated outcomes in the presence of missing data},
  author={Robins, James M and Rotnitzky, Andrea and Zhao, Lue Ping},
  journal={Journal of the American Statistical Association},
  volume={90},
  number={429},
  pages={106--121},
  year={1995},
  publisher={Taylor \& Francis}
}

@article{rosenbaum1987model,
  title={Model-based direct adjustment},
  author={Rosenbaum, Paul R},
  journal={Journal of the American Statistical Association},
  volume={82},
  number={398},
  pages={387--394},
  year={1987},
  publisher={Taylor \& Francis}
}

@article{van2007super,
  title={Super learner},
  author={Van der Laan, Mark J and Polley, Eric C and Hubbard, Alan E},
  journal={Statistical applications in genetics and molecular biology},
  volume={6},
  number={1},
  year={2007},
  pages={1--23},
  publisher={De Gruyter}
}

@article{barnhart2025sample,
  title={Sample Size and Power Calculations With Win Measures Based on Hierarchical Endpoints},
  author={Barnhart, Huiman and Lokhnygina, Yuliya and Matsouaka, Roland and Halabi, Susan and Yanez, David and Mentz, Robert J and Rockhold, Frank},
  journal={Statistics in Medicine},
  volume={44},
  number={10-12},
  pages={e70096},
  year={2025}
}

@article{wang2025restricted,
  title={Restricted Time Win Ratio: From Estimands to Estimation},
  author={Wang, Tuo and Li, Ying and Qu, Yongming},
  journal={Statistics in Biopharmaceutical Research},
  volume={17},
  number={1},
  pages={136--148},
  year={2025},
  publisher={Taylor \& Francis}
}

@article{cui2025wins,
  title={WINS: The R WINS Package},
  author={Cui, Ying and Huang, Bo},
  year={2025},
  journal={The Comprehensive R Archive Network (CRAN)},
  url={https://cran.r-project.org/web/packages/WINS/}
}

@article{liu2025coadvise,
  title={COADVISE: Covariate Adjustment with Variable Selection and Missing Data Imputation in Randomized Controlled Trials},
  author={Liu, Yi and Zhu, Ke and Han, Larry and Yang, Shu},
  journal={arXiv preprint arXiv:2501.08945},
  year={2025}
}

@article{li2024elusiveness,
  title={The Elusiveness of the Win Ratio Parameter in the Presence of Missing Data},
  author={Li, Heng and Chen, Wei-Chen and Lu, Nelson and Tang, Rong and Zhao, Yu},
  journal={Therapeutic Innovation \& Regulatory Science},
  volume={58},
  number={3},
  pages={431--432},
  year={2024},
  publisher={Springer}
}

@article{barnhart2025trial,
  title={Trial design with win statistics for multiple time-to-event endpoints with hierarchy},
  author={Barnhart, Huiman X and Lokhnygina, Yuliya and Matsouaka, Roland A and Rockhold, Frank W},
  journal={Statistics in Biopharmaceutical Research},
  volume={17},
  number={2},
  pages={197--210},
  year={2025},
  publisher={Taylor \& Francis}
}

@article{zhuang2025assessment,
  title={Assessment of treatment effect heterogeneity for multiregional randomized clinical trials},
  author={Zhuang, Haotian and Wang, Xiaofei and George, Stephen L},
  journal={Statistics in Biopharmaceutical Research},
  volume={17},
  number={3},
  pages={315--322},
  year={2025},
  publisher={Taylor \& Francis}
}

@article{dong2023win,
  title={Win statistics (win ratio, win odds, and net benefit) can complement one another to show the strength of the treatment effect on time-to-event outcomes},
  author={Dong, Gaohong and Huang, Bo and Verbeeck, Johan and Cui, Ying and Song, James and Gamalo-Siebers, Margaret and Wang, Duolao and Hoaglin, David C and Seifu, Yodit and M{\"u}tze, Tobias and others},
  journal={Pharmaceutical Statistics},
  volume={22},
  number={1},
  pages={20--33},
  year={2023},
  publisher={Wiley Online Library}
}

@article{westling2024inference,
  title={Inference for treatment-specific survival curves using machine learning},
  author={Westling, Ted and Luedtke, Alex and Gilbert, Peter B and Carone, Marco},
  journal={Journal of the American Statistical Association},
  volume={119},
  number={546},
  pages={1541--1553},
  year={2024},
  publisher={Taylor \& Francis}
}

@article{pocock2012win,
  title={The win ratio: a new approach to the analysis of composite endpoints in clinical trials based on clinical priorities},
  author={Pocock, Stuart J and Ariti, Cono A and Collier, Timothy J and Wang, Duolao},
  journal={European Heart Journal},
  volume={33},
  number={2},
  pages={176--182},
  year={2012},
  publisher={Oxford University Press}
}

@article{rubin1976inference,
  title={Inference and missing data},
  author={Rubin, Donald B},
  journal={Biometrika},
  volume={63},
  number={3},
  pages={581--592},
  year={1976},
  publisher={Oxford University Press}
}

@misc{hajek1971comment,
	title={Comment on a paper by D. Basu In: Godambe {VP} and Sprott {DA} (eds) Foundations of Statistical Inference},
	author={H{\'a}jek, J},
	year={1971},
	publisher={Toronto: Holt, Rinehart and Winston}
}

@article{hirano2003efficient,
	title={Efficient estimation of average treatment effects using the estimated propensity score},
	author={Hirano, Keisuke and Imbens, Guido W and Ridder, Geert},
	journal={Econometrica},
	volume={71},
	number={4},
	pages={1161--1189},
	year={2003},
	publisher={Wiley Online Library}
}

@article{li2018balancing,
    title={Balancing covariates via propensity score weighting},
    author={Li, Fan and Morgan, Kari Lock and Zaslavsky, Alan M},
    journal={Journal of the American Statistical Association},
    volume={113},
    number={521},
    pages={390--400},
    year={2018},
    publisher={Taylor \& Francis}
}

@article{robins1994estimation,
	title={Estimation of regression coefficients when some regressors are not always observed},
	author={Robins, James M and Rotnitzky, Andrea and Zhao, Lue Ping},
	journal={Journal of the American Statistical Association},
	volume={89},
	number={427},
	pages={846--866},
	year={1994},
	publisher={Taylor \& Francis}
}

@article{rubin1974estimating,
  title={Estimating causal effects of treatments in randomized and nonrandomized studies.},
  author={Rubin, Donald B},
  journal={Journal of Educational Psychology},
  volume={66},
  number={5},
  pages={688},
  year={1974},
  publisher={American Psychological Association}
}

@book{tsiatis2007semiparametric,
	title={Semiparametric theory and missing data},
	author={Tsiatis, Anastasios},
	year={2007},
	publisher={Springer Science \& Business Media}
}

@article{liu2024multi,
  title={Multi-source conformal inference under distribution shift},
  author={Liu, Yi and Levis, Alexander W and Normand, Sharon-Lise and Han, Larry},
  journal={Proceedings of Machine Learning Research},
  volume={235},
  pages={31344--31382},
  year={2024}
}

@article{horvitz1952generalization,
  title={A generalization of sampling without replacement from a finite universe},
  author={Horvitz, Daniel G and Thompson, Donovan J},
  journal={Journal of the American Statistical Association},
  volume={47},
  number={260},
  pages={663--685},
  year={1952},
  publisher={Taylor \& Francis}
}

@inproceedings{rubin1978multiple,
  title={Multiple imputations in sample surveys-a phenomenological Bayesian approach to nonresponse},
  author={Rubin, Donald B},
  booktitle={Proceedings of the survey research methods section of the American Statistical Association},
  volume={1},
  pages={20--34},
  year={1978},
  organization={American Statistical Association Alexandria, VA, USA}
}

@article{chernozhukov2018double,
  title={Double/debiased machine learning for treatment and structural parameters},
  author={Chernozhukov, Victor and Chetverikov, Denis and Demirer, Mert and Duflo, Esther and Hansen, Christian and Newey, Whitney and Robins, James},
  year={2018},
    journal={The Econometrics Journal},
    volume={21},
    number={1},
    pages={C1--C68},
  publisher={Oxford University Press Oxford, UK}
}
\bibliographystyle{plainnat}

\newpage

\appendix

\setcounter{section}{0}
\counterwithin{equation}{section}
\counterwithin{table}{section}
\counterwithin{figure}{section}
\counterwithin{remark}{section}
\counterwithin{assumption}{section}
\counterwithin{theorem}{section}

\begin{center}
    \LARGE Appendix
\end{center}

\section{Technical Proofs}\label{app:proof}

This appendix collects the large-sample theory for the proposed estimators. After stating common regularity conditions and a marginalization result, we present the Horvitz--Thompson IPW, H\'ajek IPW, and AIPW estimators in that order. For each estimator, we give consistency and influence-function-based variance estimation; the semiparametric efficiency result is presented with the AIPW theory.

\subsection{Regularity conditions}\label{subapp:regularity}

We impose the following conditions for the consistency and asymptotic linearity results stated in the main text and derived in the subsequent appendices.

\begin{enumerate}[label=(R\arabic*)]

\item The observations are independent and identically distributed (i.i.d.), and the number of endpoint layers and categories at each layer is finite and fixed. The treatment allocation probability satisfies
\[
0<e_a:=\Pr(A=a)<1.
\]
Under complete randomization, treatment assignment is independent of the baseline covariates and potential outcomes. If the allocation probability is estimated by the sample proportion, then
\[
\widehat e_a=\frac{n_a}{n}\xrightarrow{p}e_a,
\qquad
\sqrt{n}(\widehat e_a-e_a)
=
\frac{1}{\sqrt n}\sum_{i=1}^n
\{I(A_i=a)-e_a\}+o_p(1).
\]
When $e_a$ is known by design, the corresponding estimation term is omitted.

\item For each layer $k$ and treatment level $a$, there exists
$\epsilon>0$ such that
\[
\inf_{\mb x}\pi_k(a,\mb x)\geq\epsilon,
\qquad
e_a\geq\epsilon.
\]
With probability tending to one,
\[
\inf_{\mb x}\widehat\pi_k(a,\mb x)\geq\epsilon,
\qquad
\widehat e_a\geq\epsilon.
\]
Consequently, all inverse-probability weights are bounded in probability.

\item The nuisance estimators converge uniformly to their probability limits. Under correct specification,
\[
\sup_{\mb x}
\left|\widehat\pi_k(a,\mb x)-\pi_k(a,\mb x)\right|
\xrightarrow{p}0,
\]
and
\[
\sup_{\mb x}
\left|\widehat\mu_k(a,\mb x;i_1,\ldots,i_k)
-\mu_k(a,\mb x;i_1,\ldots,i_k)\right|
\xrightarrow{p}0.
\]
Under misspecification, the estimators converge uniformly to well-defined probability limits. Because the number of layers and categories is finite, these conditions hold jointly over all layers, treatment groups, and cells.

\item The nuisance models are finite-dimensional and twice continuously differentiable in neighborhoods of their probability limits. The corresponding population information or Jacobian matrices are finite and nonsingular, and the nuisance estimators admit joint first-order expansions. In particular,
\[
\sqrt n(\widehat{\bd\beta}_k^{(a)}-\bd\beta_{k,0}^{(a)})
=
\frac{1}{\sqrt n}\sum_{i=1}^n
\psi_{\beta_k^{(a)}}(\mc O_i)+o_p(1),
\]
and
\[
\sqrt n(\widehat{\bd\gamma}^{(a)}-\bd\gamma_0^{(a)})
=
\frac{1}{\sqrt n}\sum_{i=1}^n
\psi_{\gamma^{(a)}}(\mc O_i)+o_p(1).
\]
The corresponding score and influence-function components have finite second moments, and these expansions hold jointly across all layers and treatment groups.

\item The denominators used in the win measures are bounded away from zero:
\[
p_W>0,\qquad p_L>0,\qquad
p_W+\tfrac12p_T>0,\qquad
p_L+\tfrac12p_T>0.
\]
When the truncated estimator $p_T=\max\{0,1-p_W-p_L\}$ is used, we additionally assume $p_T>0$, so that the truncation does not impact the true tie probability estimation asymptotically.

\end{enumerate}

\subsection{Consistency of marginalizing higher-order cell probability estimators}
\label{app:marginal-consistency}

We provide a simple justification showing that a lower-order joint cell probability can be consistently estimated by marginalizing a consistent estimator of a higher-order joint cell probability. This result is useful for clarifying the internal coherence of the proposed estimators across different levels of the endpoint hierarchy.

By definition, for any $k\in\{2,\dots,K\}$,
$$
P_a^{1:k-1}(i_1,\ldots,i_{k-1})
= \sum_{i_k=1}^{\ell_k}
P_a^{1:k}(i_1,\ldots,i_k).
$$
Suppose that $\widehat P_a^{1:k}(i_1,\ldots,i_k)$ is a consistent estimator of
$P_a^{1:k}(i_1,\ldots,i_k)$ for every $(i_1,\ldots,i_k)$ in the support of
$(Y_1,\ldots,Y_k)$. Define the following estimator for the lower level joint cell probabilities:
$$
\widetilde P_a^{1:k-1}(i_1,\ldots,i_{k-1})
=
\sum_{i_k=1}^{\ell_k}
\widehat P_a^{1:k}(i_1,\ldots,i_k),
$$
which is by marginalizing over the support of $Y_k$. In the following, we call this estimator a ``marginal estimator'' for simplicity. Then,
$$
\widetilde P_a^{1:k-1}(i_1,\ldots,i_{k-1})
\xrightarrow{p}
P_a^{1:k-1}(i_1,\ldots,i_{k-1}).
$$
Since the support of $Y_k$ is finite and
$\widehat P_a^{1:k}(i_1,\ldots,i_k)
\xrightarrow{p}
P_a^{1:k}(i_1,\ldots,i_k)$
for each $i_k=1,\ldots,\ell_k$, it is straightforward that 
$$
\sum_{i_k=1}^{\ell_k}
\widehat P_a^{1:k}(i_1,\ldots,i_k)
-
\sum_{i_k=1}^{\ell_k}
P_a^{1:k}(i_1,\ldots,i_k)
=
\sum_{i_k=1}^{\ell_k}
\left\{
\widehat P_a^{1:k}(i_1,\ldots,i_k)
-
P_a^{1:k}(i_1,\ldots,i_k)
\right\}
\xrightarrow{p} 0.
$$
Therefore,
$$
\widetilde P_a^{1:k-1}(i_1,\ldots,i_{k-1})
\xrightarrow{p}
\sum_{i_k=1}^{\ell_k}
P_a^{1:k}(i_1,\ldots,i_k)
=
P_a^{1:k-1}(i_1,\ldots,i_{k-1}).
$$

This argument applies to both the proposed IPW and AIPW estimators, provided that the corresponding estimator of $P_a^{1:k}$ is consistent. For example, under the correctly specified non-missingness model, the IPW estimator $\widehat P_a^{1:k,\mathrm{ipw}}$ is consistent for $P_a^{1:k}$; hence the resulting marginal estimator $\widetilde P_a^{1:k-1}$ is consistent for $P_a^{1:k-1}$. Similarly, under the double-robust conditions for the AIPW estimator, namely that either the non-missingness propensity score model or the outcome model is correctly specified, $\widehat P_a^{1:k,\mathrm{aipw}}$ is consistent for $P_a^{1:k}$, and therefore the marginal estimator is consistent for $P_a^{1:k-1}$.

It is important to note, however, that this marginal estimator is generally not identical to the estimator obtained by directly estimating $P_a^{1:k-1}$. The direct estimator of $P_a^{1:k-1}$ uses the non-missingness indicator $\widetilde R_{1:k-1}$ and the corresponding propensity score $\pi_{k-1}(a,\mb X)$, whereas the marginal estimator based on $\widehat P_a^{1:k}$ uses $\widetilde R_{1:k}$ and $\pi_k(a,\mb X)$. Since $\widetilde R_{1:k}=1$ implies $\widetilde R_{1:k-1}=1$, but not conversely, the direct estimator can use participants observed through the $(k-1)$th endpoint even when the $k$th endpoint is missing. Therefore, while both estimators are consistent under their respective conditions, the direct level-specific estimator may be more efficient because it uses more of the partially observed data available at each hierarchy level.

\subsection{Theory of the Horvitz--Thompson IPW estimator}
\label{subapp:theory-ipw-HT}

This subsection establishes the consistency and influence-function-based
variance estimation for the Horvitz--Thompson (HT) IPW estimator introduced
in Section~\ref{subsec:ipw}.

\subsubsection{Consistency}

The consistency of the HT-IPW estimators of the win measures follows from
the consistency of the corresponding estimators of the joint cell
probabilities. It therefore suffices to show that
$\widehat P_a^{1:k,\mathrm{HT}}(i_1,\dots,i_k)$ consistently estimates
$P_a^{1:k}(i_1,\dots,i_k)$ for every $k=1,\dots,K$, $a=0,1$, and joint
cell $(i_1,\dots,i_k)$.

Under the regularity conditions in Appendix~\ref{subapp:regularity},
\[
\widehat{\Pr}(A=a)\xrightarrow{p}\Pr(A=a),
\]
and, under correct specification of the non-missingness model,
\[
\sup_{\mb x}
\left|
\widehat\pi_k(a,\mb x)-\pi_k(a,\mb x)
\right|
\xrightarrow{p}0,
\qquad
k=1,\dots,K.
\]
Therefore, by the uniform law of large numbers, Slutsky's theorem, and
Assumption~\ref{asp:miss-pos},
\begin{align*}
\widehat P_a^{1:k,\mathrm{HT}}(i_1,\dots,i_k)
&\xrightarrow{p}
\Ex\left\{
\frac{I(A=a)}{\Pr(A=a)}
\frac{\widetilde R_{1:k}}{\pi_k(a,\mb X)}
I(Y_1=i_1,\dots,Y_k=i_k)
\right\}
\\
&=
\Ex\left[
\frac{I(A=a)}{\Pr(A=a)}
\frac{1}{\pi_k(a,\mb X)}
\Ex\left\{
\widetilde R_{1:k}
I(Y_1=i_1,\dots,Y_k=i_k)
\mid A,\mb X
\right\}
\right]
\\
&=
\Ex\left[
\frac{I(A=a)}{\Pr(A=a)}
\frac{\Ex(\widetilde R_{1:k}\mid A=a,\mb X)}
{\pi_k(a,\mb X)}
\right.
\\[-1mm]
&\hspace{38mm}\left.
\times
\Ex\left\{
I(Y_1=i_1,\dots,Y_k=i_k)
\mid A=a,\mb X
\right\}
\right]
\\
&=
\Ex\left[
\frac{I(A=a)}{\Pr(A=a)}
\Ex\left\{
I(Y_1=i_1,\dots,Y_k=i_k)
\mid A=a,\mb X
\right\}
\right]
\\
&=
\Pr(Y_1=i_1,\dots,Y_k=i_k\mid A=a)
\\
&=
P_a^{1:k}(i_1,\dots,i_k).
\end{align*}
Hence,
$\widehat P_a^{1:k,\mathrm{HT}}(i_1,\dots,i_k)$
is consistent. By continuous mapping, the resulting HT-IPW estimators
$\widehat p_W^{\mathrm{HT}}$,
$\widehat p_L^{\mathrm{HT}}$, and
$\widehat p_T^{\mathrm{HT}}$, as well as the corresponding smooth win
measures, are consistent under the denominator conditions in
Appendix~\ref{subapp:regularity}.

\subsubsection{Influence function and variance estimation}

For a fixed hierarchy level $k$ and joint cell
$(i_1,\dots,i_k)$, define
\[
Z_k(i_1,\dots,i_k)
=
I(Y_1=i_1,\dots,Y_k=i_k).
\]
For notational simplicity, write
\[
e_a=\Pr(A=a),
\qquad
\widehat e_a=\widehat{\Pr}(A=a)=n_a/n.
\]
As defined in Section~\ref{subsec:ipw}, the population HT-IPW weight is
\[
\omega_k(a,\mb X,\widetilde R_{1:k})
=
\frac{I(A=a)}{e_a}
\frac{\widetilde R_{1:k}}{\pi_k(a,\mb X)}.
\]

The HT-IPW estimator satisfies
\[
\frac{1}{n}\sum_{j=1}^n
\left[
\widehat\omega_k(a,\mb X_j,\widetilde R_{j,1:k})
Z_{k,j}(i_1,\dots,i_k)
-
\widehat P_a^{1:k,\mathrm{HT}}(i_1,\dots,i_k)
\right]
=0.
\]
Because $e_a$ is estimated by $\widehat e_a=n_a/n$, its estimation
contributes to the first-order expansion. The influence function of
$\widehat P_a^{1:k,\mathrm{HT}}(i_1,\dots,i_k)$ is
\begin{align}
\psi^{\mathrm{HT}}
\{\mc O;P_a^{1:k}(i_1,\dots,i_k)\}
&=
\omega_k(a,\mb X,\widetilde R_{1:k})
Z_k(i_1,\dots,i_k)
-
P_a^{1:k}(i_1,\dots,i_k)
\nonumber\\
&\quad
-
\Gamma_{e,a,k}(i_1,\dots,i_k)
\psi(\mc O;e_a)
\nonumber\\
&\quad
-
\Gamma_{a,k}(i_1,\dots,i_k)^\top
\psi(\mc O;\bd\beta_k^{(a)}),
\label{eq:psi-M-HT}
\end{align}
where
\[
\Gamma_{e,a,k}(i_1,\dots,i_k)
=
\frac{P_a^{1:k}(i_1,\dots,i_k)}{e_a},
\qquad
\psi(\mc O;e_a)
=
I(A=a)-e_a,
\]
and
\[
\Gamma_{a,k}(i_1,\dots,i_k)
=
\Ex\left[
\frac{
\partial
\omega_k(a,\mb X,\widetilde R_{1:k})
}{
\partial\bd\beta_k^{(a)}
}
Z_k(i_1,\dots,i_k)
\right].
\]
The term involving $\psi(\mc O;e_a)$ accounts for estimation of the
treatment allocation probability, while the term involving
$\psi(\mc O;\bd\beta_k^{(a)})$ accounts for estimation of the
non-missingness model.

For the treatment-specific logistic model
\[
\pi_k(a,\mb X;\bd\beta_k^{(a)})
=
\{1+\exp(-\mb X^\top\bd\beta_k^{(a)})\}^{-1},
\]
$\bd\beta_k^{(a)}$ solves
\[
\frac{1}{n}\sum_{j=1}^n
I(A_j=a)\mb X_j
\left\{
\widetilde R_{j,1:k}
-
\pi_k(a,\mb X_j;\bd\beta_k^{(a)})
\right\}
=0.
\]
Let
\[
J_k^{(a)}
=
\Ex\left[
I(A=a)\mb X\mb X^\top
\pi_k(a,\mb X)
\{1-\pi_k(a,\mb X)\}
\right].
\]
Under the regularity conditions in
Appendix~\ref{subapp:regularity},
\[
\psi(\mc O;\bd\beta_k^{(a)})
=
{J_k^{(a)}}^{-1}
I(A=a)\mb X
\{\widetilde R_{1:k}-\pi_k(a,\mb X)\}.
\]
Moreover,
\[
\frac{
\partial
\omega_k(a,\mb X,\widetilde R_{1:k};
\bd\beta_k^{(a)})
}{
\partial\bd\beta_k^{(a)}
}
=
-
\omega_k(a,\mb X,\widetilde R_{1:k};
\bd\beta_k^{(a)})
\{1-\pi_k(a,\mb X;\bd\beta_k^{(a)})\}
\mb X.
\]

The influence function of the HT-IPW estimator of the win probability is
obtained by applying the product rule to \eqref{eq:win-pr}. Specifically,
\begin{align}
\psi^{\mathrm{HT}}(\mc O;p_W)
&=
\sum_{i_1>i_1'}
\left\{
\psi^{\mathrm{HT}}(\mc O;P_1^1(i_1))
P_0^1(i_1')
+
P_1^1(i_1)
\psi^{\mathrm{HT}}(\mc O;P_0^1(i_1'))
\right\}
\nonumber\\
&\quad+
\sum_{i_1}\sum_{i_2>i_2'}
\left\{
\psi^{\mathrm{HT}}(\mc O;P_1^{1:2}(i_1,i_2))
P_0^{1:2}(i_1,i_2')
\right.
\nonumber\\[-1mm]
&\hspace{35mm}\left.
+
P_1^{1:2}(i_1,i_2)
\psi^{\mathrm{HT}}(\mc O;P_0^{1:2}(i_1,i_2'))
\right\}
+\cdots
\nonumber\\
&\quad+
\sum_{i_1=\cdots=i_{K-1}}
\sum_{i_K>i_K'}
\left\{
\psi^{\mathrm{HT}}
(\mc O;P_1^{1:K}(i_1,\dots,i_K))
P_0^{1:K}(i_1,\dots,i_K')
\right.
\nonumber\\[-1mm]
&\hspace{35mm}\left.
+
P_1^{1:K}(i_1,\dots,i_K)
\psi^{\mathrm{HT}}
(\mc O;P_0^{1:K}(i_1,\dots,i_K'))
\right\}.
\label{eq:psi-pW-HT}
\end{align}
The corresponding influence function for $p_L$ is obtained by replacing
$>$ with $<$, and
\[
\psi^{\mathrm{HT}}(\mc O;p_T)
=
-
\psi^{\mathrm{HT}}(\mc O;p_W)
-
\psi^{\mathrm{HT}}(\mc O;p_L).
\]

Let
\[
\widehat{\bd\psi}^{\mathrm{HT}}_i
=
\begin{pmatrix}
\widehat\psi^{\mathrm{HT}}(\mc O_i;p_W)\\
\widehat\psi^{\mathrm{HT}}(\mc O_i;p_L)\\
\widehat\psi^{\mathrm{HT}}(\mc O_i;p_T)
\end{pmatrix}
\]
denote the empirical influence-function vector obtained by replacing
population quantities in the expressions above with their sample analogs.
Then
\[
\widehat{\Vx}
\begin{pmatrix}
\widehat p_W^{\mathrm{HT}}\\
\widehat p_L^{\mathrm{HT}}\\
\widehat p_T^{\mathrm{HT}}
\end{pmatrix}
=
\frac{1}{n^2}
\sum_{i=1}^n
\widehat{\bd\psi}^{\mathrm{HT}}_i
\widehat{\bd\psi}^{\mathrm{HT}\top}_i.
\]
For any smooth win measure
$\theta=g(p_W,p_L,p_T)$,
the corresponding variance estimator follows from the delta method as
described in Section~\ref{subsec:varest}.

\subsection{Theory of the H\'ajek IPW estimator}
\label{subapp:theory-ipw-Hajek}

This subsection establishes consistency and influence-function-based
variance estimation for the H\'ajek-normalized IPW estimator proposed in
Section~\ref{subsec:ipw}. We retain the notation introduced in the main
text. In particular,
\[
\widehat\omega_{k,\text{H\'ajek}}
(a,\mb X_i,\widetilde R_{i,1:k})
=
\frac{
\widehat\omega_k(a,\mb X_i,\widetilde R_{i,1:k})
}{
n^{-1}\sum_{j=1}^n
\widehat\omega_k(a,\mb X_j,\widetilde R_{j,1:k})
},
\]
and
\begin{align}
\widehat P_a^{1:k,\mathrm{ipw}}(i_1,\dots,i_k)
&=
\frac{1}{n}\sum_{j=1}^n
\widehat\omega_{k,\text{H\'ajek}}
(a,\mb X_j,\widetilde R_{j,1:k})
Z_{k,j}(i_1,\dots,i_k),
\label{eq:hajek-cell}
\end{align}
where
\[
Z_k(i_1,\dots,i_k)
=
I(Y_1=i_1,\dots,Y_k=i_k).
\]

\subsubsection{Consistency}

Under the regularity conditions in
Appendix~\ref{subapp:regularity} and correct specification of the
non-missingness model,
\[
\widehat\pi_k(a,\mb X)
\xrightarrow{p}
\pi_k(a,\mb X).
\]
The denominator used to normalize the HT weights satisfies
\begin{align*}
\frac{1}{n}\sum_{j=1}^n
\widehat\omega_k(a,\mb X_j,\widetilde R_{j,1:k})
&\xrightarrow{p}
\Ex\left\{
\frac{I(A=a)}{e_a}
\frac{\widetilde R_{1:k}}{\pi_k(a,\mb X)}
\right\}
=
1.
\end{align*}
Likewise,
\begin{align*}
\frac{1}{n}\sum_{j=1}^n
\widehat\omega_k(a,\mb X_j,\widetilde R_{j,1:k})
Z_{k,j}(i_1,\dots,i_k)
&\xrightarrow{p}
\Ex\left\{
\frac{I(A=a)}{e_a}
\frac{\widetilde R_{1:k}}{\pi_k(a,\mb X)}
Z_k(i_1,\dots,i_k)
\right\}
\\
&=
P_a^{1:k}(i_1,\dots,i_k).
\end{align*}
Therefore, by Slutsky's theorem,
\[
\widehat P_a^{1:k,\mathrm{ipw}}(i_1,\dots,i_k)
\xrightarrow{p}
P_a^{1:k}(i_1,\dots,i_k).
\]
Consequently,
$\widehat p_W^{\mathrm{ipw}}$,
$\widehat p_L^{\mathrm{ipw}}$, and
$\widehat p_T^{\mathrm{ipw}}$,
and any continuous win measure constructed from them, are consistent under
the corresponding denominator conditions in
Appendix~\ref{subapp:regularity}.

An important distinction from the HT-IPW estimator is that the treatment
allocation probability cancels from the H\'ajek normalization:
\[
\widehat\omega_{k,\text{H\'ajek}}
(a,\mb X_i,\widetilde R_{i,1:k})
=
\frac{
\dfrac{I(A_i=a)\widetilde R_{i,1:k}}
{\widehat\pi_k(a,\mb X_i)}
}{
n^{-1}\sum_{j=1}^n
\dfrac{I(A_j=a)\widetilde R_{j,1:k}}
{\widehat\pi_k(a,\mb X_j)}
}.
\]
Hence, estimating $e_a$ by $\widehat e_a=n_a/n$ does not generate a
separate first-order contribution to the influence function of the proposed
H\'ajek IPW estimator.

\subsubsection{Influence function and variance estimation}

For notational simplicity, write
\[
P=P_a^{1:k}(i_1,\dots,i_k).
\]
Because the H\'ajek estimator is a ratio of two weighted empirical means,
its leading influence-function contribution can be written as
\[
\omega_k(a,\mb X,\widetilde R_{1:k})
\{Z_k(i_1,\dots,i_k)-P\},
\]
where
\[
\omega_k(a,\mb X,\widetilde R_{1:k})
=
\frac{I(A=a)}{e_a}
\frac{\widetilde R_{1:k}}{\pi_k(a,\mb X)}
\]
is the population HT-IPW weight defined in Section~\ref{subsec:ipw}.

Accounting for estimation of the non-missingness-model parameter
$\bd\beta_k^{(a)}$, the influence function of the proposed H\'ajek IPW
cell-probability estimator is
\begin{align}
\psi^{\mathrm{ipw}}
\{\mc O;P_a^{1:k}(i_1,\dots,i_k)\}
&=
\omega_k(a,\mb X,\widetilde R_{1:k})
\left\{
Z_k(i_1,\dots,i_k)
-
P_a^{1:k}(i_1,\dots,i_k)
\right\}
\nonumber\\
&\quad
-
B_{a,k}^{\mathrm{ipw}}(i_1,\dots,i_k)^\top
\psi(\mc O;\bd\beta_k^{(a)}),
\label{eq:psi-M-hajek}
\end{align}
where
\begin{align}
B_{a,k}^{\mathrm{ipw}}(i_1,\dots,i_k)
&=
\Ex\left[
\omega_k(a,\mb X,\widetilde R_{1:k})
\{1-\pi_k(a,\mb X)\}
\right.
\nonumber\\[-1mm]
&\hspace{25mm}\left.
\times
\left\{
Z_k(i_1,\dots,i_k)
-
P_a^{1:k}(i_1,\dots,i_k)
\right\}
\mb X
\right].
\label{eq:B-hajek}
\end{align}
The second term in \eqref{eq:psi-M-hajek} accounts for estimation of the
non-missingness-model parameter. Unlike the HT influence function in
Appendix~\ref{subapp:theory-ipw-HT}, there is no separate term involving
$\psi(\mc O;e_a)$.

For the logistic non-missingness model, the nuisance-parameter influence
function is the same as that given in
Appendix~\ref{subapp:theory-ipw-HT}:
\[
\psi(\mc O;\bd\beta_k^{(a)})
=
{J_k^{(a)}}^{-1}
I(A=a)\mb X
\{\widetilde R_{1:k}-\pi_k(a,\mb X)\}.
\]

The corresponding empirical influence function is
\begin{align}
\widehat\psi^{\mathrm{ipw}}
\{\mc O_i;P_a^{1:k}(i_1,\dots,i_k)\}
&=
\widehat\omega_{k,\text{H\'ajek}}
(a,\mb X_i,\widetilde R_{i,1:k})
\left\{
Z_{k,i}(i_1,\dots,i_k)
-
\widehat P_a^{1:k,\mathrm{ipw}}(i_1,\dots,i_k)
\right\}
\nonumber\\
&\quad
-
\widehat B_{a,k}^{\mathrm{ipw}}(i_1,\dots,i_k)^\top
\widehat\psi(\mc O_i;\bd\beta_k^{(a)}),
\label{eq:emp-psi-M-hajek}
\end{align}
where
\begin{align}
\widehat B_{a,k}^{\mathrm{ipw}}(i_1,\dots,i_k)
&=
\frac{1}{n}\sum_{\ell=1}^n
\widehat\omega_{k,\text{H\'ajek}}
(a,\mb X_\ell,\widetilde R_{\ell,1:k})
\{1-\widehat\pi_k(a,\mb X_\ell)\}
\nonumber\\[-1mm]
&\qquad\times
\left\{
Z_{k,\ell}(i_1,\dots,i_k)
-
\widehat P_a^{1:k,\mathrm{ipw}}(i_1,\dots,i_k)
\right\}
\mb X_\ell.
\label{eq:Bhat-hajek}
\end{align}

The influence functions of
$\widehat p_W^{\mathrm{ipw}}$ and
$\widehat p_L^{\mathrm{ipw}}$
follow from the same product-rule calculations as in
Appendix~\ref{subapp:theory-ipw-HT}, replacing the HT cell-probability
influence functions by
$\psi^{\mathrm{ipw}}\{\mc O;P_a^{1:k}\}$.
Specifically,
\begin{align}
\psi^{\mathrm{ipw}}(\mc O;p_W)
&=
\sum_{i_1>i_1'}
\left\{
\psi^{\mathrm{ipw}}(\mc O;P_1^1(i_1))
P_0^1(i_1')
+
P_1^1(i_1)
\psi^{\mathrm{ipw}}(\mc O;P_0^1(i_1'))
\right\}
\nonumber\\
&\quad+
\sum_{i_1}\sum_{i_2>i_2'}
\left\{
\psi^{\mathrm{ipw}}(\mc O;P_1^{1:2}(i_1,i_2))
P_0^{1:2}(i_1,i_2')
\right.
\nonumber\\[-1mm]
&\hspace{35mm}\left.
+
P_1^{1:2}(i_1,i_2)
\psi^{\mathrm{ipw}}(\mc O;P_0^{1:2}(i_1,i_2'))
\right\}
+\cdots,
\label{eq:psi-pW-hajek}
\end{align}
with the final hierarchy-level term defined analogously to
\eqref{eq:psi-pW-HT}. The influence function for $p_L$ is obtained by
replacing $>$ with $<$, and
\[
\psi^{\mathrm{ipw}}(\mc O;p_T)
=
-
\psi^{\mathrm{ipw}}(\mc O;p_W)
-
\psi^{\mathrm{ipw}}(\mc O;p_L).
\]

Let
\[
\widehat{\bd\psi}^{\mathrm{ipw}}_i
=
\begin{pmatrix}
\widehat\psi^{\mathrm{ipw}}(\mc O_i;p_W)\\
\widehat\psi^{\mathrm{ipw}}(\mc O_i;p_L)\\
\widehat\psi^{\mathrm{ipw}}(\mc O_i;p_T)
\end{pmatrix}.
\]
The joint covariance matrix is estimated by
\[
\widehat{\Vx}
\begin{pmatrix}
\widehat p_W^{\mathrm{ipw}}\\
\widehat p_L^{\mathrm{ipw}}\\
\widehat p_T^{\mathrm{ipw}}
\end{pmatrix}
=
\frac{1}{n^2}
\sum_{i=1}^n
\widehat{\bd\psi}^{\mathrm{ipw}}_i
\widehat{\bd\psi}^{\mathrm{ipw}\top}_i.
\]
The variance estimators for WR, NB, WO, and DOOR then follow from the delta
method described in Section~\ref{subsec:varest}.

\subsection{Theory of the AIPW estimator}\label{subapp:theory-aipw}\label{subapp:cons-aipw}

This subsection establishes the double-robust consistency of the AIPW estimator and gives its influence-function-based variance estimator.

\subsubsection{Consistency and double robustness}

Under the regularity conditions in Appendix~\ref{subapp:regularity},
$\widehat{\Pr}(A=a)\overset{p}{\to}\Pr(A=a)$. We first consider the case in which both nuisance models are correctly specified, such that $
\sup_{\mb x}
\left|
\widehat\pi_k(a,\mb x)-\pi_k(a,\mb x)
\right|
\xrightarrow{p}0 $
and
$
\sup_{\mb x}
\left|
\widehat\mu_k(a,\mb x;i_1,\dots,i_k)
-
\mu_k(a,\mb x;i_1,\dots,i_k)
\right|
\xrightarrow{p}0.
$
It then follows from the uniform law of large numbers and Slutsky's theorem that 
\begin{align}
    & \widehat P_a^{1:k,\text{aipw}}(i_1,\dots,i_k)\nonumber \\
    & \xrightarrow{p} \Ex\left[\frac{I(A=a)}{\Pr(A=a)}\cdot\frac{\widetilde R_{1:k}}{\pi_k(a,\mb X)}\{I(Y_1=i_1,\dots,Y_k=i_k)-\mu_k(a,\mb X;i_1,\dots,i_k)\} + \mu_k(a,\mb X;i_1,\dots,i_k)\right]\nonumber \\
    & = \underbrace{\Ex\left[\frac{I(A=a)}{\Pr(A=a)}\cdot\frac{\widetilde R_{1:k}}{\pi_k(a,\mb X)}I(Y_1=i_1,\dots,Y_k=i_k)\right]}_{=~P_a^{1:k}(i_1,\dots,i_k)}\nonumber \\
    & \qquad -\Ex\left[\frac{I(A=a)}{\Pr(A=a)}\cdot\frac{\widetilde R_{1:k}}{\pi_k(a,\mb X)}\mu_k(a,\mb X;i_1,\dots,i_k) - \mu_k(a,\mb X;i_1,\dots,i_k)\right]\nonumber \\
    & = P_a^{1:k}(i_1,\dots,i_k) - \Ex\left[\underbrace{\Ex\left\{\frac{I(A=a)}{\Pr(A=a)}\cdot\frac{\widetilde R_{1:k}}{\pi_k(a,\mb X)} -1 ~\bigg|~ A=a, \mb X\right\}}_{ = 0}\mu_k(a,\mb X;i_1,\dots,i_k)\right] \label{eq:aipw-zeroeq} \\
    & = P_a^{1:k}(i_1,\dots,i_k). \nonumber
\end{align}
Thus, the AIPW estimator is consistent if both propensity score and outcome models are correctly specified. 

To prove the double robustness, we consider the following two cases of model specifications. 

Case (1): we assume $\widehat\pi_k(a,\mb X)\xrightarrow{p} \pi_k(a,\mb X)$ and $\widehat\mu_k(a,\mb X;i_1,\dots,i_k)\xrightarrow{p}\widetilde\mu_k(a,\mb X;i_1,\dots,i_k)$ (a general probability limit of the estimated outcome). In this case, it is obvious that the only change in the above proof is replacing the true outcome model to be this general limit, and so the term \eqref{eq:aipw-zeroeq} becomes 
\begin{align*}
    P_a^{1:k}(i_1,\dots,i_k) - \Ex\left[\Ex\left\{\frac{I(A=a)}{\Pr(A=a)}\cdot\frac{\widetilde R_{1:k}}{\pi_k(a,\mb X)} -1 ~\bigg|~ A=a, \mb X\right\}\widetilde\mu_k(a,\mb X;i_1,\dots,i_k)\right] = P_a^{1:k}(i_1,\dots,i_k). 
\end{align*}
The second term is still 0 since the propensity score model is correctly specified. Hence the AIPW estimator is still consistent. 

Case (2): we assume $\widehat\pi_k(a,\mb X)\xrightarrow{p} \widetilde\pi_k(a,\mb X)$ (a general limit of the estimated propensity score) and $\widehat\mu_k(a,\mb X;i_1,\dots,i_k)\xrightarrow{p}\mu_k(a,\mb X;i_1,\dots,i_k)$. In this case, the limit of the AIPW estimator above is replaced by 
\begin{align}
    & \Ex\left[\frac{I(A=a)}{\Pr(A=a)}\cdot\frac{\widetilde R_{1:k}}{\widetilde\pi_k(a,\mb X)}\{I(Y_1=i_1,\dots,Y_k=i_k)-\mu_k(a,\mb X;i_1,\dots,i_k)\} + \mu_k(a,\mb X;i_1,\dots,i_k)\right]\nonumber \\
    & = \Ex\left[\frac{I(A=a)}{\Pr(A=a)}\cdot\frac{\widetilde R_{1:k}}{\widetilde\pi_k(a,\mb X)}\{I(Y_1=i_1,\dots,Y_k=i_k)-\mu_k(a,\mb X;i_1,\dots,i_k)\}\right] + \Ex\{\mu_k(a,\mb X;i_1,\dots,i_k)\} \nonumber \\
    & = \Ex\left[\frac{I(A=a)}{\Pr(A=a)}\cdot\frac{\widetilde R_{1:k}}{\widetilde\pi_k(a,\mb X)}\underbrace{\Ex\{I(Y_1=i_1,\dots,Y_k=i_k)-\mu_k(a,\mb X;i_1,\dots,i_k)\mid A=a, \mb X\}}_{=~0}\right] \nonumber \\
    & \qquad + P_a^{1:k}(i_1,\dots,i_k) \label{eq:aipw-outresidual} \\
    &  = P_a^{1:k}(i_1,\dots,i_k). \nonumber
\end{align}
Therefore, in this case, the AIPW estimator is also consistent regardless of the value of $\widetilde\pi_k(a,\mb X)$. 

However, when both models are misspecified, the AIPW estimator is not consistent in general, because when both $\pi_k(a,\mb X)$ and $\mu_k(a,\mb X;i_1,\dots,i_k)$ are replaced by general limits $\widetilde\pi_k(a,\mb X)$ and $\widetilde\mu_k(a,\mb X;i_1,\dots,i_k)$, obviously the term \eqref{eq:aipw-outresidual} becomes 
\begin{align*}
    \Ex\left[\frac{I(A=a)}{\Pr(A=a)}\cdot\frac{\widetilde R_{1:k}}{\widetilde\pi_k(a,\mb X)}\underbrace{\Ex\{I(Y_1=i_1,\dots,Y_k=i_k)-\widetilde\mu_k(a,\mb X;i_1,\dots,i_k)\mid A=a, \mb X\}}_{\not =~0}\right] + P_a^{1:k}(i_1,\dots,i_k), 
\end{align*}
which is not always equal to the true $P_a^{1:k}(i_1,\dots,i_k)$, and the first term above quantifies the bias by misspecifying both models. As a remark, when $\widetilde\pi_k(a,\mb X)$ is closer to zero (violation of positivity by model misspecification), the bias term above becomes larger. 

\subsubsection{Influence function and variance estimation}

We next derive the influence function for the AIPW estimator, using the
notation above. For $k=1,\ldots,K$ and $a=0,1$, the AIPW estimator solves
\begin{align*}
\frac1n\sum_{j=1}^n\bigg[&\widehat\omega_k(a,\mb X_j,A_j,\widetilde R_{1:k,j})
\{Z_{k,j}-\widehat\mu_k(a,\mb X_j;i_1,\ldots,i_k)\} +\widehat\mu_k(a,\mb X_j;i_1,\ldots,i_k)
 -\widehat P_a^{1:k,\mathrm{aipw}}(i_1,\ldots,i_k)\bigg]=0.
\end{align*}
Its influence function is
\begin{align}
\psi^{\mathrm{aipw}}(\mc O;P_a^{1:k})
&=\omega_k(a,\mb X,A,\widetilde R_{1:k})
 \{Z_k-\mu_k(a,\mb X;i_1,\ldots,i_k)\} \nonumber \\
&\quad+\mu_k(a,\mb X;i_1,\ldots,i_k)-P_a^{1:k}(i_1,\ldots,i_k) \nonumber \\
&\quad-\Lambda_{e,a,k}(i_1,\ldots,i_k)\psi(\mc O;e_a) \nonumber \\
&\quad-\Lambda_{a,k}(i_1,\ldots,i_k)^\top
 \psi(\mc O;\bd\beta_k^{(a)}) \nonumber \\
&\quad-\Delta_{a,k}(i_1,\ldots,i_k)^\top
 \psi(\mc O;\bd\gamma^{(a)}),
\label{eq:psi-M-aipw}
\end{align}
where
\begin{align*}
\Lambda_{e,a,k}(i_1,\ldots,i_k)
&=\frac1{e_a}\Ex\left[\omega_k(a,\mb X,A,\widetilde R_{1:k})
 \{Z_k-\mu_k(a,\mb X;i_1,\ldots,i_k)\}\right],\\
\Lambda_{a,k}(i_1,\ldots,i_k)
&=\Ex\left[\frac{\partial\omega_k(a,\mb X,A,\widetilde R_{1:k})}
 {\partial\bd\beta_k^{(a)}}
 \{Z_k-\mu_k(a,\mb X;i_1,\ldots,i_k)\}\right],\\
\Delta_{a,k}(i_1,\ldots,i_k)
&=\Ex\left[\frac{\partial\mu_k(a,\mb X;i_1,\ldots,i_k)}
 {\partial\bd\gamma^{(a)}}
 \{1-\omega_k(a,\mb X,A,\widetilde R_{1:k})\}\right].
\end{align*}
Here $\bd\gamma^{(a)}$ denotes the full parameter vector of the conditional
outcome model, rather than a collection of unrelated cell-specific
parameters.

The term involving $\Lambda_{e,a,k}$ accounts for estimating $e_a$ by
$n_a/n$. If the outcome regression is correctly specified, then
\[
\Ex\left[\omega_k(a,\mb X,A,\widetilde R_{1:k})
\{Z_k-\mu_k(a,\mb X;i_1,\ldots,i_k)\}\right]=0,
\]
so that $\Lambda_{e,a,k}=0$. Thus, estimating the treatment allocation
probability has no first-order contribution to the AIPW influence function
when the outcome model is correctly specified.

The propensity-score derivative and influence function are those given in
Appendix~\ref{subapp:theory-ipw-HT}. For the multinomial log-linear outcome model
in \eqref{eq:loglin}, let $c=(i_1,\ldots,i_k)$ index a joint cell and write
\[
\mu_{k,c}(a,\mb X)
 =\frac{\exp(-\mb X^\top\bd\gamma_c^{(a)})}
 {\sum_{d}\exp(-\mb X^\top\bd\gamma_d^{(a)})}.
\]
The block derivative is
\[
\frac{\partial\mu_{k,c}(a,\mb X)}
 {\partial\bd\gamma_d^{(a)}}
 =-\mu_{k,c}(a,\mb X)\{I(c=d)-\mu_{k,d}(a,\mb X)\}\mb X.
\]
Let $U_{\gamma}^{(a)}(\mc O)$ and $J_{\gamma}^{(a)}$ denote the full score
vector and its population information matrix for the fitted multinomial
model, respectively. Then
\[
\psi(\mc O;\bd\gamma^{(a)})
 ={J_{\gamma}^{(a)}}^{-1}U_{\gamma}^{(a)}(\mc O).
\]
This full-vector formulation accounts for the covariance among the fitted
cell probabilities. Substituting these expressions into
\eqref{eq:psi-M-aipw}, and then applying the same product-rule calculations as in
Appendix~\ref{subapp:theory-ipw-HT}, gives the influence functions for
$(p_W,p_L,p_T)$. Let $\widehat{\bd\psi}^{\mathrm{aipw}}_i$ stack the resulting empirical
influence functions for $(p_W,p_L,p_T)$. Then
\[
\widehat\Sigma_{\mathrm{aipw}}
=\frac1n\sum_{i=1}^n
\widehat{\bd\psi}^{\mathrm{aipw}}_i
\widehat{\bd\psi}^{\mathrm{aipw}\top}_i,
\qquad
\widehat{\Vx}\begin{pmatrix}\widehat p_W\\\widehat p_L\\\widehat p_T\end{pmatrix}
=\frac1n\widehat\Sigma_{\mathrm{aipw}}.
\]
For any smooth win measure $\theta=g(p_W,p_L,p_T)$, its variance is estimated by the delta method, with the AIPW covariance estimator substituted for the HT covariance estimator.

\subsubsection{Semiparametric efficiency for joint cell probabilities}
\label{subapp:semieff}

In this section, we establish semiparametric efficiency of the proposed AIPW estimator for each joint cell probability under a corresponding reduced observed-data model. Under the regularity conditions in Appendix~\ref{subapp:regularity}, the estimator attains the efficiency bound for this model when both the missingness and outcome models are correctly specified \citep{tsiatis2007semiparametric, kennedy2016semiparametric}. The distinction between this reduced model and the full observed-data model is discussed at the end of this section.

Fix a treatment group $a$, a hierarchy level $k$, and a cell $(i_1,\dots,i_k)$. Define
\[
Z_k=\prod_{j=1}^k I(Y_j=i_j),
\qquad
R_k^\star=\widetilde R_{1:k},
\]
where $R_k^\star$ indicates that all first $k$ endpoints are observed. For this derivation, we retain only the reduced observed data
\[
\mc O_k=(\mb X,A,R_k^\star,R_k^\star Z_k).
\]
Thus, $Z_k$ is recorded when $R_k^\star=1$, while individual endpoint values are not retained when $R_k^\star=0$.

Let
\[
\mu_k(a,\mb X)
=
\Ex(Z_k\mid A=a,\mb X),
\qquad
\pi_k(a,\mb X)
=
\Pr(R_k^\star=1\mid A=a,\mb X),
\]
and write
\[
e(a,\mb X)=\Pr(A=a\mid\mb X).
\]
Under randomized treatment assignment, $e(a,\mb X)=e_a$ is a known constant and $A$ is independent of $\mb X$. Consequently, the target joint cell probability can be expressed as
\[
\lambda
:=
P_a^{1:k}(i_1,\dots,i_k)
=
\Ex\{\mu_k(a,\mb X)\},
\]
where the expectation is taken over the marginal distribution of $\mb X$.

Under Assumption~\ref{asp:miss-pos},
\[
R_k^\star \independent Z_k\mid(A,\mb X),
\]
and the observation probabilities are bounded away from zero. Treatment positivity also holds by randomization. The likelihood for the reduced observed data therefore factorizes as
\[
f(\mc O_k)
=
f_{\mb X}(\mb X)
f_{A\mid\mb X}(A\mid\mb X)
f_{R_k^\star\mid A,\mb X}(R_k^\star\mid A,\mb X)
\left\{
f_{Z_k\mid A,\mb X}(Z_k\mid A,\mb X)
\right\}^{R_k^\star}.
\]
Here, the treatment assignment mechanism is fixed by design, while the covariate distribution, missingness mechanism, and conditional distribution of $Z_k$ are otherwise unrestricted within this reduced model.

We derive the efficient influence function (EIF) using a regular parametric submodel
$\{f_\epsilon:\epsilon\in(-\delta,\delta)\}$ through the true distribution at $\epsilon=0$, following \cite{kennedy2016semiparametric}. Let
\[
\ell(\mc O_k)
=
\left.
\frac{\partial}{\partial\epsilon}
\log f_\epsilon(\mc O_k)
\right|_{\epsilon=0}
\]
denote its score, and let $\lambda(\epsilon)$ denote the target parameter along the submodel. A gradient for $\lambda$ must satisfy
\[
\left.
\frac{\partial\lambda(\epsilon)}{\partial\epsilon}
\right|_{\epsilon=0}
=
\Ex\{\phi(\mc O_k)\ell(\mc O_k)\}.
\]

By the chain rule,
\begin{align*}
\left.
\frac{\partial\lambda(\epsilon)}{\partial\epsilon}
\right|_{\epsilon=0}
&=
\left.
\frac{\partial}{\partial\epsilon}
\Ex_\epsilon\{\mu_{k,\epsilon}(a,\mb X)\}
\right|_{\epsilon=0}
=
\left.
\frac{\partial}{\partial\epsilon}
\Ex_\epsilon\{\mu_k(a,\mb X)\}
\right|_{\epsilon=0}
+
\Ex\left\{
\left.
\frac{\partial\mu_{k,\epsilon}(a,\mb X)}
{\partial\epsilon}
\right|_{\epsilon=0}
\right\}.
\end{align*}

For the first term, let $\ell_{\mb X}(\mb X)$ denote the score for the marginal distribution of $\mb X$. Since
$\Ex\{\ell_{\mb X}(\mb X)\}=0$ and
$\Ex\{\ell(\mc O_k)\mid\mb X\}=\ell_{\mb X}(\mb X)$,
\begin{align*}
\left.
\frac{\partial}{\partial\epsilon}
\Ex_\epsilon\{\mu_k(a,\mb X)\}
\right|_{\epsilon=0}
&=
\Ex\left[
\{\mu_k(a,\mb X)-\lambda\}
\ell_{\mb X}(\mb X)
\right]
=
\Ex\left[
\{\mu_k(a,\mb X)-\lambda\}
\ell(\mc O_k)
\right].
\end{align*}

For the second term, let
$\ell_Z(Z_k\mid A,\mb X)$ denote the conditional outcome score. Conditional independence of $R_k^\star$ and $Z_k$ implies that this is also the outcome score among participants with $R_k^\star=1$. Therefore,
\begin{align*}
&\Ex\left\{
\left.
\frac{\partial\mu_{k,\epsilon}(a,\mb X)}
{\partial\epsilon}
\right|_{\epsilon=0}
\right\}
\\
&\quad=
\Ex\left[
\Ex\left\{
\{Z_k-\mu_k(a,\mb X)\}
\ell_Z(Z_k\mid a,\mb X)
\mid A=a,\mb X,R_k^\star=1
\right\}
\right]
\\
&\quad=
\Ex\left[
\frac{I(A=a)}{e(a,\mb X)}
\frac{R_k^\star}{\pi_k(a,\mb X)}
\{Z_k-\mu_k(a,\mb X)\}
\ell_Z(Z_k\mid A,\mb X)
\right]
\\
&\quad=
\Ex\left[
\frac{I(A=a)}{e(a,\mb X)}
\frac{R_k^\star}{\pi_k(a,\mb X)}
\{Z_k-\mu_k(a,\mb X)\}
\ell(\mc O_k)
\right].
\end{align*}
The last equality follows because the weighted residual has conditional mean zero given $(\mb X,A,R_k^\star)$ and is therefore orthogonal to the remaining score components.

Combining these expressions gives
\begin{equation}
\phi_{\mathrm{eff}}(\mc O_k)
=
\frac{I(A=a)}{e(a,\mb X)}
\frac{R_k^\star}{\pi_k(a,\mb X)}
\{Z_k-\mu_k(a,\mb X)\}
+
\mu_k(a,\mb X)-\lambda.
\label{eq:eif_Mak}
\end{equation}
This function has mean zero and represents the pathwise derivative of $\lambda$. Moreover, its two terms belong to the outcome and covariate components of the tangent space, respectively. It is therefore the canonical gradient, or EIF, under the reduced observed-data model. The corresponding semiparametric efficiency bound is
\[
\Ex\{\phi_{\mathrm{eff}}(\mc O_k)^2\}.
\]

Under the stated regularity conditions, when both nuisance models are correctly specified, the influence function of the proposed AIPW estimator reduces to \eqref{eq:eif_Mak}. Its asymptotic variance therefore attains this efficiency bound. Although the treatment allocation probability is fixed by design, our implementation uses its empirical counterpart $\widehat e_a=n_a/n$. As shown in Appendix~\ref{subapp:theory-aipw}, estimating $e_a$ has no first-order contribution when the outcome model is correctly specified, because the corresponding sensitivity term vanishes.

\begin{remark}
The same construction can be extended to observational studies by replacing the treatment-group proportion with an estimated treatment propensity score $\widehat e(a,\mb X)$. In that setting, the target is the standardized joint cell probability
\[
\lambda_a
=
\Ex\{\mu_k(a,\mb X)\},
\]
which generally differs from
$\Pr(Y_1=i_1,\dots,Y_k=i_k\mid A=a)$.
Under consistency, treatment positivity, and conditional exchangeability of the potential outcomes,
\[
\{Y_1(a),\dots,Y_k(a)\}\independent A\mid\mb X,
\]
this standardized probability identifies
$\Pr\{Y_1(a)=i_1,\dots,Y_k(a)=i_k\}$
\citep{rubin1974estimating}.
Under the stated missingness assumption, the resulting AIPW estimator is consistent if either (i) the outcome model is correctly specified or (ii) both the treatment propensity score and the missingness model are correctly specified. When all three nuisance models are correctly specified and the required regularity conditions hold, its influence function equals the EIF under the corresponding reduced observed-data model with an unknown treatment assignment mechanism.
\end{remark}

\begin{remark}
The efficiency result above is specific to the reduced observed-data structure
$\mc O_k=(\mb X,A,R_k^\star,R_k^\star Z_k)$
for a fixed hierarchy level and cell. It does not establish efficiency under the full observed-data structure, which retains all observed endpoint values and their individual missingness indicators. For example, when $k=2$, a participant with $Y_1$ observed but $Y_2$ missing has $R_2^\star=0$. The reduced data do not retain this participant's observed $Y_1$, although it may contain information about the joint cell probability.

Furthermore, the efficiency calculations are performed separately for the reduced model associated with each layer and cell. They therefore do not establish joint efficiency of the collection of cell probability estimators under the full observed-data model. Consequently, we make no claim that the resulting plug-in estimators of $(p_W,p_L,p_T)$ or the corresponding win measures attain the semiparametric efficiency bound under that model. Deriving efficient estimators that fully exploit the information available across missingness patterns requires a separate analysis and remains a direction for future research.
\end{remark}

\subsection{Covariate-adjusted treatment propensity estimation under randomization}
\label{subapp:treatment-ps}

In the randomized-trial setting considered in this paper, the treatment allocation probability is known by design. Nevertheless, one may estimate a working treatment propensity score using baseline covariates as an additional form of covariate adjustment. Estimating the propensity score, rather than directly using the known or true propensity score, can reduce asymptotic variance under suitable conditions by exploiting chance covariate imbalance between treatment groups \citep{rosenbaum1987model,robins1995analysis,hirano2003efficient,tsiatis2007semiparametric}. Here, we study this optional extension for the proposed H\'ajek IPW and AIPW estimators. The known randomization probability remains the primary implementation used in this paper.

Let
\[
e_1(\mb X;\bd\alpha)=\Pr(A=1\mid \mb X;\bd\alpha), \qquad e_0(\mb X;\bd\alpha)=1-e_1(\mb X;\bd\alpha),
\]
denote a working treatment propensity score model. For concreteness, consider the logistic model
\[
e_1(\mb X;\bd\alpha)=\left[1+\exp\{-\mb h(\mb X)^\top\bd\alpha\}\right]^{-1},
\]
where $\mb h(\mb X)$ is a prespecified vector of functions of the baseline covariates that includes an intercept.

Under simple randomization,
\[
\Pr(A=1\mid\mb X)=e,
\]
where $e$ is the known randomization probability. We assume that the working propensity model contains the true randomization mechanism. Thus, at the true parameter value $\bd\alpha_0$, the intercept equals $\operatorname{logit}(e)$ and the coefficients corresponding to nonconstant functions of $\mb X$ are zero.

The score for $\bd\alpha$ is
\[
S_{\alpha}(\mc O)=\mb h(\mb X)\{A-e_1(\mb X;\bd\alpha_0)\},
\]
with information matrix
\[
J_{\alpha}=\Ex\left[S_{\alpha}(\mc O)S_{\alpha}(\mc O)^\top\right].
\]
When the treatment allocation probability is treated as known, $\bd\alpha$ is not estimated and the treatment-score term does not enter the influence-function expansion. Under simple randomization,
\[
J_{\alpha}=e(1-e)\Ex\left[\mb h(\mb X)\mb h(\mb X)^\top\right].
\]

For notational convenience, define
\[
Z_k(i_1,\ldots,i_k)=I(Y_1=i_1,\ldots,Y_k=i_k).
\]

\subsubsection{H\'ajek IPW estimator}

When the treatment propensity score is modeled using baseline covariates, define
\[
\omega_k^{e}(a,\mb X,\widetilde R_{1:k})
=
\frac{I(A=a)}{e_a(\mb X;\bd\alpha_0)}
\frac{\widetilde R_{1:k}}{\pi_k(a,\mb X)}.
\]
Its empirical counterpart replaces $\bd\alpha_0$ and the parameters of the non-missingness model by their estimators. The corresponding H\'ajek-normalized weight is
\[
\widehat\omega_{k,\text{H\'ajek}}^{e}(a,\mb X_i,\widetilde R_{i,1:k})
=
\frac{\widehat\omega_k^{e}(a,\mb X_i,\widetilde R_{i,1:k})}
{n^{-1}\sum_{j=1}^n\widehat\omega_k^{e}(a,\mb X_j,\widetilde R_{j,1:k})},
\]
yielding
\[
\widehat P_a^{1:k,\mathrm{ipw},e}(i_1,\ldots,i_k)
=
\frac{1}{n}\sum_{j=1}^n
\widehat\omega_{k,\text{H\'ajek}}^{e}(a,\mb X_j,\widetilde R_{j,1:k})
Z_{k,j}(i_1,\ldots,i_k).
\]

Because the working treatment propensity model contains the true randomization mechanism, this estimator has the same probability limit as the H\'ajek IPW estimator based directly on the known randomization probability. Estimation of $\bd\alpha$, however, can alter its first-order influence function and hence its asymptotic variance.

Let
\[
\psi^{\mathrm{ipw},0}\{\mc O;P_a^{1:k}(i_1,\ldots,i_k)\}
\]
denote the influence function derived in Appendix~\ref{subapp:theory-ipw-Hajek} when the treatment allocation probability is treated as known. This influence function already accounts for estimation of the non-missingness-model parameter $\bd\beta_k^{(a)}$. Define
\[
\mb C_{a,k}^{\mathrm{ipw}}(i_1,\ldots,i_k)
=
\Ex\left[
S_{\alpha}(\mc O)
\psi^{\mathrm{ipw},0}\{\mc O;P_a^{1:k}(i_1,\ldots,i_k)\}
\right].
\]

\begin{proposition}
\label{prop:estimated-treatment-ps-ipw}
Suppose that the working treatment propensity score model is correctly specified and contains the true randomization mechanism, and that the regularity conditions in Appendix~\ref{subapp:regularity} hold. Then $\widehat P_a^{1:k,\mathrm{ipw},e}(i_1,\ldots,i_k)$ is asymptotically linear with influence function
\begin{align}
\psi^{\mathrm{ipw},e}\{\mc O;P_a^{1:k}(i_1,\ldots,i_k)\}
&=
\psi^{\mathrm{ipw},0}\{\mc O;P_a^{1:k}(i_1,\ldots,i_k)\}
-
\mb C_{a,k}^{\mathrm{ipw}}(i_1,\ldots,i_k)^\top
J_{\alpha}^{-1}
S_{\alpha}(\mc O).
\label{eq:if-est-treatment-ipw}
\end{align}
Consequently,
\begin{align}
\Vx\left[
\psi^{\mathrm{ipw},e}\{\mc O;P_a^{1:k}(i_1,\ldots,i_k)\}
\right]
&=
\Vx\left[
\psi^{\mathrm{ipw},0}\{\mc O;P_a^{1:k}(i_1,\ldots,i_k)\}
\right]
\nonumber\\
&\quad-
\mb C_{a,k}^{\mathrm{ipw}}(i_1,\ldots,i_k)^\top
J_{\alpha}^{-1}
\mb C_{a,k}^{\mathrm{ipw}}(i_1,\ldots,i_k).
\label{eq:var-est-treatment-ipw}
\end{align}
Thus, under correct specification of the working treatment propensity model, estimating the treatment propensity score using baseline covariates cannot increase the first-order asymptotic variance of the H\'ajek IPW cell-probability estimator.
\end{proposition}

\begin{proof}
The H\'ajek estimator can be represented as the solution to
\[
\mathbb P_n\left[
\omega_k^{e}(a,\mb X,\widetilde R_{1:k})
\{Z_k(i_1,\ldots,i_k)-P_a^{1:k}(i_1,\ldots,i_k)\}
\right]=0.
\]
When the treatment propensity score is treated as known, its first-order expansion, including the contribution from estimating the non-missingness model, yields $\psi^{\mathrm{ipw},0}$ as derived in Appendix~\ref{subapp:theory-ipw-Hajek}.

When $\bd\alpha$ is estimated by maximum likelihood,
\[
\sqrt n(\widehat{\bd\alpha}-\bd\alpha_0)
=
\frac{1}{\sqrt n}\sum_{i=1}^nJ_{\alpha}^{-1}S_{\alpha}(\mc O_i)+o_p(1).
\]
Differentiating the estimating equation with respect to $\bd\alpha$ at $\bd\alpha_0$ gives
\[
-\mb C_{a,k}^{\mathrm{ipw}}(i_1,\ldots,i_k)^\top.
\]
Moreover, the treatment propensity score is orthogonal to the score for the correctly specified non-missingness model. In particular,
\[
\Ex\left[
S_{\alpha}(\mc O)
\left\{
I(A=a)\mb X
\bigl(\widetilde R_{1:k}-\pi_k(a,\mb X)\bigr)
\right\}^{\top}
\right]
=
\mb 0,
\]
by iterated expectations conditional on $(A,\mb X)$. Therefore, the additional first-order contribution from estimating $\bd\alpha$ is
\[
-
\mb C_{a,k}^{\mathrm{ipw}}(i_1,\ldots,i_k)^\top
J_{\alpha}^{-1}
S_{\alpha}(\mc O),
\]
which establishes \eqref{eq:if-est-treatment-ipw}.

For notational simplicity, write
\[
\mb C=\mb C_{a,k}^{\mathrm{ipw}}(i_1,\ldots,i_k).
\]
Since
\[
\Ex\left[
S_{\alpha}(\mc O)
\psi^{\mathrm{ipw},0}\{\mc O;P_a^{1:k}(i_1,\ldots,i_k)\}
\right]
=
\mb C,
\]
we have
\begin{align*}
\Vx\left(\psi^{\mathrm{ipw},e}\right)
&=
\Vx\left(
\psi^{\mathrm{ipw},0}
-
\mb C^\top J_{\alpha}^{-1}S_{\alpha}
\right)\\
&=
\Vx\left(\psi^{\mathrm{ipw},0}\right)
-
2\mb C^\top J_{\alpha}^{-1}\mb C
+
\mb C^\top J_{\alpha}^{-1}
\Ex(S_{\alpha}S_{\alpha}^\top)
J_{\alpha}^{-1}\mb C\\
&=
\Vx\left(\psi^{\mathrm{ipw},0}\right)
-
\mb C^\top J_{\alpha}^{-1}\mb C.
\end{align*}
Because $J_{\alpha}$ is positive definite, $\mb C^\top J_{\alpha}^{-1}\mb C\geq0$, which proves \eqref{eq:var-est-treatment-ipw}.
\end{proof}

Proposition~\ref{prop:estimated-treatment-ps-ipw} also characterizes when an efficiency gain occurs. If
\[
\mb C_{a,k}^{\mathrm{ipw}}(i_1,\ldots,i_k)=\mb 0,
\]
estimating the treatment propensity score has no first-order effect on the variance. Otherwise, the variance is reduced by the projection of the original influence function onto the treatment-score space. Thus, covariate adjustment through the treatment propensity score can improve precision when prognostic baseline covariates exhibit chance imbalance between randomized treatment groups.

The intercept component alone does not produce such a gain. Under simple randomization,
\[
\Ex\left[
\psi^{\mathrm{ipw},0}\{\mc O;P_a^{1:k}(i_1,\ldots,i_k)\}
\{I(A=a)-e_a\}
\right]
=0.
\]
This agrees with the exact cancellation of a constant treatment allocation factor under H\'ajek normalization: replacing the known randomization probability by the empirical treatment proportion $n_a/n$ does not change the H\'ajek IPW estimator. Any additional first-order efficiency gain therefore arises from incorporating nonconstant functions of baseline covariates into the working treatment propensity model.

\subsubsection{AIPW estimator}

A similar argument applies to the AIPW estimator. With a modeled treatment propensity score, consider
\begin{align}
\widehat P_a^{1:k,\mathrm{aipw},e}(i_1,\ldots,i_k)
=
\frac{1}{n}\sum_{j=1}^n
\Bigg[
&
\frac{I(A_j=a)}
{\widehat e_a(\mb X_j)}
\frac{\widetilde R_{j,1:k}}
{\widehat\pi_k(a,\mb X_j)}
\left\{
Z_{k,j}(i_1,\ldots,i_k)
-
\widehat\mu_k(a,\mb X_j;i_1,\ldots,i_k)
\right\}
\nonumber\\
&+
\widehat\mu_k(a,\mb X_j;i_1,\ldots,i_k)
\Bigg].
\label{eq:aipw-est-treatment-ps}
\end{align}

Let
\[
\psi^{\mathrm{aipw},0}\{\mc O;P_a^{1:k}(i_1,\ldots,i_k)\}
\]
denote the influence function in Appendix~\ref{subapp:theory-aipw} when the treatment allocation probability is treated as known, and define
\[
\mb C_{a,k}^{\mathrm{aipw}}(i_1,\ldots,i_k)
=
\Ex\left[
S_{\alpha}(\mc O)
\psi^{\mathrm{aipw},0}\{\mc O;P_a^{1:k}(i_1,\ldots,i_k)\}
\right].
\]
Under the same regularity conditions,
\[
\psi^{\mathrm{aipw},e}
=
\psi^{\mathrm{aipw},0}
-
\mb C_{a,k}^{\mathrm{aipw}\top}
J_{\alpha}^{-1}
S_{\alpha},
\]
so that
\[
\Vx\left(\psi^{\mathrm{aipw},e}\right)
=
\Vx\left(\psi^{\mathrm{aipw},0}\right)
-
\mb C_{a,k}^{\mathrm{aipw}\top}
J_{\alpha}^{-1}
\mb C_{a,k}^{\mathrm{aipw}}.
\]

The following special case is particularly relevant to the proposed AIPW estimator.

\begin{proposition}
\label{prop:aipw-treatment-ps-no-gain}
Suppose that the conditional outcome regression model $\mu_k(a,\mb X;i_1,\ldots,i_k)$ is correctly specified. Then
\[
\mb C_{a,k}^{\mathrm{aipw}}(i_1,\ldots,i_k)=\mb 0.
\]
Consequently,
\[
\psi^{\mathrm{aipw},e}\{\mc O;P_a^{1:k}(i_1,\ldots,i_k)\}
=
\psi^{\mathrm{aipw},0}\{\mc O;P_a^{1:k}(i_1,\ldots,i_k)\}
\]
up to first order. Thus, when the outcome regression model is correctly specified, estimating the treatment propensity score using baseline covariates provides no additional first-order efficiency gain for the AIPW cell-probability estimator.
\end{proposition}

\begin{proof}
Under correct specification of the outcome regression model, the leading AIPW influence-function contribution is
\[
\frac{I(A=a)}{e_a}
\frac{\widetilde R_{1:k}}{\pi_k(a,\mb X)}
\left[
Z_k(i_1,\ldots,i_k)
-
\mu_k(a,\mb X;i_1,\ldots,i_k)
\right]
+
\mu_k(a,\mb X;i_1,\ldots,i_k)
-
P_a^{1:k}(i_1,\ldots,i_k).
\]
The first term has conditional mean zero given $(A,\mb X)$ under the missingness assumption and correct specification of the outcome regression. In addition,
\[
\Ex\{S_{\alpha}(\mc O)\mid\mb X\}=\mb 0.
\]
Therefore, by iterated expectations, the leading AIPW contribution is orthogonal to $S_{\alpha}(\mc O)$. The first-order nuisance-estimation contributions under the correctly specified models are likewise orthogonal to the treatment score. Hence
\[
\Ex\left[
S_{\alpha}(\mc O)
\psi^{\mathrm{aipw},0}\{\mc O;P_a^{1:k}(i_1,\ldots,i_k)\}
\right]
=
\mb 0,
\]
which proves the result.
\end{proof}

Proposition~\ref{prop:aipw-treatment-ps-no-gain} shows that when the outcome regression correctly incorporates the prognostic baseline covariates, additional estimation of the treatment propensity score does not further reduce the first-order asymptotic variance. If the outcome regression is misspecified while the treatment and non-missingness models remain correctly specified, this orthogonality need not hold, and estimating the treatment propensity score may still yield an efficiency gain.

\section{Complete Simulation Results}\label{app:simu}

\subsection{Simulation results under setting I}\label{subapp:I-results}

\begin{table}[H]
\centering
\small
\caption{Simulation results for WR under setting I and no treatment effect (true WR = 1).}
\label{tab:sim-WR-WO-I}
\begin{tabular}{lrcccc}
\toprule
\textbf{Missing data scenario} & \textbf{Method} &
\textbf{Bias} & \textbf{RMSE} & \textbf{CP} & \textbf{CIW} \\
\midrule
 No missing data & Standard & 0.012 & 0.132 & 0.956 & 0.519 \\
  & CC & 0.012 & 0.132 & 0.956 & 0.519 \\
  & MI & 0.012 & 0.132 & 0.956 & 0.519 \\
  & IPW & 0.012 & 0.132 & 0.956 & 0.519 \\
\addlinespace
 Homo, $Y_1$ only (20\%, both groups) & Standard & 0.466 & 0.508 & \textbf{0.203} & 0.778 \\
  & CC & 0.009 & 0.150 & 0.947 & 0.580 \\
  & MI & -0.001 & 0.147 & 0.945 & 0.562 \\
  & IPW & 0.009 & 0.150 & 0.947 & 0.580 \\
\addlinespace
 Homo, $Y_2$ only (20\%, both groups) & Standard & -0.174 & 0.211 & \textbf{0.700} & 0.460 \\
  & CC & 0.014 & 0.149 & 0.954 & 0.584 \\
  & MI & 0.001 & 0.136 & 0.947 & 0.525 \\
  & IPW & 0.013 & 0.138 & 0.950 & 0.531 \\
\addlinespace
 Homo, $Y_1$ and $Y_2$ (20\%, both groups) & Standard & 0.157 & 0.229 & \textbf{0.844} & 0.661 \\
  & CC & 0.011 & 0.166 & 0.943 & 0.653 \\
  & MI & -0.010 & 0.147 & 0.940 & 0.569 \\
  & IPW & 0.010 & 0.151 & 0.943 & 0.595 \\
\addlinespace
 Hete, $Y_1$ only (10\% control, 30\% treated) & Standard & 0.480 & 0.519 & \textbf{0.174} & 0.791 \\
  & CC & 0.007 & 0.149 & 0.950 & 0.600 \\
  & MI & -0.009 & 0.144 & 0.944 & 0.571 \\
  & IPW & 0.007 & 0.149 & 0.950 & 0.600 \\
\addlinespace
 Hete, $Y_2$ only (10\% control, 30\% treated) & Standard & -0.186 & 0.217 & \textbf{0.687} & 0.456 \\
  & CC & 0.008 & 0.148 & 0.952 & 0.599 \\
  & MI & -0.013 & 0.130 & 0.957 & 0.519 \\
  & IPW & 0.006 & 0.130 & 0.960 & 0.529 \\
\addlinespace
 Hete, $Y_1$ and $Y_2$ (10\% control, 30\% treated) & Standard & 0.161 & 0.233 & \textbf{0.845} & 0.673 \\
  & CC & 0.015 & 0.180 & 0.945 & 0.707 \\
  & MI & -0.019 & 0.152 & \textbf{0.933} & 0.578 \\
  & IPW & 0.011 & 0.157 & 0.951 & 0.618 \\
\addlinespace
\bottomrule
\end{tabular}
\begin{tablenotes}\scriptsize
\item RMSE: root mean square error; CP: coverage probability, with values outside the range [0.94, 0.96] shown in bold; CIW: confidence interval width; Homo: the two treatment groups have the same (homogeneous) marginal missing data rate; Hete: the two treatment groups have different (heterogeneous) marginal missing data rates; CC: complete-case analysis; MI: multiple imputation.
\end{tablenotes}
\end{table}

\begin{table}[H]
\centering
\small
\caption{Simulation results for WO under setting I and no treatment effect (true WO = 1).}
\label{tab:sim-WO-I}
\begin{tabular}{lrcccc}
\toprule
\textbf{Missing data scenario} & \textbf{Method} &
\textbf{Bias} & \textbf{RMSE} & \textbf{CP} & \textbf{CIW} \\
\midrule
 No missing data & Standard & 0.009 & 0.106 & 0.956 & 0.417 \\
  & CC & 0.009 & 0.106 & 0.956 & 0.417 \\
  & MI & 0.009 & 0.106 & 0.956 & 0.417 \\
  & IPW & 0.009 & 0.106 & 0.956 & 0.417 \\
\addlinespace
 Homo, $Y_1$ only (20\%, both groups) & Standard & 0.313 & 0.338 & \textbf{0.204} & 0.497 \\
  & CC & 0.005 & 0.120 & 0.947 & 0.465 \\
  & MI & -0.003 & 0.118 & 0.945 & 0.449 \\
  & IPW & 0.005 & 0.120 & 0.947 & 0.465 \\
\addlinespace
 Homo, $Y_2$ only (20\%, both groups) & Standard & -0.151 & 0.183 & \textbf{0.700} & 0.401 \\
  & CC & 0.010 & 0.120 & 0.954 & 0.467 \\
  & MI & -0.000 & 0.109 & 0.947 & 0.419 \\
  & IPW & 0.009 & 0.111 & 0.950 & 0.426 \\
\addlinespace
 Homo, $Y_1$ and $Y_2$ (20\%, both groups) & Standard & 0.115 & 0.168 & \textbf{0.844} & 0.484 \\
  & CC & 0.007 & 0.133 & 0.943 & 0.522 \\
  & MI & -0.009 & 0.118 & 0.940 & 0.453 \\
  & IPW & 0.006 & 0.121 & 0.943 & 0.477 \\
\addlinespace
 Hete, $Y_1$ only (10\% control, 30\% treated) & Standard & 0.320 & 0.344 & \textbf{0.176} & 0.501 \\
  & CC & 0.004 & 0.119 & 0.950 & 0.481 \\
  & MI & -0.009 & 0.115 & 0.943 & 0.455 \\
  & IPW & 0.004 & 0.119 & 0.950 & 0.481 \\
\addlinespace
 Hete, $Y_2$ only (10\% control, 30\% treated) & Standard & -0.162 & 0.189 & \textbf{0.688} & 0.399 \\
  & CC & 0.005 & 0.119 & 0.951 & 0.480 \\
  & MI & -0.011 & 0.104 & 0.957 & 0.415 \\
  & IPW & 0.003 & 0.104 & 0.960 & 0.425 \\
\addlinespace
 Hete, $Y_1$ and $Y_2$ (10\% control, 30\% treated) & Standard & 0.119 & 0.171 & \textbf{0.844} & 0.491 \\
  & CC & 0.010 & 0.144 & 0.946 & 0.564 \\
  & MI & -0.017 & 0.121 & \textbf{0.933} & 0.459 \\
  & IPW & 0.007 & 0.126 & 0.952 & 0.495 \\
\addlinespace
\bottomrule
\end{tabular}
\begin{tablenotes}\scriptsize
\item RMSE: root mean square error; CP: coverage probability, with values outside the range [0.94, 0.96] shown in bold; CIW: confidence interval width; Homo: the two treatment groups have the same (homogeneous) marginal missing data rate; Hete: the two treatment groups have different (heterogeneous) marginal missing data rates; CC: complete-case analysis; MI: multiple imputation.
\end{tablenotes}
\end{table}

\begin{table}[H]
\centering
\small
\caption{Simulation results for NB under setting I and no treatment effect (true NB = 0).}
\label{tab:sim-NB-DOOR-I}
\begin{tabular}{lrcccc}
\toprule
\textbf{Missing data scenario} & \textbf{Method} &
\textbf{Bias} & \textbf{RMSE} & \textbf{CP} & \textbf{CIW} \\
\midrule
 No missing data & Standard & 0.001 & 0.052 & 0.954 & 0.205 \\
  & CC & 0.001 & 0.052 & 0.954 & 0.205 \\
  & MI & 0.001 & 0.052 & 0.954 & 0.205 \\
  & IPW & 0.001 & 0.052 & 0.954 & 0.205 \\
\addlinespace
 Homo, $Y_1$ only (20\%, both groups) & Standard & 0.132 & 0.141 & \textbf{0.202} & 0.184 \\
  & CC & -0.001 & 0.060 & 0.946 & 0.229 \\
  & MI & -0.005 & 0.059 & 0.942 & 0.222 \\
  & IPW & -0.001 & 0.060 & 0.946 & 0.229 \\
\addlinespace
 Homo, $Y_2$ only (20\%, both groups) & Standard & -0.071 & 0.086 & \textbf{0.695} & 0.192 \\
  & CC & 0.001 & 0.059 & 0.952 & 0.229 \\
  & MI & -0.003 & 0.054 & 0.946 & 0.208 \\
  & IPW & 0.002 & 0.054 & 0.949 & 0.209 \\
\addlinespace
 Homo, $Y_1$ and $Y_2$ (20\%, both groups) & Standard & 0.039 & 0.057 & \textbf{0.841} & 0.163 \\
  & CC & -0.001 & 0.065 & \textbf{0.940} & 0.256 \\
  & MI & -0.008 & 0.059 & \textbf{0.938} & 0.226 \\
  & IPW & -0.000 & 0.060 & 0.943 & 0.234 \\
\addlinespace
 Hete, $Y_1$ only (10\% control, 30\% treated) & Standard & 0.135 & 0.143 & \textbf{0.173} & 0.185 \\
  & CC & -0.001 & 0.059 & 0.947 & 0.236 \\
  & MI & -0.008 & 0.058 & 0.941 & 0.227 \\
  & IPW & -0.001 & 0.059 & 0.947 & 0.236 \\
\addlinespace
 Hete, $Y_2$ only (10\% control, 30\% treated) & Standard & -0.075 & 0.089 & \textbf{0.681} & 0.192 \\
  & CC & -0.001 & 0.059 & 0.949 & 0.236 \\
  & MI & -0.008 & 0.053 & 0.956 & 0.208 \\
  & IPW & -0.001 & 0.052 & 0.959 & 0.210 \\
\addlinespace
 Hete, $Y_1$ and $Y_2$ (10\% control, 30\% treated) & Standard & 0.040 & 0.057 & \textbf{0.840} & 0.163 \\
  & CC & 0.000 & 0.071 & 0.943 & 0.274 \\
  & MI & -0.012 & 0.062 & \textbf{0.931} & 0.231 \\
  & IPW & -0.000 & 0.062 & 0.950 & 0.242 \\
\addlinespace
\bottomrule
\end{tabular}
\begin{tablenotes}\scriptsize
\item RMSE: root mean square error; CP: coverage probability, with values outside the range [0.94, 0.96] shown in bold; CIW: confidence interval width; Homo: the two treatment groups have the same (homogeneous) marginal missing data rate; Hete: the two treatment groups have different (heterogeneous) marginal missing data rates; CC: complete-case analysis; MI: multiple imputation.
\end{tablenotes}
\end{table}

\begin{table}[H]
\centering
\small
\caption{Simulation results for DOOR under setting I and no treatment effect (true DOOR = 0.5).}
\label{tab:sim-DOOR-I}
\begin{tabular}{lrcccc}
\toprule
\textbf{Missing data scenario} & \textbf{Method} &
\textbf{Bias} & \textbf{RMSE} & \textbf{CP} & \textbf{CIW} \\
\midrule
 No missing data & Standard & 0.001 & 0.026 & 0.954 & 0.102 \\
  & CC & 0.001 & 0.026 & 0.954 & 0.102 \\
  & MI & 0.001 & 0.026 & 0.954 & 0.102 \\
  & IPW & 0.001 & 0.026 & 0.954 & 0.102 \\
\addlinespace
 Homo, $Y_1$ only (20\%, both groups) & Standard & 0.066 & 0.070 & \textbf{0.202} & 0.092 \\
  & CC & -0.000 & 0.030 & 0.946 & 0.114 \\
  & MI & -0.002 & 0.029 & 0.942 & 0.111 \\
  & IPW & -0.000 & 0.030 & 0.946 & 0.114 \\
\addlinespace
 Homo, $Y_2$ only (20\%, both groups) & Standard & -0.035 & 0.043 & \textbf{0.695} & 0.096 \\
  & CC & 0.001 & 0.029 & 0.952 & 0.114 \\
  & MI & -0.002 & 0.027 & 0.946 & 0.104 \\
  & IPW & 0.001 & 0.027 & 0.949 & 0.105 \\
\addlinespace
 Homo, $Y_1$ and $Y_2$ (20\%, both groups) & Standard & 0.020 & 0.028 & \textbf{0.841} & 0.081 \\
  & CC & -0.000 & 0.033 & \textbf{0.940} & 0.128 \\
  & MI & -0.004 & 0.030 & \textbf{0.938} & 0.113 \\
  & IPW & -0.000 & 0.030 & 0.943 & 0.117 \\
\addlinespace
 Hete, $Y_1$ only (10\% control, 30\% treated) & Standard & 0.068 & 0.072 & \textbf{0.173} & 0.092 \\
  & CC & -0.001 & 0.030 & 0.947 & 0.118 \\
  & MI & -0.004 & 0.029 & 0.941 & 0.113 \\
  & IPW & -0.001 & 0.030 & 0.947 & 0.118 \\
\addlinespace
 Hete, $Y_2$ only (10\% control, 30\% treated) & Standard & -0.037 & 0.044 & \textbf{0.681} & 0.096 \\
  & CC & -0.001 & 0.029 & 0.949 & 0.118 \\
  & MI & -0.004 & 0.026 & 0.956 & 0.104 \\
  & IPW & -0.000 & 0.026 & 0.959 & 0.105 \\
\addlinespace
 Hete, $Y_1$ and $Y_2$ (10\% control, 30\% treated) & Standard & 0.020 & 0.029 & \textbf{0.840} & 0.081 \\
  & CC & 0.000 & 0.035 & 0.943 & 0.137 \\
  & MI & -0.006 & 0.031 & \textbf{0.931} & 0.115 \\
  & IPW & -0.000 & 0.031 & 0.950 & 0.121 \\
\addlinespace
\bottomrule
\end{tabular}
\begin{tablenotes}\scriptsize
\item RMSE: root mean square error; CP: coverage probability, with values outside the range [0.94, 0.96] shown in bold; CIW: confidence interval width; Homo: the two treatment groups have the same (homogeneous) marginal missing data rate; Hete: the two treatment groups have different (heterogeneous) marginal missing data rates; CC: complete-case analysis; MI: multiple imputation.
\end{tablenotes}
\end{table}

\begin{table}[H]
\centering
\small
\caption{Simulation results for WR under setting I and significant treatment effect (true WR = 1.69).}
\label{tab:sim-WR-WO-I-2}
\begin{tabular}{lrcccc}
\toprule
\textbf{Missing data scenario} & \textbf{Method} &
\textbf{Bias} & \textbf{RMSE} & \textbf{CP} & \textbf{CIW} \\
\midrule
 No missing data & Standard & 0.031 & 0.247 & 0.944 & 0.945 \\
  & CC & 0.031 & 0.247 & 0.944 & 0.945 \\
  & MI & 0.031 & 0.247 & 0.944 & 0.945 \\
  & IPW & 0.031 & 0.247 & 0.944 & 0.945 \\
\addlinespace
 Homo, $Y_1$ only (20\%, both groups) & Standard & 0.480 & 0.581 & \textbf{0.624} & 1.258 \\
  & CC & 0.028 & 0.277 & 0.941 & 1.059 \\
  & MI & 0.020 & 0.269 & 0.943 & 1.007 \\
  & IPW & 0.028 & 0.277 & 0.941 & 1.059 \\
\addlinespace
 Homo, $Y_2$ only (20\%, both groups) & Standard & -0.174 & 0.290 & \textbf{0.858} & 0.896 \\
  & CC & 0.026 & 0.277 & \textbf{0.940} & 1.057 \\
  & MI & 0.013 & 0.251 & 0.941 & 0.946 \\
  & IPW & 0.022 & 0.249 & 0.941 & 0.957 \\
\addlinespace
 Homo, $Y_1$ and $Y_2$ (20\%, both groups) & Standard & 0.185 & 0.339 & \textbf{0.917} & 1.159 \\
  & CC & 0.033 & 0.298 & 0.956 & 1.193 \\
  & MI & 0.016 & 0.260 & 0.947 & 1.022 \\
  & IPW & 0.031 & 0.268 & 0.957 & 1.081 \\
\addlinespace
 Hete, $Y_1$ only (10\% control, 30\% treated) & Standard & 0.476 & 0.572 & \textbf{0.626} & 1.253 \\
  & CC & 0.011 & 0.270 & 0.952 & 1.052 \\
  & MI & 0.017 & 0.263 & 0.946 & 1.004 \\
  & IPW & 0.011 & 0.270 & 0.952 & 1.052 \\
\addlinespace
 Hete, $Y_2$ only (10\% control, 30\% treated) & Standard & -0.181 & 0.290 & \textbf{0.864} & 0.894 \\
  & CC & 0.026 & 0.271 & 0.954 & 1.061 \\
  & MI & 0.026 & 0.249 & 0.947 & 0.955 \\
  & IPW & 0.020 & 0.244 & 0.952 & 0.958 \\
\addlinespace
 Hete, $Y_1$ and $Y_2$ (10\% control, 30\% treated) & Standard & 0.180 & 0.351 & \textbf{0.911} & 1.160 \\
  & CC & 0.024 & 0.314 & 0.942 & 1.207 \\
  & MI & 0.030 & 0.280 & \textbf{0.932} & 1.030 \\
  & IPW & 0.019 & 0.282 & 0.940 & 1.082 \\
\addlinespace
\bottomrule
\end{tabular}
\begin{tablenotes}\scriptsize
\item RMSE: root mean square error; CP: coverage probability, with values outside the range [0.94, 0.96] shown in bold; CIW: confidence interval width; Homo: the two treatment groups have the same (homogeneous) marginal missing data rate; Hete: the two treatment groups have different (heterogeneous) marginal missing data rates; CC: complete-case analysis; MI: multiple imputation.
\end{tablenotes}
\end{table}

\begin{table}[H]
\centering
\small
\caption{Simulation results for WO under setting I and significant treatment effect (true WO = 1.46).}
\label{tab:sim-WO-I-2}
\begin{tabular}{lrcccc}
\toprule
\textbf{Missing data scenario} & \textbf{Method} &
\textbf{Bias} & \textbf{RMSE} & \textbf{CP} & \textbf{CIW} \\
\midrule
 No missing data & Standard & 0.015 & 0.153 & 0.945 & 0.585 \\
  & CC & 0.015 & 0.153 & 0.945 & 0.585 \\
  & MI & 0.015 & 0.153 & 0.945 & 0.585 \\
  & IPW & 0.015 & 0.153 & 0.945 & 0.585 \\
\addlinespace
 Homo, $Y_1$ only (20\%, both groups) & Standard & 0.188 & 0.248 & \textbf{0.759} & 0.621 \\
  & CC & 0.012 & 0.171 & 0.943 & 0.654 \\
  & MI & 0.017 & 0.171 & 0.942 & 0.636 \\
  & IPW & 0.012 & 0.171 & 0.943 & 0.654 \\
\addlinespace
 Homo, $Y_2$ only (20\%, both groups) & Standard & -0.080 & 0.185 & \textbf{0.903} & 0.638 \\
  & CC & 0.011 & 0.172 & \textbf{0.939} & 0.654 \\
  & MI & 0.014 & 0.162 & 0.940 & 0.605 \\
  & IPW & 0.010 & 0.156 & 0.942 & 0.596 \\
\addlinespace
 Homo, $Y_1$ and $Y_2$ (20\%, both groups) & Standard & 0.091 & 0.189 & \textbf{0.931} & 0.673 \\
  & CC & 0.014 & 0.185 & 0.953 & 0.734 \\
  & MI & 0.022 & 0.171 & 0.945 & 0.658 \\
  & IPW & 0.014 & 0.167 & 0.958 & 0.670 \\
\addlinespace
 Hete, $Y_1$ only (10\% control, 30\% treated) & Standard & 0.183 & 0.241 & \textbf{0.776} & 0.617 \\
  & CC & 0.001 & 0.168 & 0.951 & 0.655 \\
  & MI & 0.017 & 0.168 & 0.949 & 0.638 \\
  & IPW & 0.001 & 0.168 & 0.951 & 0.655 \\
\addlinespace
 Hete, $Y_2$ only (10\% control, 30\% treated) & Standard & -0.084 & 0.184 & \textbf{0.905} & 0.640 \\
  & CC & 0.011 & 0.169 & 0.953 & 0.659 \\
  & MI & 0.024 & 0.161 & 0.947 & 0.611 \\
  & IPW & 0.008 & 0.152 & 0.954 & 0.597 \\
\addlinespace
 Hete, $Y_1$ and $Y_2$ (10\% control, 30\% treated) & Standard & 0.088 & 0.197 & \textbf{0.922} & 0.675 \\
  & CC & 0.009 & 0.196 & 0.941 & 0.750 \\
  & MI & 0.033 & 0.184 & \textbf{0.928} & 0.667 \\
  & IPW & 0.007 & 0.177 & 0.942 & 0.677 \\
\addlinespace
\bottomrule
\end{tabular}
\begin{tablenotes}\scriptsize
\item RMSE: root mean square error; CP: coverage probability, with values outside the range [0.94, 0.96] shown in bold; CIW: confidence interval width; Homo: the two treatment groups have the same (homogeneous) marginal missing data rate; Hete: the two treatment groups have different (heterogeneous) marginal missing data rates; CC: complete-case analysis; MI: multiple imputation.
\end{tablenotes}
\end{table}

\begin{table}[H]
\centering
\small
\caption{Simulation results for NB under setting I and significant treatment effect (true NB = 0.19).}
\label{tab:sim-NB-DOOR-I-2}
\begin{tabular}{lrcccc}
\toprule
\textbf{Missing data scenario} & \textbf{Method} &
\textbf{Bias} & \textbf{RMSE} & \textbf{CP} & \textbf{CIW} \\
\midrule
 No missing data & Standard & 0.002 & 0.049 & 0.941 & 0.189 \\
  & CC & 0.002 & 0.049 & 0.941 & 0.189 \\
  & MI & 0.002 & 0.049 & 0.941 & 0.189 \\
  & IPW & 0.002 & 0.049 & 0.941 & 0.189 \\
\addlinespace
 Homo, $Y_1$ only (20\%, both groups) & Standard & 0.055 & 0.072 & \textbf{0.751} & 0.175 \\
  & CC & 0.000 & 0.055 & 0.941 & 0.211 \\
  & MI & 0.002 & 0.055 & \textbf{0.939} & 0.205 \\
  & IPW & 0.000 & 0.055 & 0.941 & 0.211 \\
\addlinespace
 Homo, $Y_2$ only (20\%, both groups) & Standard & -0.060 & 0.077 & \textbf{0.740} & 0.182 \\
  & CC & -0.000 & 0.056 & \textbf{0.939} & 0.211 \\
  & MI & 0.001 & 0.052 & \textbf{0.939} & 0.195 \\
  & IPW & 0.000 & 0.051 & 0.940 & 0.193 \\
\addlinespace
 Homo, $Y_1$ and $Y_2$ (20\%, both groups) & Standard & -0.027 & 0.047 & \textbf{0.914} & 0.156 \\
  & CC & 0.000 & 0.060 & 0.951 & 0.236 \\
  & MI & 0.003 & 0.054 & 0.944 & 0.211 \\
  & IPW & 0.001 & 0.054 & 0.955 & 0.216 \\
\addlinespace
 Hete, $Y_1$ only (10\% control, 30\% treated) & Standard & 0.054 & 0.070 & \textbf{0.761} & 0.175 \\
  & CC & -0.003 & 0.055 & 0.950 & 0.213 \\
  & MI & 0.002 & 0.054 & 0.945 & 0.205 \\
  & IPW & -0.003 & 0.055 & 0.950 & 0.213 \\
\addlinespace
 Hete, $Y_2$ only (10\% control, 30\% treated) & Standard & -0.062 & 0.078 & \textbf{0.736} & 0.182 \\
  & CC & -0.000 & 0.055 & 0.949 & 0.213 \\
  & MI & 0.004 & 0.052 & 0.947 & 0.196 \\
  & IPW & -0.000 & 0.050 & 0.954 & 0.194 \\
\addlinespace
 Hete, $Y_1$ and $Y_2$ (10\% control, 30\% treated) & Standard & -0.030 & 0.051 & \textbf{0.878} & 0.155 \\
  & CC & -0.002 & 0.064 & \textbf{0.940} & 0.242 \\
  & MI & 0.006 & 0.058 & \textbf{0.924} & 0.212 \\
  & IPW & -0.002 & 0.057 & 0.942 & 0.219 \\
\addlinespace
\bottomrule
\end{tabular}
\begin{tablenotes}\scriptsize
\item RMSE: root mean square error; CP: coverage probability, with values outside the range [0.94, 0.96] shown in bold; CIW: confidence interval width; Homo: the two treatment groups have the same (homogeneous) marginal missing data rate; Hete: the two treatment groups have different (heterogeneous) marginal missing data rates; CC: complete-case analysis; MI: multiple imputation.
\end{tablenotes}
\end{table}

\begin{table}[H]
\centering
\small
\caption{Simulation results for DOOR under setting I and significant treatment effect (true DOOR = 0.59).}
\label{tab:sim-DOOR-I-2}
\begin{tabular}{lrcccc}
\toprule
\textbf{Missing data scenario} & \textbf{Method} &
\textbf{Bias} & \textbf{RMSE} & \textbf{CP} & \textbf{CIW} \\
\midrule
 No missing data & Standard & 0.001 & 0.025 & 0.941 & 0.094 \\
  & CC & 0.001 & 0.025 & 0.941 & 0.094 \\
  & MI & 0.001 & 0.025 & 0.941 & 0.094 \\
  & IPW & 0.001 & 0.025 & 0.941 & 0.094 \\
\addlinespace
 Homo, $Y_1$ only (20\%, both groups) & Standard & 0.027 & 0.036 & \textbf{0.751} & 0.088 \\
  & CC & 0.000 & 0.028 & 0.941 & 0.106 \\
  & MI & 0.001 & 0.027 & \textbf{0.939} & 0.102 \\
  & IPW & 0.000 & 0.028 & 0.941 & 0.106 \\
\addlinespace
 Homo, $Y_2$ only (20\%, both groups) & Standard & -0.030 & 0.038 & \textbf{0.740} & 0.091 \\
  & CC & -0.000 & 0.028 & \textbf{0.939} & 0.106 \\
  & MI & 0.001 & 0.026 & \textbf{0.939} & 0.098 \\
  & IPW & 0.000 & 0.025 & 0.940 & 0.097 \\
\addlinespace
 Homo, $Y_1$ and $Y_2$ (20\%, both groups) & Standard & -0.014 & 0.024 & \textbf{0.914} & 0.078 \\
  & CC & 0.000 & 0.030 & 0.951 & 0.118 \\
  & MI & 0.002 & 0.027 & 0.944 & 0.105 \\
  & IPW & 0.000 & 0.027 & 0.955 & 0.108 \\
\addlinespace
 Hete, $Y_1$ only (10\% control, 30\% treated) & Standard & 0.027 & 0.035 & \textbf{0.761} & 0.087 \\
  & CC & -0.002 & 0.028 & 0.950 & 0.107 \\
  & MI & 0.001 & 0.027 & 0.945 & 0.103 \\
  & IPW & -0.002 & 0.028 & 0.950 & 0.107 \\
\addlinespace
 Hete, $Y_2$ only (10\% control, 30\% treated) & Standard & -0.031 & 0.039 & \textbf{0.736} & 0.091 \\
  & CC & -0.000 & 0.027 & 0.949 & 0.106 \\
  & MI & 0.002 & 0.026 & 0.947 & 0.098 \\
  & IPW & -0.000 & 0.025 & 0.954 & 0.097 \\
\addlinespace
 Hete, $Y_1$ and $Y_2$ (10\% control, 30\% treated) & Standard & -0.015 & 0.025 & \textbf{0.878} & 0.077 \\
  & CC & -0.001 & 0.032 & \textbf{0.940} & 0.121 \\
  & MI & 0.003 & 0.029 & \textbf{0.924} & 0.106 \\
  & IPW & -0.001 & 0.029 & 0.942 & 0.110 \\
\addlinespace
\bottomrule
\end{tabular}
\begin{tablenotes}\scriptsize
\item RMSE: root mean square error; CP: coverage probability, with values outside the range [0.94, 0.96] shown in bold; CIW: confidence interval width; Homo: the two treatment groups have the same (homogeneous) marginal missing data rate; Hete: the two treatment groups have different (heterogeneous) marginal missing data rates; CC: complete-case analysis; MI: multiple imputation.
\end{tablenotes}
\end{table}

\subsection{Simulation results under setting II}\label{subapp:II-results}

\begin{table}[H]
\centering
\footnotesize
\caption{Simulation results for WR and WO under Setting II: bias and RMSE. The true values are WR = 1.29 and WO = 1.19.}
\label{tab:sim-WR-WO-II-bias-rmse}
\setlength{\tabcolsep}{2.8pt}
\renewcommand{\arraystretch}{1.02}
\begin{tabular}{l p{6.5cm} cccccccc}
\toprule
 & & \multicolumn{8}{c}{\textbf{Bias}} \\
\cmidrule(lr){3-10}
\textbf{Estimand} & \makecell{\textbf{Missing data}\\\textbf{scenario}} & \textbf{Standard} & \textbf{CC} & \textbf{MI} & \multicolumn{2}{c}{\textbf{IPW}} & \multicolumn{3}{c}{\textbf{AIPW}} \\
\cmidrule(lr){6-7}\cmidrule(lr){8-10}
 & & & & & \textbf{A} & \textbf{B} & \textbf{A} & \textbf{B} & \textbf{C} \\
\midrule
\multirow{7}{*}{WR} & No missing data & 0.004 & 0.004 & 0.004 & 0.004 & 0.004 & 0.001 & 0.001 & 0.004 \\
 & Homo: $Y_1$ only (20\%, both) & 0.157 & 0.157 & 0.004 & 0.002 & 0.148 & 0.002 & 0.002 & 0.004 \\
 & Homo: $Y_2$ only (20\%, both) & 0.100 & 0.161 & 0.006 & 0.002 & 0.072 & 0.005 & 0.005 & 0.006 \\
 & Homo: $Y_1$ and $Y_2$ (20\%, both) & 0.154 & 0.154 & 0.002 & 0.003 & 0.147 & 0.002 & 0.002 & 0.003 \\
 & Hete: $Y_1$ only (10\% control, 30\% treated) & -0.075 & -0.075 & 0.006 & 0.004 & -0.024 & 0.003 & 0.003 & 0.003 \\
 & Hete: $Y_2$ only (10\% control, 30\% treated) & 0.013 & -0.076 & 0.001 & 0.003 & -0.011 & -0.001 & -0.001 & -0.000 \\
 & Hete: $Y_1$ and $Y_2$ (10\% control, 30\% treated) & -0.068 & -0.068 & 0.012 & 0.004 & -0.026 & 0.006 & 0.006 & 0.009 \\
\addlinespace
\multirow{7}{*}{WO} & No missing data & 0.002 & 0.002 & 0.002 & 0.002 & 0.002 & 0.001 & 0.001 & 0.002 \\
 & Homo: $Y_1$ only (20\%, both) & -0.026 & 0.081 & 0.002 & 0.001 & 0.080 & 0.001 & 0.001 & 0.002 \\
 & Homo: $Y_2$ only (20\%, both) & 0.012 & 0.084 & 0.003 & 0.000 & 0.045 & 0.003 & 0.003 & 0.003 \\
 & Homo: $Y_1$ and $Y_2$ (20\%, both) & -0.027 & 0.079 & 0.001 & 0.001 & 0.080 & 0.001 & 0.001 & 0.001 \\
 & Hete: $Y_1$ only (10\% control, 30\% treated) & -0.110 & -0.058 & 0.003 & 0.002 & -0.023 & 0.002 & 0.002 & 0.001 \\
 & Hete: $Y_2$ only (10\% control, 30\% treated) & -0.033 & -0.058 & 0.000 & 0.002 & -0.008 & -0.001 & -0.001 & -0.001 \\
 & Hete: $Y_1$ and $Y_2$ (10\% control, 30\% treated) & -0.108 & -0.053 & 0.007 & 0.002 & -0.024 & 0.004 & 0.004 & 0.006 \\
\midrule
 & & \multicolumn{8}{c}{\textbf{RMSE}} \\
\cmidrule(lr){3-10}
\textbf{Estimand} & \makecell{\textbf{Missing data}\\\textbf{scenario}} & \textbf{Standard} & \textbf{CC} & \textbf{MI} & \multicolumn{2}{c}{\textbf{IPW}} & \multicolumn{3}{c}{\textbf{AIPW}} \\
\cmidrule(lr){6-7}\cmidrule(lr){8-10}
 & & & & & \textbf{A} & \textbf{B} & \textbf{A} & \textbf{B} & \textbf{C} \\
\midrule
\multirow{7}{*}{WR} & No missing data & 0.088 & 0.088 & 0.088 & 0.089 & 0.089 & 0.066 & 0.066 & 0.088 \\
 & Homo: $Y_1$ only (20\%, both) & 0.195 & 0.195 & 0.096 & 0.098 & 0.188 & 0.074 & 0.074 & 0.098 \\
 & Homo: $Y_2$ only (20\%, both) & 0.147 & 0.201 & 0.094 & 0.093 & 0.123 & 0.073 & 0.072 & 0.096 \\
 & Homo: $Y_1$ and $Y_2$ (20\%, both) & 0.193 & 0.193 & 0.095 & 0.095 & 0.185 & 0.075 & 0.074 & 0.098 \\
 & Hete: $Y_1$ only (10\% control, 30\% treated) & 0.124 & 0.124 & 0.098 & 0.100 & 0.106 & 0.076 & 0.076 & 0.098 \\
 & Hete: $Y_2$ only (10\% control, 30\% treated) & 0.099 & 0.124 & 0.091 & 0.094 & 0.095 & 0.070 & 0.069 & 0.092 \\
 & Hete: $Y_1$ and $Y_2$ (10\% control, 30\% treated) & 0.121 & 0.121 & 0.099 & 0.097 & 0.104 & 0.078 & 0.078 & 0.101 \\
\addlinespace
\multirow{7}{*}{WO} & No missing data & 0.058 & 0.058 & 0.058 & 0.058 & 0.058 & 0.043 & 0.043 & 0.058 \\
 & Homo: $Y_1$ only (20\%, both) & 0.048 & 0.105 & 0.063 & 0.064 & 0.106 & 0.048 & 0.048 & 0.064 \\
 & Homo: $Y_2$ only (20\%, both) & 0.055 & 0.109 & 0.062 & 0.061 & 0.078 & 0.048 & 0.047 & 0.063 \\
 & Homo: $Y_1$ and $Y_2$ (20\%, both) & 0.048 & 0.104 & 0.062 & 0.062 & 0.105 & 0.049 & 0.049 & 0.064 \\
 & Hete: $Y_1$ only (10\% control, 30\% treated) & 0.116 & 0.085 & 0.064 & 0.065 & 0.069 & 0.050 & 0.050 & 0.064 \\
 & Hete: $Y_2$ only (10\% control, 30\% treated) & 0.060 & 0.085 & 0.060 & 0.061 & 0.062 & 0.046 & 0.046 & 0.060 \\
 & Hete: $Y_1$ and $Y_2$ (10\% control, 30\% treated) & 0.114 & 0.083 & 0.065 & 0.063 & 0.068 & 0.051 & 0.051 & 0.066 \\
\bottomrule
\end{tabular}
\begin{tablenotes}\scriptsize
\item RMSE: root mean square error.
\item Homo: the two treatment groups have the same (homogeneous) marginal missing-data rate; Hete: the two treatment groups have different (heterogeneous) marginal missing-data rates.
\item Standard: missing comparisons are treated as ties; CC: complete-case analysis; MI: multiple imputation using 25 imputed data sets; IPW: H\'ajek-normalized inverse probability weighting; IPW-A: correctly specified missingness model; IPW-B: misspecified missingness model; AIPW-A: both missingness and outcome models correctly specified; AIPW-B: misspecified missingness model; AIPW-C: misspecified outcome model.
\end{tablenotes}
\end{table}

\begin{table}[H]
\centering
\footnotesize
\caption{Simulation results for WR and WO under Setting II: coverage probability and confidence-interval width.}
\label{tab:sim-WR-WO-II-cp-ciw}
\setlength{\tabcolsep}{2.8pt}
\renewcommand{\arraystretch}{1.02}
\begin{tabular}{l p{6.5cm} cccccccc}
\toprule
 & & \multicolumn{8}{c}{\textbf{CP}} \\
\cmidrule(lr){3-10}
\textbf{Estimand} & \makecell{\textbf{Missing data}\\\textbf{scenario}} & \textbf{Standard} & \textbf{CC} & \textbf{MI} & \multicolumn{2}{c}{\textbf{IPW}} & \multicolumn{3}{c}{\textbf{AIPW}} \\
\cmidrule(lr){6-7}\cmidrule(lr){8-10}
 & & & & & \textbf{A} & \textbf{B} & \textbf{A} & \textbf{B} & \textbf{C} \\
\midrule
\multirow{7}{*}{WR} & No missing data & 0.950 & 0.950 & 0.950 & 0.950 & 0.950 & 0.951 & 0.951 & 0.950 \\
 & Homo: $Y_1$ only (20\%, both) & \textbf{0.719} & \textbf{0.719} & 0.950 & 0.951 & \textbf{0.730} & 0.955 & 0.956 & \textbf{0.973} \\
 & Homo: $Y_2$ only (20\%, both) & \textbf{0.845} & \textbf{0.698} & 0.946 & 0.949 & \textbf{0.895} & 0.945 & 0.947 & 0.949 \\
 & Homo: $Y_1$ and $Y_2$ (20\%, both) & \textbf{0.712} & \textbf{0.712} & 0.957 & \textbf{0.962} & \textbf{0.749} & 0.953 & 0.952 & \textbf{0.972} \\
 & Hete: $Y_1$ only (10\% control, 30\% treated) & \textbf{0.871} & \textbf{0.871} & 0.946 & \textbf{0.938} & \textbf{0.928} & 0.948 & 0.948 & \textbf{0.964} \\
 & Hete: $Y_2$ only (10\% control, 30\% treated) & 0.957 & \textbf{0.874} & 0.952 & 0.953 & 0.945 & 0.952 & 0.958 & 0.959 \\
 & Hete: $Y_1$ and $Y_2$ (10\% control, 30\% treated) & \textbf{0.887} & \textbf{0.887} & \textbf{0.939} & 0.954 & 0.940 & 0.943 & 0.945 & \textbf{0.961} \\
\addlinespace
\multirow{7}{*}{WO} & No missing data & 0.952 & 0.952 & 0.952 & 0.952 & 0.952 & 0.953 & 0.953 & 0.951 \\
 & Homo: $Y_1$ only (20\%, both) & \textbf{0.892} & \textbf{0.776} & 0.947 & 0.951 & \textbf{0.776} & 0.957 & 0.958 & \textbf{0.973} \\
 & Homo: $Y_2$ only (20\%, both) & 0.947 & \textbf{0.756} & 0.944 & 0.951 & \textbf{0.895} & 0.942 & 0.946 & 0.948 \\
 & Homo: $Y_1$ and $Y_2$ (20\%, both) & \textbf{0.894} & \textbf{0.766} & 0.958 & \textbf{0.961} & \textbf{0.787} & 0.952 & 0.951 & \textbf{0.972} \\
 & Hete: $Y_1$ only (10\% control, 30\% treated) & \textbf{0.180} & \textbf{0.845} & 0.945 & \textbf{0.938} & \textbf{0.920} & 0.947 & 0.948 & \textbf{0.965} \\
 & Hete: $Y_2$ only (10\% control, 30\% treated) & \textbf{0.896} & \textbf{0.833} & 0.951 & 0.954 & 0.947 & 0.953 & 0.957 & 0.959 \\
 & Hete: $Y_1$ and $Y_2$ (10\% control, 30\% treated) & \textbf{0.201} & \textbf{0.855} & 0.940 & 0.952 & \textbf{0.928} & 0.945 & 0.947 & \textbf{0.962} \\
\midrule
 & & \multicolumn{8}{c}{\textbf{CIW}} \\
\cmidrule(lr){3-10}
\textbf{Estimand} & \makecell{\textbf{Missing data}\\\textbf{scenario}} & \textbf{Standard} & \textbf{CC} & \textbf{MI} & \multicolumn{2}{c}{\textbf{IPW}} & \multicolumn{3}{c}{\textbf{AIPW}} \\
\cmidrule(lr){6-7}\cmidrule(lr){8-10}
 & & & & & \textbf{A} & \textbf{B} & \textbf{A} & \textbf{B} & \textbf{C} \\
\midrule
\multirow{7}{*}{WR} & No missing data & 0.349 & 0.349 & 0.349 & 0.349 & 0.349 & 0.257 & 0.257 & 0.349 \\
 & Homo: $Y_1$ only (20\%, both) & 0.455 & 0.455 & 0.374 & 0.380 & 0.448 & 0.292 & 0.292 & 0.424 \\
 & Homo: $Y_2$ only (20\%, both) & 0.415 & 0.457 & 0.358 & 0.364 & 0.393 & 0.279 & 0.281 & 0.373 \\
 & Homo: $Y_1$ and $Y_2$ (20\%, both) & 0.453 & 0.453 & 0.374 & 0.381 & 0.449 & 0.293 & 0.292 & 0.423 \\
 & Hete: $Y_1$ only (10\% control, 30\% treated) & 0.379 & 0.379 & 0.375 & 0.379 & 0.392 & 0.295 & 0.296 & 0.412 \\
 & Hete: $Y_2$ only (10\% control, 30\% treated) & 0.388 & 0.379 & 0.358 & 0.365 & 0.367 & 0.278 & 0.280 & 0.369 \\
 & Hete: $Y_1$ and $Y_2$ (10\% control, 30\% treated) & 0.381 & 0.381 & 0.376 & 0.379 & 0.392 & 0.295 & 0.297 & 0.415 \\
\addlinespace
\multirow{7}{*}{WO} & No missing data & 0.228 & 0.228 & 0.228 & 0.228 & 0.228 & 0.169 & 0.169 & 0.228 \\
 & Homo: $Y_1$ only (20\%, both) & 0.156 & 0.267 & 0.244 & 0.248 & 0.269 & 0.191 & 0.191 & 0.276 \\
 & Homo: $Y_2$ only (20\%, both) & 0.207 & 0.268 & 0.235 & 0.238 & 0.251 & 0.183 & 0.184 & 0.243 \\
 & Homo: $Y_1$ and $Y_2$ (20\%, both) & 0.156 & 0.266 & 0.245 & 0.249 & 0.269 & 0.192 & 0.191 & 0.276 \\
 & Hete: $Y_1$ only (10\% control, 30\% treated) & 0.144 & 0.239 & 0.246 & 0.249 & 0.249 & 0.194 & 0.195 & 0.271 \\
 & Hete: $Y_2$ only (10\% control, 30\% treated) & 0.199 & 0.239 & 0.234 & 0.239 & 0.241 & 0.183 & 0.185 & 0.242 \\
 & Hete: $Y_1$ and $Y_2$ (10\% control, 30\% treated) & 0.144 & 0.240 & 0.246 & 0.248 & 0.249 & 0.194 & 0.196 & 0.272 \\
\bottomrule
\end{tabular}
\begin{tablenotes}\scriptsize
\item CP: coverage probability; CIW: average 95\% confidence-interval width. CP values outside the range [0.94, 0.96] are shown in bold.
\item Homo: the two treatment groups have the same (homogeneous) marginal missing-data rate; Hete: the two treatment groups have different (heterogeneous) marginal missing-data rates.
\item Standard: missing comparisons are treated as ties; CC: complete-case analysis; MI: multiple imputation using 25 imputed data sets; IPW: H\'ajek-normalized inverse probability weighting; IPW-A: correctly specified missingness model; IPW-B: misspecified missingness model; AIPW-A: both missingness and outcome models correctly specified; AIPW-B: misspecified missingness model; AIPW-C: misspecified outcome model.
\end{tablenotes}
\end{table}

\begin{table}[H]
\centering
\footnotesize
\caption{Simulation results for NB and DOOR under Setting II: bias and RMSE. The true values are NB = 0.09 and DOOR = 0.54.}
\label{tab:sim-NB-DOOR-combined}
\label{tab:sim-NB-DOOR-II-bias-rmse}
\setlength{\tabcolsep}{2.8pt}
\renewcommand{\arraystretch}{1.02}
\begin{tabular}{l p{6.5cm} cccccccc}
\toprule
 & & \multicolumn{8}{c}{\textbf{Bias}} \\
\cmidrule(lr){3-10}
\textbf{Estimand} & \makecell{\textbf{Missing data}\\\textbf{scenario}} & \textbf{Standard} & \textbf{CC} & \textbf{MI} & \multicolumn{2}{c}{\textbf{IPW}} & \multicolumn{3}{c}{\textbf{AIPW}} \\
\cmidrule(lr){6-7}\cmidrule(lr){8-10}
 & & & & & \textbf{A} & \textbf{B} & \textbf{A} & \textbf{B} & \textbf{C} \\
\midrule
\multirow{7}{*}{NB} & No missing data & 0.000 & 0.000 & 0.000 & 0.000 & 0.000 & 0.000 & 0.000 & 0.000 \\
 & Homo: $Y_1$ only (20\%, both) & -0.011 & 0.032 & 0.000 & 0.000 & 0.032 & 0.000 & 0.000 & 0.000 \\
 & Homo: $Y_2$ only (20\%, both) & 0.004 & 0.033 & 0.001 & 0.000 & 0.018 & 0.001 & 0.001 & 0.001 \\
 & Homo: $Y_1$ and $Y_2$ (20\%, both) & -0.012 & 0.031 & -0.000 & 0.000 & 0.031 & 0.000 & 0.000 & -0.000 \\
 & Hete: $Y_1$ only (10\% control, 30\% treated) & -0.048 & -0.025 & 0.001 & 0.000 & -0.010 & 0.000 & 0.000 & -0.000 \\
 & Hete: $Y_2$ only (10\% control, 30\% treated) & -0.014 & -0.026 & -0.000 & 0.000 & -0.004 & -0.001 & -0.001 & -0.001 \\
 & Hete: $Y_1$ and $Y_2$ (10\% control, 30\% treated) & -0.047 & -0.023 & 0.002 & 0.000 & -0.011 & 0.001 & 0.001 & 0.002 \\
\addlinespace
\multirow{7}{*}{DOOR} & No missing data & 0.000 & 0.000 & 0.000 & 0.000 & 0.000 & 0.000 & 0.000 & 0.000 \\
 & Homo: $Y_1$ only (20\%, both) & -0.006 & 0.016 & 0.000 & 0.000 & 0.016 & 0.000 & 0.000 & 0.000 \\
 & Homo: $Y_2$ only (20\%, both) & 0.002 & 0.016 & 0.000 & 0.000 & 0.009 & 0.000 & 0.000 & 0.000 \\
 & Homo: $Y_1$ and $Y_2$ (20\%, both) & -0.006 & 0.016 & -0.000 & 0.000 & 0.016 & 0.000 & 0.000 & -0.000 \\
 & Hete: $Y_1$ only (10\% control, 30\% treated) & -0.024 & -0.013 & 0.000 & 0.000 & -0.005 & 0.000 & 0.000 & -0.000 \\
 & Hete: $Y_2$ only (10\% control, 30\% treated) & -0.007 & -0.013 & -0.000 & 0.000 & -0.002 & -0.000 & -0.000 & -0.000 \\
 & Hete: $Y_1$ and $Y_2$ (10\% control, 30\% treated) & -0.024 & -0.012 & 0.001 & 0.000 & -0.005 & 0.001 & 0.001 & 0.001 \\
\midrule
 & & \multicolumn{8}{c}{\textbf{RMSE}} \\
\cmidrule(lr){3-10}
\textbf{Estimand} & \makecell{\textbf{Missing data}\\\textbf{scenario}} & \textbf{Standard} & \textbf{CC} & \textbf{MI} & \multicolumn{2}{c}{\textbf{IPW}} & \multicolumn{3}{c}{\textbf{AIPW}} \\
\cmidrule(lr){6-7}\cmidrule(lr){8-10}
 & & & & & \textbf{A} & \textbf{B} & \textbf{A} & \textbf{B} & \textbf{C} \\
\midrule
\multirow{7}{*}{NB} & No missing data & 0.024 & 0.024 & 0.024 & 0.024 & 0.024 & 0.018 & 0.018 & 0.024 \\
 & Homo: $Y_1$ only (20\%, both) & 0.020 & 0.041 & 0.026 & 0.026 & 0.041 & 0.020 & 0.020 & 0.026 \\
 & Homo: $Y_2$ only (20\%, both) & 0.023 & 0.042 & 0.025 & 0.025 & 0.031 & 0.020 & 0.020 & 0.026 \\
 & Homo: $Y_1$ and $Y_2$ (20\%, both) & 0.020 & 0.041 & 0.026 & 0.026 & 0.041 & 0.020 & 0.020 & 0.027 \\
 & Hete: $Y_1$ only (10\% control, 30\% treated) & 0.051 & 0.037 & 0.026 & 0.027 & 0.030 & 0.021 & 0.021 & 0.027 \\
 & Hete: $Y_2$ only (10\% control, 30\% treated) & 0.026 & 0.037 & 0.025 & 0.025 & 0.026 & 0.019 & 0.019 & 0.025 \\
 & Hete: $Y_1$ and $Y_2$ (10\% control, 30\% treated) & 0.050 & 0.036 & 0.027 & 0.026 & 0.029 & 0.021 & 0.021 & 0.027 \\
\addlinespace
\multirow{7}{*}{DOOR} & No missing data & 0.012 & 0.012 & 0.012 & 0.012 & 0.012 & 0.009 & 0.009 & 0.012 \\
 & Homo: $Y_1$ only (20\%, both) & 0.010 & 0.021 & 0.013 & 0.013 & 0.021 & 0.010 & 0.010 & 0.013 \\
 & Homo: $Y_2$ only (20\%, both) & 0.011 & 0.021 & 0.013 & 0.013 & 0.016 & 0.010 & 0.010 & 0.013 \\
 & Homo: $Y_1$ and $Y_2$ (20\%, both) & 0.010 & 0.020 & 0.013 & 0.013 & 0.020 & 0.010 & 0.010 & 0.013 \\
 & Hete: $Y_1$ only (10\% control, 30\% treated) & 0.026 & 0.019 & 0.013 & 0.014 & 0.015 & 0.010 & 0.010 & 0.013 \\
 & Hete: $Y_2$ only (10\% control, 30\% treated) & 0.013 & 0.019 & 0.012 & 0.013 & 0.013 & 0.010 & 0.010 & 0.013 \\
 & Hete: $Y_1$ and $Y_2$ (10\% control, 30\% treated) & 0.025 & 0.018 & 0.013 & 0.013 & 0.015 & 0.011 & 0.011 & 0.014 \\
\bottomrule
\end{tabular}
\begin{tablenotes}\scriptsize
\item RMSE: root mean square error.
\item Homo: the two treatment groups have the same (homogeneous) marginal missing-data rate; Hete: the two treatment groups have different (heterogeneous) marginal missing-data rates.
\item Standard: missing comparisons are treated as ties; CC: complete-case analysis; MI: multiple imputation using 25 imputed data sets; IPW: H\'ajek-normalized inverse probability weighting; IPW-A: correctly specified missingness model; IPW-B: misspecified missingness model; AIPW-A: both missingness and outcome models correctly specified; AIPW-B: misspecified missingness model; AIPW-C: misspecified outcome model.
\end{tablenotes}
\end{table}

\begin{table}[H]
\centering
\footnotesize
\caption{Simulation results for NB and DOOR under Setting II: coverage probability and confidence-interval width.}
\label{tab:sim-NB-DOOR-II-cp-ciw}
\setlength{\tabcolsep}{2.8pt}
\renewcommand{\arraystretch}{1.02}
\begin{tabular}{l p{6.5cm} cccccccc}
\toprule
 & & \multicolumn{8}{c}{\textbf{CP}} \\
\cmidrule(lr){3-10}
\textbf{Estimand} & \makecell{\textbf{Missing data}\\\textbf{scenario}} & \textbf{Standard} & \textbf{CC} & \textbf{MI} & \multicolumn{2}{c}{\textbf{IPW}} & \multicolumn{3}{c}{\textbf{AIPW}} \\
\cmidrule(lr){6-7}\cmidrule(lr){8-10}
 & & & & & \textbf{A} & \textbf{B} & \textbf{A} & \textbf{B} & \textbf{C} \\
\midrule
\multirow{7}{*}{NB} & No missing data & 0.952 & 0.952 & 0.952 & 0.951 & 0.951 & 0.952 & 0.952 & 0.951 \\
 & Homo: $Y_1$ only (20\%, both) & \textbf{0.896} & \textbf{0.773} & 0.948 & 0.950 & \textbf{0.770} & 0.957 & 0.956 & \textbf{0.973} \\
 & Homo: $Y_2$ only (20\%, both) & 0.946 & \textbf{0.753} & 0.944 & 0.950 & \textbf{0.891} & 0.942 & 0.947 & 0.947 \\
 & Homo: $Y_1$ and $Y_2$ (20\%, both) & \textbf{0.897} & \textbf{0.764} & 0.958 & \textbf{0.961} & \textbf{0.787} & 0.951 & 0.951 & \textbf{0.972} \\
 & Hete: $Y_1$ only (10\% control, 30\% treated) & \textbf{0.185} & \textbf{0.846} & 0.946 & \textbf{0.936} & \textbf{0.919} & 0.947 & 0.948 & \textbf{0.964} \\
 & Hete: $Y_2$ only (10\% control, 30\% treated) & \textbf{0.897} & \textbf{0.835} & 0.951 & 0.954 & 0.947 & 0.953 & 0.957 & 0.959 \\
 & Hete: $Y_1$ and $Y_2$ (10\% control, 30\% treated) & \textbf{0.204} & \textbf{0.859} & 0.941 & 0.953 & \textbf{0.927} & 0.944 & 0.947 & \textbf{0.961} \\
\addlinespace
\multirow{7}{*}{DOOR} & No missing data & 0.952 & 0.952 & 0.952 & 0.951 & 0.951 & 0.952 & 0.952 & 0.951 \\
 & Homo: $Y_1$ only (20\%, both) & \textbf{0.896} & \textbf{0.773} & 0.948 & 0.950 & \textbf{0.770} & 0.957 & 0.956 & \textbf{0.973} \\
 & Homo: $Y_2$ only (20\%, both) & 0.946 & \textbf{0.753} & 0.944 & 0.950 & \textbf{0.891} & 0.942 & 0.947 & 0.947 \\
 & Homo: $Y_1$ and $Y_2$ (20\%, both) & \textbf{0.897} & \textbf{0.764} & 0.958 & \textbf{0.961} & \textbf{0.787} & 0.951 & 0.951 & \textbf{0.972} \\
 & Hete: $Y_1$ only (10\% control, 30\% treated) & \textbf{0.185} & \textbf{0.846} & 0.946 & \textbf{0.936} & \textbf{0.919} & 0.947 & 0.948 & \textbf{0.964} \\
 & Hete: $Y_2$ only (10\% control, 30\% treated) & \textbf{0.897} & \textbf{0.835} & 0.951 & 0.954 & 0.947 & 0.953 & 0.957 & 0.959 \\
 & Hete: $Y_1$ and $Y_2$ (10\% control, 30\% treated) & \textbf{0.204} & \textbf{0.859} & 0.941 & 0.953 & \textbf{0.927} & 0.944 & 0.947 & \textbf{0.961} \\
\midrule
 & & \multicolumn{8}{c}{\textbf{CIW}} \\
\cmidrule(lr){3-10}
\textbf{Estimand} & \makecell{\textbf{Missing data}\\\textbf{scenario}} & \textbf{Standard} & \textbf{CC} & \textbf{MI} & \multicolumn{2}{c}{\textbf{IPW}} & \multicolumn{3}{c}{\textbf{AIPW}} \\
\cmidrule(lr){6-7}\cmidrule(lr){8-10}
 & & & & & \textbf{A} & \textbf{B} & \textbf{A} & \textbf{B} & \textbf{C} \\
\midrule
\multirow{7}{*}{NB} & No missing data & 0.095 & 0.095 & 0.095 & 0.095 & 0.095 & 0.070 & 0.070 & 0.095 \\
 & Homo: $Y_1$ only (20\%, both) & 0.066 & 0.103 & 0.101 & 0.103 & 0.104 & 0.080 & 0.080 & 0.114 \\
 & Homo: $Y_2$ only (20\%, both) & 0.085 & 0.103 & 0.097 & 0.099 & 0.100 & 0.076 & 0.076 & 0.101 \\
 & Homo: $Y_1$ and $Y_2$ (20\%, both) & 0.066 & 0.103 & 0.101 & 0.103 & 0.104 & 0.080 & 0.080 & 0.114 \\
 & Hete: $Y_1$ only (10\% control, 30\% treated) & 0.066 & 0.105 & 0.102 & 0.103 & 0.106 & 0.081 & 0.081 & 0.112 \\
 & Hete: $Y_2$ only (10\% control, 30\% treated) & 0.085 & 0.105 & 0.097 & 0.099 & 0.101 & 0.076 & 0.077 & 0.101 \\
 & Hete: $Y_1$ and $Y_2$ (10\% control, 30\% treated) & 0.066 & 0.105 & 0.101 & 0.103 & 0.106 & 0.081 & 0.081 & 0.112 \\
\addlinespace
\multirow{7}{*}{DOOR} & No missing data & 0.047 & 0.047 & 0.047 & 0.047 & 0.047 & 0.035 & 0.035 & 0.047 \\
 & Homo: $Y_1$ only (20\%, both) & 0.033 & 0.052 & 0.051 & 0.052 & 0.052 & 0.040 & 0.040 & 0.057 \\
 & Homo: $Y_2$ only (20\%, both) & 0.043 & 0.052 & 0.049 & 0.049 & 0.050 & 0.038 & 0.038 & 0.050 \\
 & Homo: $Y_1$ and $Y_2$ (20\%, both) & 0.033 & 0.052 & 0.051 & 0.052 & 0.052 & 0.040 & 0.040 & 0.057 \\
 & Hete: $Y_1$ only (10\% control, 30\% treated) & 0.033 & 0.052 & 0.051 & 0.052 & 0.053 & 0.040 & 0.041 & 0.056 \\
 & Hete: $Y_2$ only (10\% control, 30\% treated) & 0.043 & 0.052 & 0.049 & 0.050 & 0.050 & 0.038 & 0.039 & 0.050 \\
 & Hete: $Y_1$ and $Y_2$ (10\% control, 30\% treated) & 0.033 & 0.052 & 0.051 & 0.052 & 0.053 & 0.040 & 0.041 & 0.056 \\
\bottomrule
\end{tabular}
\begin{tablenotes}\scriptsize
\item CP: coverage probability; CIW: average 95\% confidence-interval width. CP values outside the range [0.94, 0.96] are shown in bold.
\item Homo: the two treatment groups have the same (homogeneous) marginal missing-data rate; Hete: the two treatment groups have different (heterogeneous) marginal missing-data rates.
\item Standard: missing comparisons are treated as ties; CC: complete-case analysis; MI: multiple imputation using 25 imputed data sets; IPW: H\'ajek-normalized inverse probability weighting; IPW-A: correctly specified missingness model; IPW-B: misspecified missingness model; AIPW-A: both missingness and outcome models correctly specified; AIPW-B: misspecified missingness model; AIPW-C: misspecified outcome model.
\end{tablenotes}
\end{table}

\subsection{Simulation comparison of Horvitz--Thompson and H\'ajek weighting} \label{subapp:hajek}

We compare the finite-sample performance of the Horvitz--Thompson and H\'ajek versions of the proposed IPW estimator under two missingness scenarios from Simulation Setting~II: homogeneous missingness, with a 20\% missingness rate in both treatment groups, and heterogeneous missingness, with missingness rates of 30\% in the control group and 10\% in the treated group. IPW-A uses the correctly specified missingness model including $X_1$ and $X_2$, whereas IPW-B uses a misspecified model that omits $X_1$. Each scenario included sample size $n=1,000$ and $2,000$ Monte Carlo replicates.

\begin{table}[H]
\centering
\small
\caption{Simulation results comparing Horvitz--Thompson and H\'ajek weighting under selected scenarios from Setting II.}
\label{tab:hajek-settingII}
\setlength{\tabcolsep}{3.2pt}
\renewcommand{\arraystretch}{1.02}
\begin{tabular}{p{6cm} cccc}
\toprule
& \makecell{\textbf{IPW-A}\\Horvitz--Thompson}
& \makecell{\textbf{IPW-A}\\H\'ajek}
& \makecell{\textbf{IPW-B}\\Horvitz--Thompson}
& \makecell{\textbf{IPW-B}\\H\'ajek} \\
\midrule

& \multicolumn{4}{c}{\textbf{Bias}} \\
\cmidrule(lr){2-5}
Homo: both 20\%, WR
& 0.0082 & 0.0082 & 0.1526 & 0.1526 \\
Homo: both 20\%, WO
& 0.0038 & 0.0039 & 0.0825 & 0.0825 \\
Homo: both 20\%, NB
& 0.0002 & 0.0002 & 0.0316 & 0.0316 \\
Homo: both 20\%, DOOR
& 0.0001 & 0.0001 & 0.0158 & 0.0158 \\
Hete: control 30\%, treated 10\%, WR
& 0.0088 & 0.0088 & -0.0202 & -0.0202 \\
Hete: control 30\%, treated 10\%, WO
& 0.0045 & 0.0045 & -0.0214 & -0.0214 \\
Hete: control 30\%, treated 10\%, NB
& 0.0005 & 0.0005 & -0.0105 & -0.0105 \\
Hete: control 30\%, treated 10\%, DOOR
& 0.0002 & 0.0002 & -0.0052 & -0.0052 \\

\midrule
& \multicolumn{4}{c}{\textbf{RMSE}} \\
\cmidrule(lr){2-5}
Homo: both 20\%, WR
& 0.1399 & 0.1399 & 0.2239 & 0.2239 \\
Homo: both 20\%, WO
& 0.0906 & 0.0908 & 0.1277 & 0.1277 \\
Homo: both 20\%, NB
& 0.0374 & 0.0374 & 0.0491 & 0.0491 \\
Homo: both 20\%, DOOR
& 0.0187 & 0.0187 & 0.0245 & 0.0245 \\
Hete: control 30\%, treated 10\%, WR
& 0.1390 & 0.1390 & 0.1460 & 0.1460 \\
Hete: control 30\%, treated 10\%, WO
& 0.0906 & 0.0906 & 0.0937 & 0.0937 \\
Hete: control 30\%, treated 10\%, NB
& 0.0374 & 0.0374 & 0.0400 & 0.0400 \\
Hete: control 30\%, treated 10\%, DOOR
& 0.0187 & 0.0187 & 0.0200 & 0.0200 \\

\midrule
& \multicolumn{4}{c}{\textbf{CP}} \\
\cmidrule(lr){2-5}
Homo: both 20\%, WR
& \textbf{0.9625} & 0.9425 & \textbf{0.8515} & \textbf{0.8490} \\
Homo: both 20\%, WO
& \textbf{0.9620} & 0.9420 & \textbf{0.8745} & \textbf{0.8710} \\
Homo: both 20\%, NB
& \textbf{0.9615} & 0.9420 & \textbf{0.8700} & \textbf{0.8655} \\
Homo: both 20\%, DOOR
& \textbf{0.9615} & 0.9420 & \textbf{0.8700} & \textbf{0.8655} \\
Hete: control 30\%, treated 10\%, WR
& \textbf{0.9605} & 0.9450 & 0.9430 & 0.9410 \\
Hete: control 30\%, treated 10\%, WO
& \textbf{0.9630} & 0.9465 & \textbf{0.9395} & \textbf{0.9385} \\
Hete: control 30\%, treated 10\%, NB
& \textbf{0.9615} & 0.9460 & \textbf{0.9395} & \textbf{0.9380} \\
Hete: control 30\%, treated 10\%, DOOR
& \textbf{0.9615} & 0.9460 & \textbf{0.9395} & \textbf{0.9380} \\

\midrule
& \multicolumn{4}{c}{\textbf{CIW}} \\
\cmidrule(lr){2-5}
Homo: both 20\%, WR
& 0.5977 & 0.5405 & 0.6437 & 0.6365 \\
Homo: both 20\%, WO
& 0.3879 & 0.3524 & 0.3849 & 0.3809 \\
Homo: both 20\%, NB
& 0.1595 & 0.1456 & 0.1487 & 0.1471 \\
Homo: both 20\%, DOOR
& 0.0798 & 0.0728 & 0.0743 & 0.0736 \\
Hete: control 30\%, treated 10\%, WR
& 0.5849 & 0.5381 & 0.5633 & 0.5569 \\
Hete: control 30\%, treated 10\%, WO
& 0.3825 & 0.3518 & 0.3564 & 0.3528 \\
Hete: control 30\%, treated 10\%, NB
& 0.1580 & 0.1459 & 0.1512 & 0.1496 \\
Hete: control 30\%, treated 10\%, DOOR
& 0.0790 & 0.0729 & 0.0756 & 0.0748 \\

\bottomrule
\end{tabular}
\begin{tablenotes}\scriptsize
\item CP values outside the range $[0.94,0.96]$ are shown in bold. 
IPW-A uses the correctly specified missingness model, whereas IPW-B uses a misspecified model that omits $X_1$. Homo: the two treatment groups have the same (homogeneous) marginal missing-data rate; Hete: the two treatment groups have different (heterogeneous) marginal missing-data rates. 
\end{tablenotes}
\end{table}

Additional diagnostics confirmed that the H\'ajek estimates summed to one within each treatment group and hierarchy level up to numerical precision. For IPW-A, the maximum absolute deviation of the Horvitz--Thompson cell-probability sums from one was $5.54\times10^{-2}$ under homogeneous missingness and $2.06\times10^{-2}$ under heterogeneous missingness, whereas the corresponding maximum deviations under H\'ajek weighting were below $2.0\times10^{-15}$ and thus effectively zero. No propensity-score monotonicity violations or negative raw tie-probability estimates were observed in any simulation scenario. These results support the use of H\'ajek weighting as a normalization strategy that ensures within-layer coherence and, in the settings considered, improves finite-sample CI coverage. 

\subsection{Additional simulations for sensitivity to missingness assumption violation}
\label{subapp:miss-vio}

We conducted an additional simulation study to assess the sensitivity of the proposed estimators to violations of Assumption~\ref{asp:miss-pos}. In particular, we considered a setting in which the missingness of $Y_2$ depends on $Y_1$. 

The data generation of covariates, treatment and $Y_1$ is the same as Setting II in Section \ref{sec:simu} of the main text. The latent variable for $Y_2$ was generated as related to $Y_1$ by
\[
Y_2^*=\eta_2(A,\mb X)+0.9Y_1+\varepsilon_2,
\qquad \varepsilon_2\sim\mathrm{Logistic}(0,1),
\]
and was categorized using cutpoints $(-1,1)$. Thus, the two endpoints were associated.

Let $\mb X=(X_1,X_2)^\top$ denote the covariates used in the missingness models. For each treatment group $a\in\{0,1\}$, the first-layer observation probability was specified as
\[
\Pr(R_1=1\mid A=a,\mb X)
=
\operatorname{expit}\left\{
\alpha_{1a}+\boldsymbol{\gamma}_{1a}^{\top}\mb X
\right\},
\]
where
\[
\boldsymbol{\gamma}_{10}
=
\boldsymbol{\gamma}_{20}
=
(0.5,1)^\top,
\qquad
\boldsymbol{\gamma}_{11}
=
\boldsymbol{\gamma}_{21}
=
(1,1)^\top.
\]
Conditional on $R_1=1$, the second endpoint was observed according to
\[
\Pr(R_{12}=1\mid R_1=1,A=a,\mb X,Y_1)
=
\operatorname{expit}\left\{
\alpha_{2a}+\boldsymbol{\gamma}_{2a}^{\top}\mb X
+1.25Y_1
\right\}.
\]
Thus, after conditioning on the observed first endpoint, the probability of observing the second endpoint depends directly on $Y_1$. The intercepts $\alpha_{1a}$ were numerically calibrated so that
\[
\Pr(R_1=1\mid A=a)=1-p_{1a},
\]
whereas $\alpha_{2a}$ were calibrated so that
\[
\Pr(R_{12}=1\mid A=a)=1-p_{2a},
\]
with $(p_{1a},p_{2a})$ determined by the missingness scenario. 

We considered two missingness scenarios, homogeneous and heterogeneous missingness. In the homogeneous scenario, the marginal missingness rate for $Y_1$ was 10\% in both treatment groups and the cumulative missingness rate through $Y_2$ was 20\% in both groups. In the heterogeneous scenario, the corresponding rates were 20\% and 30\% in the
control group and 5\% and 10\% in the treated group. The win measure truth, obtained from a super population sample $n=2\times 10^8$, was
\[
(\mathrm{WR},\mathrm{NB},\mathrm{WO},\mathrm{DOOR})
=(1.26,\ 0.07,\ 1.15,\ 0.54).
\]

We compared the Standard, complete-case (CC), multiple-imputation (MI), IPW-A, IPW-B, AIPW-A, AIPW-B, and AIPW-C methods. The MI analysis used 25 imputed data sets, with treatment assignment, baseline covariates, and observed endpoint values included in the imputation models. The IPW and AIPW specifications were defined as in Setting~II; however, none of the fitted missingness models included the outcome-dependent term involving $Y_1$.

\begin{table}[H]
\centering
\scriptsize
\caption{Simulation results under outcome-dependent missingness.}
\label{tab:sim-miss-vio}
\setlength{\tabcolsep}{2.4pt}
\renewcommand{\arraystretch}{1.02}
\begin{tabular}{p{6.3cm} cccccccc}
\toprule
& \textbf{Standard} & \textbf{CC} & \textbf{MI}
& \textbf{IPW-A} & \textbf{IPW-B}
& \textbf{AIPW-A} & \textbf{AIPW-B} & \textbf{AIPW-C} \\
\midrule

& \multicolumn{8}{c}{\textbf{Bias}} \\
\cmidrule(lr){2-9}
Homo: $Y_1$ 10\%, cumulative 20\%, WR
& 0.155 & 0.197 & 0.012 & 0.003 & 0.130 & -0.003 & 0.006 & 0.003 \\
Homo: $Y_1$ 10\%, cumulative 20\%, WO
& 0.008 & 0.068 & 0.005 & 0.005 & 0.066 & 0.002 & 0.007 & 0.005 \\
Homo: $Y_1$ 10\%, cumulative 20\%, NB
& 0.003 & 0.027 & 0.001 & 0.000 & 0.026 & 0.000 & 0.002 & 0.001 \\
Homo: $Y_1$ 10\%, cumulative 20\%, DOOR
& 0.001 & 0.013 & 0.000 & 0.000 & 0.013 & 0.000 & 0.001 & 0.000 \\
Hete: control 30\%, treated 10\%, WR
& -0.055 & -0.152 & 0.016 & 0.020 & -0.025 & 0.019 & 0.017 & 0.019 \\
Hete: control 30\%, treated 10\%, WO
& -0.074 & -0.098 & 0.007 & 0.014 & -0.019 & 0.014 & 0.012 & 0.014 \\
Hete: control 30\%, treated 10\%, NB
& -0.034 & -0.046 & 0.002 & 0.004 & -0.010 & 0.005 & 0.004 & 0.004 \\
Hete: control 30\%, treated 10\%, DOOR
& -0.017 & -0.023 & 0.001 & 0.002 & -0.005 & 0.003 & 0.002 & 0.002 \\

\midrule
& \multicolumn{8}{c}{\textbf{RMSE}} \\
\cmidrule(lr){2-9}
Homo: $Y_1$ 10\%, cumulative 20\%, WR
& 0.240 & 0.286 & 0.147 & 0.150 & 0.220 & 0.116 & 0.114 & 0.150 \\
Homo: $Y_1$ 10\%, cumulative 20\%, WO
& 0.066 & 0.115 & 0.082 & 0.088 & 0.114 & 0.068 & 0.067 & 0.088 \\
Homo: $Y_1$ 10\%, cumulative 20\%, NB
& 0.028 & 0.046 & 0.035 & 0.037 & 0.046 & 0.029 & 0.028 & 0.037 \\
Homo: $Y_1$ 10\%, cumulative 20\%, DOOR
& 0.014 & 0.023 & 0.018 & 0.019 & 0.023 & 0.015 & 0.014 & 0.019 \\
Hete: control 30\%, treated 10\%, WR
& 0.164 & 0.216 & 0.146 & 0.150 & 0.157 & 0.118 & 0.117 & 0.150 \\
Hete: control 30\%, treated 10\%, WO
& 0.094 & 0.127 & 0.082 & 0.088 & 0.089 & 0.070 & 0.069 & 0.088 \\
Hete: control 30\%, treated 10\%, NB
& 0.043 & 0.060 & 0.035 & 0.037 & 0.040 & 0.030 & 0.029 & 0.037 \\
Hete: control 30\%, treated 10\%, DOOR
& 0.022 & 0.030 & 0.018 & 0.019 & 0.020 & 0.015 & 0.015 & 0.019 \\

\midrule
& \multicolumn{8}{c}{\textbf{CP}} \\
\cmidrule(lr){2-9}
Homo: $Y_1$ 10\%, cumulative 20\%, WR
& \textbf{0.842} & \textbf{0.795} & 0.941 & 0.953 & \textbf{0.863} & 0.950 & 0.952 & 0.953 \\
Homo: $Y_1$ 10\%, cumulative 20\%, WO
& \textbf{0.936} & \textbf{0.852} & \textbf{0.939} & 0.955 & \textbf{0.867} & 0.953 & 0.951 & 0.958 \\
Homo: $Y_1$ 10\%, cumulative 20\%, NB
& \textbf{0.935} & \textbf{0.850} & \textbf{0.938} & 0.955 & \textbf{0.865} & 0.953 & 0.951 & 0.958 \\
Homo: $Y_1$ 10\%, cumulative 20\%, DOOR
& \textbf{0.935} & \textbf{0.850} & \textbf{0.938} & 0.955 & \textbf{0.865} & 0.953 & 0.951 & 0.958 \\
Hete: control 30\%, treated 10\%, WR
& \textbf{0.904} & \textbf{0.792} & 0.948 & 0.954 & \textbf{0.929} & 0.944 & 0.946 & 0.955 \\
Hete: control 30\%, treated 10\%, WO
& \textbf{0.724} & \textbf{0.727} & 0.949 & 0.954 & \textbf{0.929} & 0.943 & 0.944 & 0.956 \\
Hete: control 30\%, treated 10\%, NB
& \textbf{0.724} & \textbf{0.727} & 0.948 & 0.953 & \textbf{0.929} & 0.942 & 0.943 & 0.955 \\
Hete: control 30\%, treated 10\%, DOOR
& \textbf{0.724} & \textbf{0.727} & 0.948 & 0.953 & \textbf{0.929} & 0.942 & 0.943 & 0.955 \\

\midrule
& \multicolumn{8}{c}{\textbf{CIW}} \\
\cmidrule(lr){2-9}
Homo: $Y_1$ 10\%, cumulative 20\%, WR
& 0.666 & 0.731 & 0.551 & 0.594 & 0.646 & 0.442 & 0.442 & 0.595 \\
Homo: $Y_1$ 10\%, cumulative 20\%, WO
& 0.240 & 0.332 & 0.311 & 0.347 & 0.344 & 0.260 & 0.258 & 0.348 \\
Homo: $Y_1$ 10\%, cumulative 20\%, NB
& 0.103 & 0.135 & 0.133 & 0.148 & 0.140 & 0.112 & 0.111 & 0.149 \\
Homo: $Y_1$ 10\%, cumulative 20\%, DOOR
& 0.051 & 0.067 & 0.067 & 0.074 & 0.070 & 0.056 & 0.055 & 0.074 \\
Hete: control 30\%, treated 10\%, WR
& 0.569 & 0.551 & 0.556 & 0.599 & 0.576 & 0.452 & 0.450 & 0.598 \\
Hete: control 30\%, treated 10\%, WO
& 0.215 & 0.291 & 0.315 & 0.349 & 0.325 & 0.265 & 0.262 & 0.350 \\
Hete: control 30\%, treated 10\%, NB
& 0.099 & 0.138 & 0.134 & 0.149 & 0.143 & 0.113 & 0.112 & 0.149 \\
Hete: control 30\%, treated 10\%, DOOR
& 0.050 & 0.069 & 0.067 & 0.074 & 0.071 & 0.056 & 0.056 & 0.074 \\

\bottomrule
\end{tabular}
\begin{tablenotes}\scriptsize
\item CP values outside the range $[0.94,0.96]$ are shown in bold. 
Homo: 10\% missingness for $Y_1$ and 20\% cumulative missingness through $Y_2$ in both treatment groups. Hete: corresponding rates of 20\% and 30\% in the control group and 5\% and 10\% in the treated group. Standard: missing comparisons are treated as ties; CC: complete-case analysis; MI: multiple imputation using 25 imputed data sets; IPW-A: model including $(X_1,X_2)$; IPW-B: reduced model including $X_2$ only; AIPW-A, AIPW-B, and AIPW-C: corresponding AIPW specifications. 
\end{tablenotes}
\end{table}

\section{Additional Data Analysis Results}\label{app:diag-data}

\subsection{Non-missingness propensity score distributions by treatment group}\label{subapp:data-ps}

\begin{figure}[H]
    \centering
    \includegraphics[width=\linewidth]{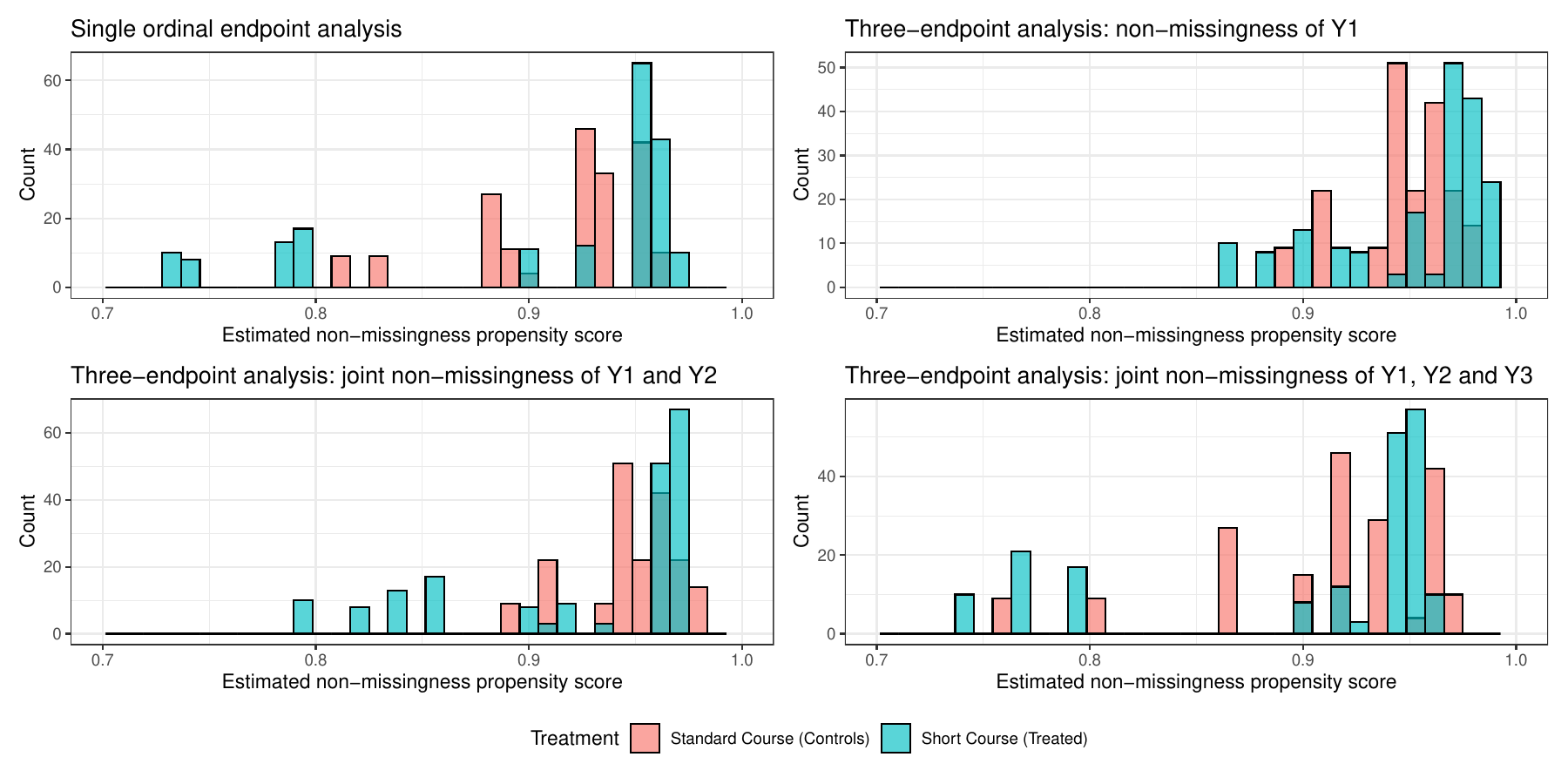}
    \caption{Histograms of non-missingness propensity scores (using covariates for modeling) by treatment groups in SCOUT-CAP trial data analysis.}
    \label{fig:ps-scoutcap}
\end{figure}

\begin{figure}[H]
    \centering
    \includegraphics[width=0.7\linewidth]{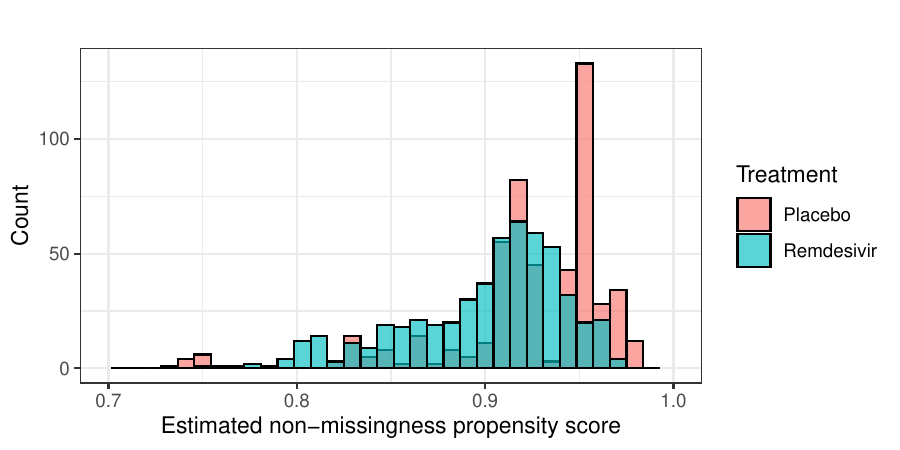}
    \caption{Histogram of the estimated non-missingness propensity scores (using covariates for modeling) by treatment group in ACTT-1 trial data analysis. }
    \label{fig:ps-actt1}
\end{figure}

\subsection{Sensitivity analyses for the ACTT-1 AIPW estimator}\label{subapp:actt1-aipw}

The covariate-adjusted AIPW analysis of ACTT-1 produced a more attenuated win measure estimate than the corresponding IPW analysis. To further investigate this difference, we conducted two sets of additional analyses. First, we examined the sensitivity of the ACTT-1 results to alternative outcome categorizations and outcome-regression specifications. Second, we conducted an ACTT-1-motivated simulation study to assess whether appreciable differences between the IPW and AIPW point estimates can arise in a single finite sample even when both nuisance models are correctly specified.

\subsubsection{Empirical sensitivity analyses}\label{subsubapp:actt1-aipw}

We examined whether the difference between the IPW and AIPW estimates in ACTT-1 could be explained by sparse outcome categories or the specification of the outcome regression (Table \ref{tab:actt1-wr-sensitivity}). Collapsing categories 3--4 or 3--7 had little impact on either estimator. For AIPW, using only baseline severity indicators in the outcome model yielded an estimated win ratio of 1.231, close to the primary estimate of 1.217, whereas excluding baseline severity and retaining only age, sex, and race increased the estimate to 1.318. These results suggest that baseline severity was an important contributor to the difference between the IPW and AIPW estimates.

\begin{table}[H]
\centering
\caption{Sensitivity analyses for the win ratio in the ACTT-1 application.}
\label{tab:actt1-wr-sensitivity}
\begin{tabular}{lcccc}
\toprule
\textbf{Analysis} & \textbf{Estimate} & \textbf{SE} & \textbf{95\% CI} & \textbf{p-value} \\
\midrule
IPW, original categories, full covariates
& 1.343 & 0.163 & (1.060, 1.704) & 0.015 \\ 
AIPW, original categories, full outcome model
& 1.217 & 0.138 & (0.974, 1.520) & 0.084 \\
\addlinespace
IPW, categories 3--4 collapsed
& 1.344 & 0.161 & (1.062, 1.700) & 0.014 \\
AIPW, categories 3--4 collapsed
& 1.215 & 0.130 & (0.986, 1.499) & 0.068 \\
IPW, categories 3--7 collapsed
& 1.349 & 0.169 & (1.056, 1.725) & 0.017 \\
AIPW, categories 3--7 collapsed
& 1.218 & 0.137 & (0.976, 1.518) & 0.080 \\
\addlinespace
AIPW, severity-only outcome model
& 1.231 & 0.138 & (0.989, 1.533) & 0.063 \\
AIPW, other-covariates-only outcome model
& 1.318 & 0.154 & (1.048, 1.656) & 0.018 \\
\bottomrule
\end{tabular}
\begin{tablenotes}
\footnotesize
\item The missingness propensity-score model was fixed at the full-covariate specification in the outcome-model sensitivity analyses.
\end{tablenotes}
\end{table}

Baseline severity was also strongly associated with the observed day-27 ordinal outcome in both treatment groups. Table~\ref{tab:actt1-severity-outcome} shows that the distribution of the day-27 outcome shifted markedly across baseline severity levels, with progressively lower proportions of participants in the best outcome category as baseline severity increased. Table~\ref{tab:actt1-severity-association} summarizes the corresponding association tests, which were significant in both the control group ($\chi^2=53.919$, $df=18$, $p<0.001$) and the treated group ($\chi^2=118.250$, $df=18$, $p<0.001$), with Cram\'er's $V$ values of 0.194 and 0.287, respectively. Overall, these diagnostics indicate that the observed IPW--AIPW difference was not driven by sparse outcome categories, but was closely related to the prognostic information contributed by baseline severity.

\begin{table}[H]
\centering
\caption{Cross-tabulation of baseline severity score and observed day-27 ordinal outcome in ACTT-1, stratified by treatment group.}
\label{tab:actt1-severity-outcome}
\small
\begin{tabular}{lrrrrrrr}
\toprule
& \multicolumn{7}{c}{\textbf{Day-27 ordinal outcome, $n$ (\%)}} \\
\cmidrule(lr){2-8}
\textbf{Baseline severity} & $8$ & $7$ & $6$ & $5$ & $4$ & $3$ & $0$ \\
\midrule
\multicolumn{8}{l}{\textit{Control}} \\
4 & 3 (5.7) & 1 (1.9) & 0 (0.0) & 1 (1.9) & 2 (3.8) & 1 (1.9) & 45 (84.9) \\
5 & 24 (12.8) & 8 (4.3) & 3 (1.6) & 7 (3.7) & 7 (3.7) & 2 (1.1) & 137 (72.9) \\
6 & 20 (21.3) & 6 (6.4) & 4 (4.3) & 5 (5.3) & 3 (3.2) & 0 (0.0) & 56 (59.6) \\
7 & 29 (20.1) & 26 (18.1) & 6 (4.2) & 9 (6.2) & 7 (4.9) & 0 (0.0) & 67 (46.5) \\
\addlinespace
\multicolumn{8}{l}{\textit{Treated}} \\
4 & 3 (4.8) & 0 (0.0) & 0 (0.0) & 0 (0.0) & 1 (1.6) & 2 (3.2) & 57 (90.5) \\
5 & 9 (4.2) & 7 (3.3) & 0 (0.0) & 5 (2.3) & 6 (2.8) & 2 (0.9) & 185 (86.4) \\
6 & 18 (21.7) & 4 (4.8) & 1 (1.2) & 5 (6.0) & 7 (8.4) & 0 (0.0) & 48 (57.8) \\
7 & 28 (23.3) & 21 (17.5) & 4 (3.3) & 10 (8.3) & 5 (4.2) & 0 (0.0) & 52 (43.3) \\
\bottomrule
\end{tabular}
\begin{tablenotes}
\footnotesize
\item Entries are numbers of participants with observed day-27 outcomes, with row percentages in parentheses. Percentages are calculated within each baseline severity level and treatment group. Larger values of the reverse-coded day-27 ordinal outcome indicate better outcomes.
\end{tablenotes}
\end{table}

\begin{table}[H]
\centering
\caption{Association between baseline severity score and observed day-27 ordinal outcome in ACTT-1.}
\label{tab:actt1-severity-association}
\begin{tabular}{lrrrr}
\toprule
\textbf{Treatment group} & \textbf{$N$} & \textbf{$\chi^2$} & \textbf{p-value} & \textbf{Cram\'er's $V$} \\
\midrule
Control & 479 & 53.919  & $<0.001$ & 0.194 \\
Treated & 480 & 118.250 & $<0.001$ & 0.287 \\
\bottomrule
\end{tabular}
\end{table}

\subsubsection{ACTT-1-motivated simulation}

To assess whether the discrepancy between the IPW and AIPW estimates observed in ACTT-1 could plausibly arise from finite-sample variation, we conducted an additional diagnostic simulation motivated by the ACTT-1 application. Our preceding diagnostics suggested that baseline severity was strongly prognostic of the ordinal outcome and was also the only baseline covariate with a statistically significant coefficient in the fitted missingness model. We therefore generated one severity-like baseline covariate that played a dominant role in both the outcome and missingness mechanisms.

We generated a randomized treatment indicator $A\sim\operatorname{Bernoulli}(0.5)$ and four baseline covariates. To mimic the baseline severity score in ACTT-1, we generated $X_1\in\{4,5,6,7\}$ with probabilities $(0.15,0.35,0.35,0.15)$ and defined $X_{1c}=X_1-5.5$. The remaining covariates were generated as $X_2=|Z_2|$ with $Z_2\sim N(0,6^2)$, $X_3\sim\operatorname{Binomial}(6,0.15)$, and $X_4=|Z_4|$ with $Z_4\sim N(0,8^2)$. A three-category ordinal outcome $Y\in\{1,2,3\}$ was generated from
\[
\log\left\{\frac{\Pr(Y=2\mid A,\mb X)}{\Pr(Y=1\mid A,\mb X)}\right\}
=
-0.6+0.5A+2.5X_{1c}+0.05X_2+0.005X_3+0.50X_4,
\]
and
\[
\log\left\{\frac{\Pr(Y=3\mid A,\mb X)}{\Pr(Y=1\mid A,\mb X)}\right\}
=
-0.2+0.5A+2.5X_{1c}+0.01X_2+0.050X_3-0.10X_4.
\]

Missingness depended only on the severity-like covariate:
\[
\Pr(R=1\mid A=a,\mb X)
=
\operatorname{expit}\left(\alpha_a-0.5X_{1c}\right),
\]
where $\alpha_0$ and $\alpha_1$ were calibrated to yield approximately 92.5\% and 90.0\% outcome observation rates, following ACTT-1. In estimation, the treatment-specific missingness models included all four covariates $X_1$--$X_4$. Thus, the fitted missingness model was overparameterized relative to the data-generating mechanism but remained correctly specified because the true model was nested within it. The AIPW estimator additionally used correctly specified treatment-specific multinomial outcome regressions including all four covariates.

We generated 1,000 independent datasets of size $n=1{,}050$, approximately matching the ACTT-1 sample size. For each replicate, we computed the H\'ajek IPW and AIPW estimators of $\theta_{\mathrm{WR}}$. We first examined the replicate-specific discrepancy
\[
\left|
\widehat{\theta}_{\mathrm{WR}}^{\mathrm{ipw}}
-
\widehat{\theta}_{\mathrm{WR}}^{\mathrm{aipw}}
\right|.
\]

\begin{figure}[H]
\centering
\begin{subfigure}[t]{0.48\textwidth}
    \centering
    \includegraphics[width=\textwidth]{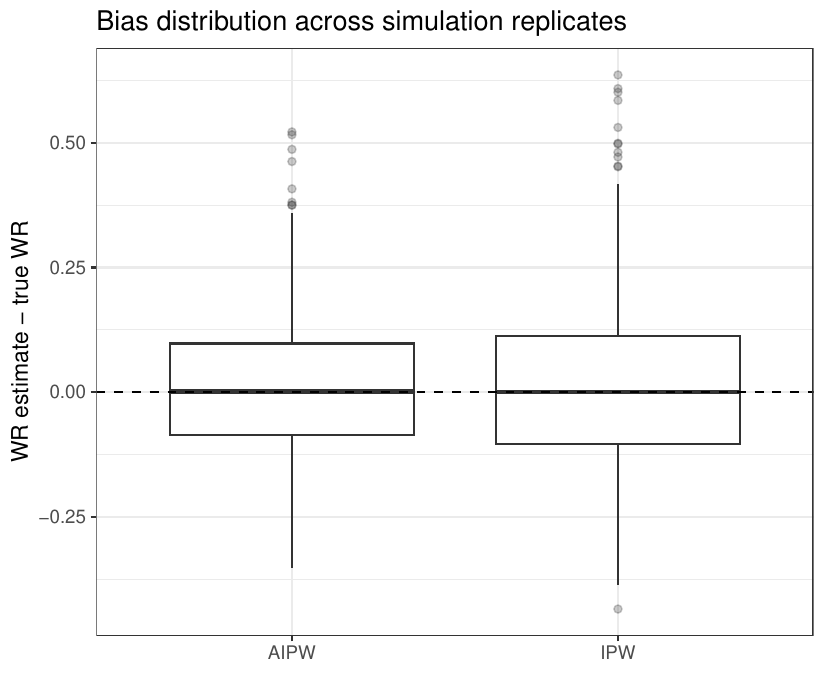}
    \caption{Bias distributions of the IPW and AIPW estimators.}
    \label{fig:actt1-sim-bias}
\end{subfigure}
\hfill
\begin{subfigure}[t]{0.48\textwidth}
    \centering
    \includegraphics[width=\textwidth]{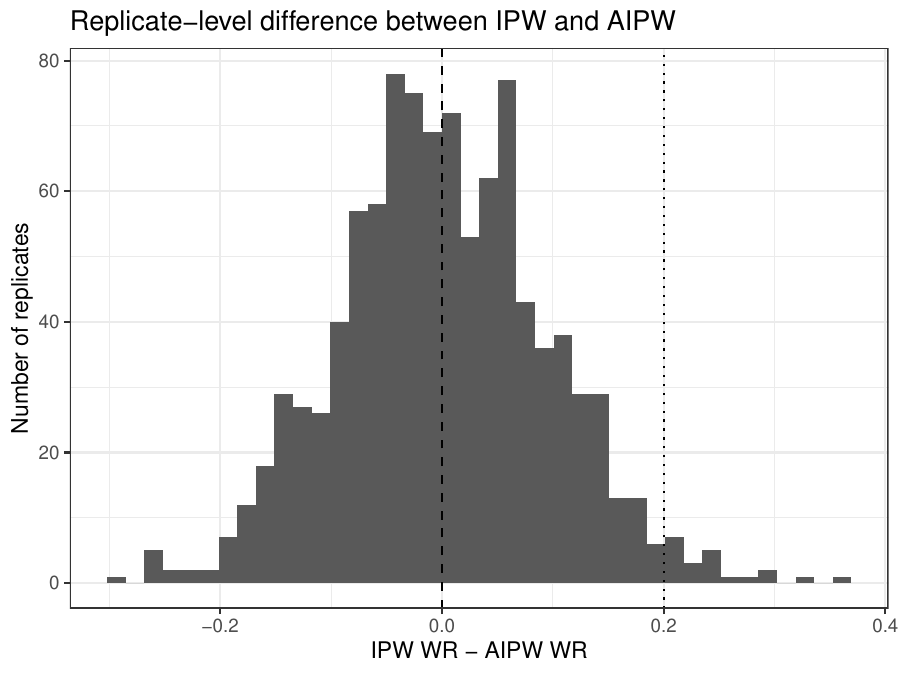}
    \caption{Replicate-specific difference between the IPW and AIPW estimators.}
    \label{fig:actt1-sim-difference}
\end{subfigure}
\caption{Finite-sample behavior of the IPW and AIPW estimators of $\theta_{\mathrm{WR}}$ in the ACTT-1-motivated simulation.}
\label{fig:actt1-sim-diagnostics}
\end{figure}

\begin{table}[H]
\centering
\caption{Replicate-specific differences between the IPW and AIPW estimators of $\theta_{\mathrm{WR}}$ in the ACTT-1-motivated simulation.}
\label{tab:actt1-sim-difference}
\begin{tabular}{lc}
\toprule
\textbf{Summary} & \textbf{Value} \\
\midrule
$\Pr\!\left(\left|\widehat{\theta}_{\mathrm{WR}}^{\mathrm{ipw}}-\widehat{\theta}_{\mathrm{WR}}^{\mathrm{aipw}}\right|\geq 0.10\right)$ & 0.285 \\
$\Pr\!\left(\left|\widehat{\theta}_{\mathrm{WR}}^{\mathrm{ipw}}-\widehat{\theta}_{\mathrm{WR}}^{\mathrm{aipw}}\right|\geq 0.15\right)$ & 0.106 \\
$\Pr\!\left(\left|\widehat{\theta}_{\mathrm{WR}}^{\mathrm{ipw}}-\widehat{\theta}_{\mathrm{WR}}^{\mathrm{aipw}}\right|\geq 0.20\right)$ & 0.033 \\
\bottomrule
\end{tabular}
\end{table}

The simulation shows that both estimators are unbiased on average. But at the individual level, the absolute difference bewteen the two estimator can exceed 0.10 in 28.5\% of the simulated data and exceeded 0.15 in 10.6\% of the simulated data. This implies that difference of 0.1 or larger can be observed in a single data set with 28.5\% chance even though they are estimating the same true WR.

We next examined whether the overall estimation behavior was comparable to that observed in ACTT-1. The true WR, obtained by large Monte Carlo integration, was $\theta_{\mathrm{WR}}=1.219$. The simulation performance is summarized in Table~\ref{tab:actt1-sim-performance}.

\begin{table}[H]
\centering
\caption{Simulation performance of the IPW and AIPW estimators of $\theta_{\mathrm{WR}}$ in the ACTT-1-motivated setting.}
\label{tab:actt1-sim-performance}
\begin{tabular}{lcccccc}
\toprule
\textbf{Method} & \textbf{Truth} & \textbf{Mean} & \textbf{Bias} & \textbf{Empirical SD} & \textbf{Mean SE} & \textbf{Coverage} \\
\midrule
IPW  & 1.219 & 1.229 & 0.010 & 0.165 & 0.164 & 0.951 \\
AIPW & 1.219 & 1.227 & 0.008 & 0.135 & 0.132 & 0.950 \\
\bottomrule
\end{tabular}
\begin{tablenotes}
\footnotesize
\item Results are based on 1,000 simulation replicates with $n=1{,}050$ participants per replicate. Coverage denotes the empirical coverage probability of the nominal 95\% confidence interval.
\end{tablenotes}
\end{table}

Both estimators had small finite-sample bias and approximately nominal coverage. Furthermore, the empirical standard deviations and mean estimated standard errors were close to the uncertainty observed in ACTT-1: approximately 0.16 for IPW and 0.13 for AIPW. These results indicate that, although the IPW and AIPW estimators target the same true WR, there is a 28.5\% probability that their estimates differ by at least 0.1 in a single sample. The AIPW estimator is more efficient, as its outcome regression component accounts for the strong association between a predictive baseline covariate and the outcome. 

\end{document}